\documentclass[modern, a4paper, times]{aastex62}
\usepackage[english]{babel}
\usepackage{amsmath}
\usepackage{graphicx}
\usepackage{natbib}

\submitjournal{ApJ}

\shorttitle{Synthesis of the distribution function of ISN He} 

\begin{document}

\title{Distribution function of neutral helium outside and inside the heliopause}

\correspondingauthor{M. Bzowski}
\email{bzowski@cbk.waw.pl} 

\author[0000-0002-5204-9645]{M. A. Kubiak}
\affil{Space Research Centre PAS (CBK PAN), Bartycka 18a, 00-716 Warsaw, Poland}

\author[0000-0003-3957-2359]{M. Bzowski}
\affil{Space Research Centre PAS (CBK PAN), Bartycka 18a, 00-716 Warsaw, Poland}

\author[0000-0002-4173-3601]{J.M. Sok{\'o}{\l}}
\affil{Space Research Centre PAS (CBK PAN), Bartycka 18a, 00-716 Warsaw, Poland}

\newcommand{\ccm}{~cm$^{-3}$}
\newcommand{\kms}{~km~s$^{-1}$}

\begin{abstract}
Interaction of the solar wind with interstellar matter involves, among other, charge exchange between interstellar neutral atoms and plasma, which results in the creation of secondary population of interstellar neutral (ISN) atoms. The secondary population of interstellar He was detected by Interstellar Boundary Explorer (IBEX), but interpretation of these measurements was mostly based on an approximation that the primary interstellar neutral population and the secondary population were non-interacting homogeneous Maxwell-Boltzmann functions in the outer heliosheath. We simulate the distribution function in the outer heliosheath and inside the heliopause using method of characteristics with statistical weights obtained from solutions of the production and loss equations for the secondary atoms due to charge-exchange collisions in the outer heliosheath. We show that the two-Maxwellian approximation for the distribution function of neutral He is not a good approximation within the outer heliosheath but a reasonable one inside the termination shock. This is due to a strong selection effect: the He atoms able to penetrate inside the termination shock are a small, peculiar subset of the entire secondary He population. Nevertheless, the two-Maxwellian approximation reproduces the density distribution of ISN He inside the termination shock well  and enables a realistic reproduction of the orientation of the plane defined by the Sun's velocity vector through the local interstellar matter and the vector of unperturbed interstellar magnetic field. 
\end{abstract}

\keywords{ISM: ions -- ISM: atoms, ISMS: clouds -- ISM: magnetic fields -- local interstellar matter -- Sun: heliosphere -- ISM: kinematics and dynamics}

\section{Introduction}
\label{sec:intro}
The heliosphere is created due to interaction between the solar wind and interstellar plasma, mediated by charge-exchange reaction between the plasma and interstellar neutral (ISN) atoms \citep{baranov_malama:93, zank:99}. The boundary of the heliosphere is the heliopause, which is unpenetrable for ions but transparent for atoms \citep{suess:90a}. As the Sun is moving through interstellar matter, a wake of perturbed plasma is formed, conventionally called the outer heliosheath (OHS), and the plasma flow becomes detached from the flow of ISN gas. As a result, charge exchange reactions between interstellar ions and atoms in the OHS create the so-called secondary population of ISN atoms. These atoms inherit the kinematic properties of their parent ions, but since they are not tied by the ambient magnetic field, they can travel large distances (on the order of 100~au) without interaction. 

Interstellar gas is mostly composed of hydrogen and helium, with relatively small admixtures of heavier species. With the total number abundance of He$^+$/H$^+$ about 0.17 and He/H about 0.10 \citep{bzowski_etal:19a}, He is an important component of the local interstellar matter, well suited to the role of information carrier from the OHS to spaceborne detectors at 1~au from the Sun. In fact, the currently most reliable estimates of the Sun's velocity vector and the temperature of the very local interstellar matter (VLISM) were obtained based on observations of ISN He from the IBEX-Lo detector \citep{bzowski_etal:15a, mccomas_etal:15b, swaczyna_etal:18a} onboard Interstellar Boundary Explorer \citep[IBEX; ][]{mccomas_etal:09a, fuselier_etal:09b, mobius_etal:09a}. IBEX observations resulted in detection of the secondary component of ISN He \citep{bzowski_etal:12a, kubiak_etal:14a}, dubbed the Warm Breeze. \citet{kubiak_etal:16a} analyzed IBEX observations of the Warm Breeze within a model of ISN He in the OHS constructed as a superposition of two non-interacting, collisionless Maxwell-Boltzmann populations with different densities, flow vectors, and temperatures. One of them corresponded to the unperturbed ISN He (the so-called primary population of ISN He), and the other one to the Warm Breeze population. These authors set the source region for these populations at 150~au ahead of the Sun. Beyond this distance, the parameters of these two populations were assumed invariable with time and homogeneous in space. With this, \citet{kubiak_etal:16a} fitted the inflow direction and speed, temperature, and density of the Warm Breeze using the nWTPM simulation model \citep{sokol_etal:15b} and data analysis method from \citet{swaczyna_etal:15a}. 

The magnitude of reduced chi-square obtained from the best-fit model was statistically too high, suggesting that the interpretation model was not fully adequate. Nevertheless, the velocity vector of the Warm Breeze turned out to be within the plane defined by the Sun's velocity vector and the vector of interstellar magnetic field (the B-V plane), determined by \citet{zirnstein_etal:16b} by fitting the center and radius of the IBEX Ribbon \citep{funsten_etal:09b,mccomas_etal:09b, funsten_etal:13a}. Hybrid MHD-kinetic models of the heliosphere interaction with interstellar matter suggest that the secondary component of ISN gas flows from the OHS towards the Sun from a direction contained in the B-V plane \citep{izmodenov_alexashov:15a, heerikhuisen_etal:10a}. Because of this, the finding by \citet{kubiak_etal:16a} was considered as a decisive evidence that the Warm Breeze is, in fact, the secondary population of ISN He. As shown already by \citet{bzowski_etal:12a}, the reaction responsible for the creation of this population is charge exchange between ISN He atoms and interstellar He$^+$ ions in the OHS. Due to the resonant character of this reaction, the cross section is much larger than the cross section for the non-resonant charge exchange reaction between the more plentiful ISN H atoms and He$^+$ ions. 

However, as pointed out by \citet{bzowski_etal:17a}, the model with two homogenous, non-interacting Maxwell-Boltzmann populations of neutral He in the OHS does not seem to be realistic. In particular, it provides neither physical relation between these two populations nor between the secondary population and the source population of the OHS plasma. \citet{bzowski_etal:17a} demonstrated  that the neutral He signal observed by IBEX-Lo can be qualitatively reproduced when one abandons the two-Maxwellian model and adopts a physically more realistic scenario. In their approach, a plasma flow around the heliopause resulting from a MHD-kinetic model of the heliosphere is adopted, with the relevant parameter values for the MHD model taken from observations. The neutral He signal observed by IBEX is then qualitatively reproduced when only one population of ISN He at more than 1000~au from the Sun is assumed. The signal is created by atoms that are produced by charge-exchange collisions between He$^+$ ions perturbed in the OHS and the unperturbed ISN He atoms and injected into trajectories leading them to the IBEX detector. Statistical weights of the atoms forming the signal are obtained by solving the production and loss balance equation along these trajectories, and the signal is integrated similarly as it was done by \citet{bzowski_etal:15a}. Throughout our paper, this method will be referred to as the distribution function synthesis method. 

Based on this insight, \citet{bzowski_etal:19a} pointed out that the simulated signal depends on the assumed density of interstellar He$^+$ in the unperturbed VLISM and fitted this density to IBEX-Lo observations while keeping the other parameters of the model unchanged and identical to those used by \citet{zirnstein_etal:16b}. With this, they obtained a fit with reduced chi-square value lower than \citet{kubiak_etal:16a}, even though the model used by \citet{bzowski_etal:19a} has fewer free parameters. This suggests that the distribution function synthesis method pioneered by \citet{bzowski_etal:17a} and continued by \citet{bzowski_etal:19a} better reproduces the physical reality than the two-Maxwellian model used by \citet{kubiak_etal:16a}.

Nevertheless, the approximation of superposition of two non-interacting Maxwellian functions for the distribution function of He in the OHS was shown to provide a reasonably good approximation for the IBEX-Lo signal, and the parameters of the Warm Breeze (secondary ISN He) population obtained from fitting this model provide a reasonable approximation for the orientation of the B-V plane in space. Here, we investigate why the two-Maxwellian approximation applied to IBEX-Lo signal is working so well. Using the methodology developed by \citet{bzowski_etal:17a} to simulate the interaction of ISN He with plasma in the OHS, we synthesize the distribution function of ISN He in selected locations in the OHS and compare the results with the two-Maxwellian approximation. Subsequently, we continue the comparisons for locations increasingly closer to the Sun, ending up at 1~au. Finally, we synthesize the distribution function of ISN He precisely for the IBEX observation conditions for selected IBEX orbits and we illustrate that the agreement between the two mentioned approaches is due to selection effects: the He atoms able to reach the IBEX-Lo detector are a small, peculiar subset of the population of ISN He created in the OHS.

\section{Modeling the distribution function of ISN He}
\label{sec:modelDifu}
Modeling of the distribution function of neutral He in the OHS and inside the heliosphere is carried out based on an appropriately modified paradigm of the hot model of ISN gas \citep{fahr:78, thomas:78} using the method of characteristics. The method of characteristics is a method used to simulate the distribution function of interstellar gas in the heliosphere since the introduction of the hot model, however the term``method of characteristics'' has seldom been used in this context. Mathematically, this is a method of solving partial differential equations (PDEs)  that involves reducing a PDE to a family of ordinary differential equations, along which the solutions can be integrated from certain initial conditions given on a suitable hypersurface.

Here, we adapt the methodology used to simulate the flux of neutral He observed by IBEX-Lo, developed by \citet{bzowski_etal:17a} and employed by \citet{bzowski_etal:19a}. 

In this method, which we call the synthesis metod, the local distribution function $f_\text{loc}(\vec{r}, \vec{v}, t)$ of neutral He for a time $t$ at a location given by a heliocentric radius-vector $\vec{r}$ for a velocity $\vec{v}$ is given by:
\begin{equation}
f_\text{loc}(\vec{r}, \vec{v}, t) = f_\text{MB}\left(\vec{v}_\text{lim}(\vec{q}_\text{loc}), n_\text{VLISM}, \vec{u}_\text{VLISM}, T_\text{VLISM}  \right)\, \omega\left(\vec{q}_\text{loc}, r_\text{lim}, t\right)
\label{eq:locDiFu}
\end{equation}
where $\vec{q}_\text{loc} = (\vec{r}_\text{loc}, \vec{v}_\text{loc})$ is the state vector of a trajectory (i.e., of the characteristic), determining the atom orbit, and $\omega$ is a statistical weight corresponding to the solution of the production and loss balance equation between $\vec{r}_\text{loc}$ and the boundary of the calculation region, defined as a heliocentric sphere with a radius $r_\text{lim}=1000$~au. The velocity of the atom at the boundary of the calculation region is $\vec{v}_\text{lim}(\vec{q}_\text{loc})$, and the unperturbed VLISM is assumed to have a temperature $T_\text{VLISM}$ and a velocity relative to the Sun $\vec{u}_\text{VLISM}$. The quantity $f_\text{MB}$ is the Maxwell-Boltzmann distribution function for a population with a number density $n_\text{VLISM}$, defined as:
\begin{equation}
f_\text{MB}\left(\vec{v}, n, \vec{u}, T \right)= n \left(\frac{m}{2 \pi k_\text{B} T}\right)^{\frac{3}{2}}\,\exp\!\left[ -\frac{m(\vec{v}-\vec{u})^2}{2 k_\text{B} T}\right],
\label{eq:MBdef}
\end{equation} 
where $k_\text{B}$ is the Boltzmann constant, $T$ the temperature, $\vec{u}$ the flow (bulk) velocity, and $m$ the mass of individual particles belonging to this population. 

The statistical weight $\omega$ in Equation~\ref{eq:locDiFu} is calculated as a numerical solution of the production and loss balance equation, as presented by \citet{bzowski_etal:17a} and \citet{bzowski_etal:19a}, with the condition at the outer boundary of the calculation region $\omega(r_\text{lim})$ defined as:
\begin{equation}
\omega\left(r_\text{lim}\right) = f_\text{MB}\left(\vec{v}_\text{lim}(\vec{q}_\text{loc}), n_{\text{He}_\text{VLISM}}, \vec{u}_\text{VLISM}, T_\text{VLISM} \right).
\label{eq:omegaLim}
\end{equation} 
The production and loss balance equation is defined (see Equation~4 in \citet{bzowski_etal:19a}:
\begin{equation}
\frac{d\omega(t)}{d t} = \beta_{\text{pr}}(\vec{r}_s(t),\vec{v}_s(t))\, f_{\text{He}^+_\text{OHS}}(\vec{r}_s(t),\vec{v}_s(t)) - \beta_{\text{loss}}(\vec{r}_s(t),\vec{v}_s(t))\, \omega(t) .
\label{eq:prodlossBalance}
\end{equation}
In this equation, $\beta_\text{pr}$ is the rate of injection of He atoms at the location $\vec{r}_s$ into the trajectory defined by $\vec{q}_\text{loc}$ due to charge exchange collisions with He$^+$ ions. The distribution function of these parent He$^+$ ions is given by $f_{\text{He}^+_\text{OHS}}(\vec{r}_s(t),\vec{v}_s(t))$, assumed to be the Maxwell-Boltzmann distribution function defined in Equation~\ref{eq:MBdef} with the parameters corresponding to the ambient plasma parameters. The definition of $\beta_{\text{pr}}$ is given in Equations 5--7 in \citet{bzowski_etal:19a}. The loss rate $\beta_\text{loss}$ is a sum of the solar photoionization rate at $\vec{r}_s$ and of the charge exchange rate between a He atom at $\vec{r}_s$ traveling at a velocity $\vec{v}_s$ and all He$^+$ ions at this location. The definition is provided in Equations 7--9 in \citet{bzowski_etal:19a}.  

In the solution of Equation~\ref{eq:prodlossBalance} it is assumed that the temperature and bulk velocity of He$^+$ are identical to the equivalent parameters of the ambient plasma, obtained from the global model of the heliosphere \citep{zirnstein_etal:16b}, and the number density is equal to the number density of protons multiplied by a constant factor equal to the ratio of He$^+$/H$^+$ densities at the boundary of the calculation region. We have adopted this quantity to be equal to 0.166, as determined by \citet{bzowski_etal:19a} from IBEX-Lo observations. Inside the heliopause, we assumed $n_{\text{He}^+}=0$ because the core solar wind features a very low density of this ion. Even though the content of pickup He$^+$ ions in the solar wind gradually increases with the solar distance up to 15--35\% of the solar wind alpha density \citep{rucinski_etal:98}, these ions have velocities outside the region in velocity space occupied by the primary and secondary populations of ISN He, so He atoms produced as a result of charge exchange reactions between these ions and ISN He atoms are outside the region of velocity space occupied by the ISN He atoms \citep{grzedzielski_etal:14a}. Consequently, for the portion of He atom orbit located inside the heliopause, $\beta_\text{pr} = 0$. 

For comparison, we also calculate the two-Maxwellian distribution function. In this case, we assume that the distribution function of ISN He at the boundary of the simulation region $f_\text{MB2}$ is a superposition of two Maxwell-Boltzmann functions with the parameters corresponding to the primary ISN He population \citep{bzowski_etal:15a} and the Warm Breeze \citep{kubiak_etal:16a}:
\begin{equation}
f_\text{MB2}(\vec{v}, r_\text{lim}) = f_\text{MB}\left(\vec{v}, n_\text{ISNHe}, \vec{u}_\text{VLISM}, T_\text{VLISM} \right)+ f_\text{MB}\left(\vec{v}, n_\text{WB}, \vec{u}_\text{WB}, T_\text{WB} \right), 
\label{eq:2maxw}
\end{equation}
where the subscript WB corresponds to the relevant parameters of the Warm Breeze, found by \citet{kubiak_etal:16a}, and the subscript VLISM to the velocity vector and temperature of ISN He from \citet{bzowski_etal:15a}. The density of ISN He in the VLISM was adopted after analysis of He$^+$ and He$^{++}$ pickup ions by \citet{gloeckler_etal:04a} and direct-sampling observations of ISN He by \citet{witte:04}. To calculate the local distribution function in this model, we adopt an appropriately modified Equation~\ref{eq:locDiFu}:
\begin{equation}
f_{\text{loc}_\text{MB2}}(\vec{r}, \vec{v}, t) =
\left(f_\text{MB}(\vec{v}_\text{lim}(\vec{q}_\text{loc}), n_\text{pri}, \vec{u}_\text{VLISM}, T_\text{VLISM})
 + 
f_\text{MB}(\vec{v}_\text{lim}(\vec{q}_\text{loc}), n_\text{WB}, \vec{u}_\text{WB}, T_\text{WB})\right)
\,\omega(\vec{q}_\text{loc}, r_\text{lim}, t),
\label{eq:2MaxwDF}
\end{equation}
where the subscripts ''pri'' and ''WB'' correspond to the primary ISN He population and the Warm Breeze (secondary population), respectively. The statistical weights $\omega$ are obtained from integration of the loss rate along a given trajectory $q_\text{loc}$, as discussed by \citet{bzowski_etal:13a}. For a given state vector $\vec{q}_\text{loc}$, they only differ by the initial condition at the boundary of the calculation region, defined in Equation~\ref{eq:omegaLim}, which is evaluated either with the parameters corresponding to the primary ISN flow or to the Warm Breeze. In this model, there is no production of the secondary atoms  by assumption; the secondary population is pre-assumed to exist at the boundary of the calculation region. The losses are only due to ionization by solar factors, represented by a time-independent rate dropping with the square of solar distance. 

In the following sections, we will discuss the distribution function of ISN He in various locations in the upwind hemisphere, both within the OHS and inside the heliopause. Since it is impossible to show a four-dimensional object, which is the distribution function, on a two-dimensional page, we will present partial integrals. With the distribution function calculated in either of the two aforementioned models, we start from calculating the directional distribution, i.e., the distribution function integrated over speed. We select spherical coordinates in the reference system with the B-V plane being the cardinal (x-y) plane and the upwind direction corresponding to +x axis. The orientation of this plane in space was determined by \citet{zirnstein_etal:16b}: the normal direction to this plane in the ecliptic coordinates is $(\lambda, \beta) = (320.5\degr, 37.8\degr)$. In this reference system, the longitude (azimuth) angle $\lambda_\text{BV}$ is counted within the B-V plane off the upwind direction, and the latitude (elevation) angle $\beta_\text{BV}$ towards the +z and -z axes, perpendicular to the B-V plane. In these coordinates, referred to as the BV coordinates, the longitude (azimuth angle) of the interstellar magnetic field direction is equal to 320.5\degr, and the (azimuth, elevation) angles of the inflow direction of the Warm Breeze obtained by \citet{kubiak_etal:16a} are equal to (352.1\degr, 1.1\degr).

The speed-integrated distribution function $F_V$ at a location $\vec{r}$ is obtained from the formula:
\begin{equation}
F_V\left(\vec{r}, \lambda_\text{BV}, \beta_\text{BV} \right) = \int\limits_0^\infty f_\text{loc}\left(\vec{r}, v, \lambda_\text{BV}, \beta_\text{BV} \right)\, v^2 dv,
\label{eq:difuVInt}
\end{equation}
where we replace the velocity vector $\vec{v}$ with its spherical coordinates $(v, \lambda_\text{BV}, \beta_\text{BV})$ in the B-V coordinate system, and $f_\text{loc}$ is defined either in Equation~\ref{eq:locDiFu} or Equation~\ref{eq:2MaxwDF}. 

The speed-integrated distribution function $F_V(\vec{r}, \lambda_\text{BV}, \beta_\text{BV})$ can be further integrated over BV-longitude or latitude, becoming a one-dimensional function of the angle:
\begin{equation}
F_{\text{V}\lambda}\left(\vec{r}, \lambda_\text{BV} \right) = \int\limits_{-\pi/2}^{\pi/2}F_V\left(\vec{r}, \lambda_\text{BV}, \beta_\text{BV} \right)\, \sin \beta_\text{BV}\, d\beta_\text{BV},  
\label{eq:difuBetaInt}
\end{equation}
\begin{equation}
F_{\text{V}\beta}\left(\vec{r}, \beta_\text{BV} \right) = \int\limits_0^{2\pi}F_V\left(\vec{r}, \lambda_\text{BV}, \beta_\text{BV} \right)\, \, d\lambda_\text{BV}.
\label{eq:difuLambdaInt}
\end{equation}
The calculations were performed using an appropriately modified version of the nWTPM model \citep{sokol_etal:15b, bzowski_etal:17a, bzowski_etal:19a}. Throughout the paper, we show these distribution functions normalized by the corresponding peak values of the respective distribution functions $F_V, F_{V \lambda}, F_{V \beta}$ at the upwind direction, 1000~au from the Sun.

\section{Distribution function of neutral He in the OHS and inside the heliopause}
\label{sec:difu}
In this section, we present the evolution of the distribution function of ISN He at selected locations in the OHS and inside the heliopause, down to 1~au. We begin with a very brief description of the plasma behavior in the OHS, obtained from the global MHD-kinetic simulation of the heliosphere. Next, we qualitatively present the evolution of the distribution function of ISN He from the boundary of the calculation region down to 1~au and approximately identify the regions of the production of secondary He atoms. Subsequently, we discuss longitude- and latitude-integrated distribution functions and compare them with the two-Maxwellian model.

\subsection{The plasma in the OHS}
\label{sec:OHSPlasma}
The global MHD-kinetic simulation of the heliosphere \citep{zirnstein_etal:16b} was performed assuming that the VLISM material is thermalized beyond the calculation boundary, partly ionized, and magnetized. Within the MHD paradigm, it was assumed that the interstellar plasma is composed of protons and electrons and that its distribution function is Maxwell-Boltzmann. This plasma interacts with neutral H atoms by resonant charge exchange. Since this reaction involves no momentum exchange between the reaction substrates, the reaction products inherit the kinematic parameters of the parent particles. The product protons are immediately thermalized with the ambient plasma, modifying its temperature and velocity. The product neutral atoms continue to be treated kinetically, i.e., they may enter in charge-exchange reactions just as the original interstellar Ha atoms do.
\begin{figure*}
\plottwo{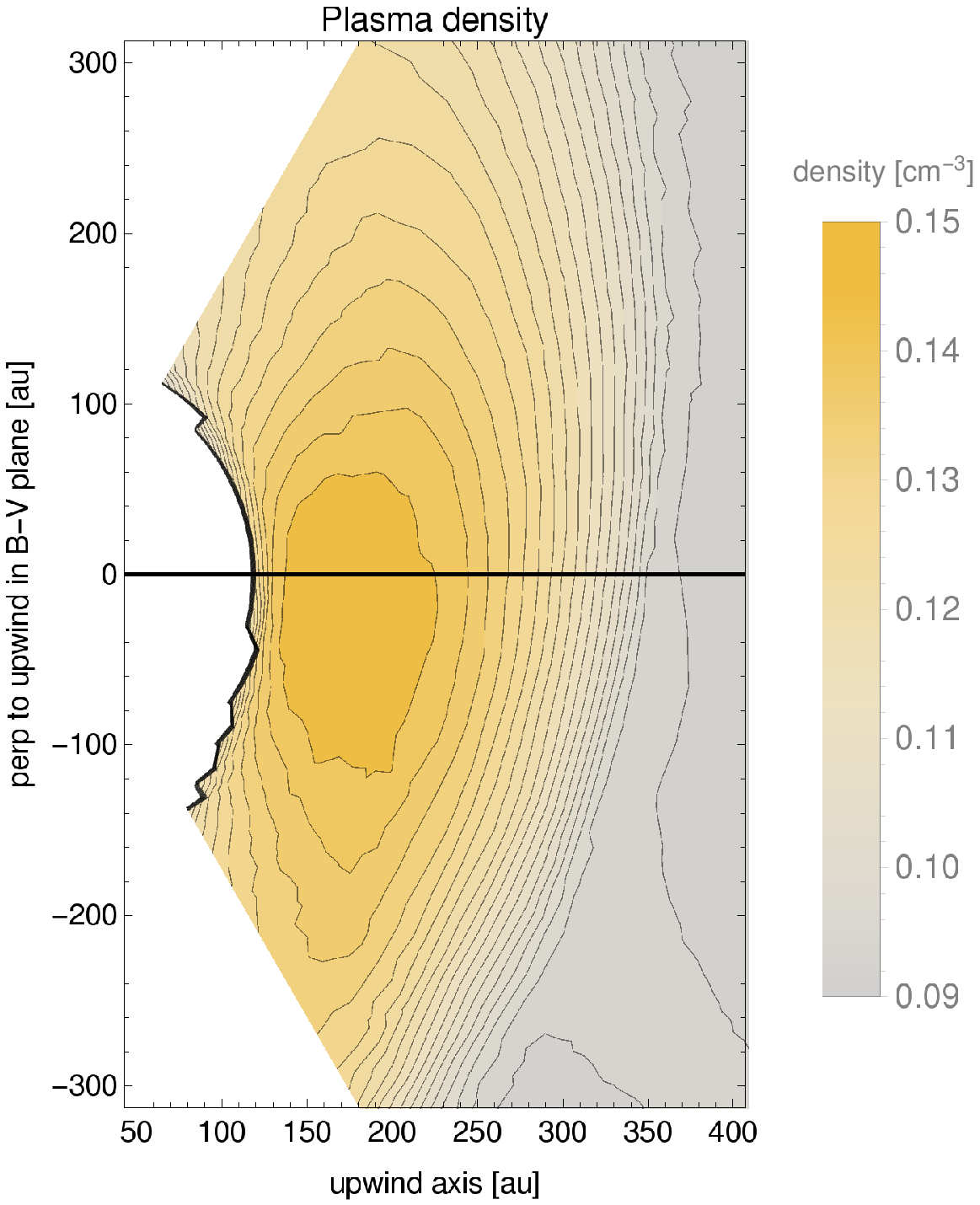}{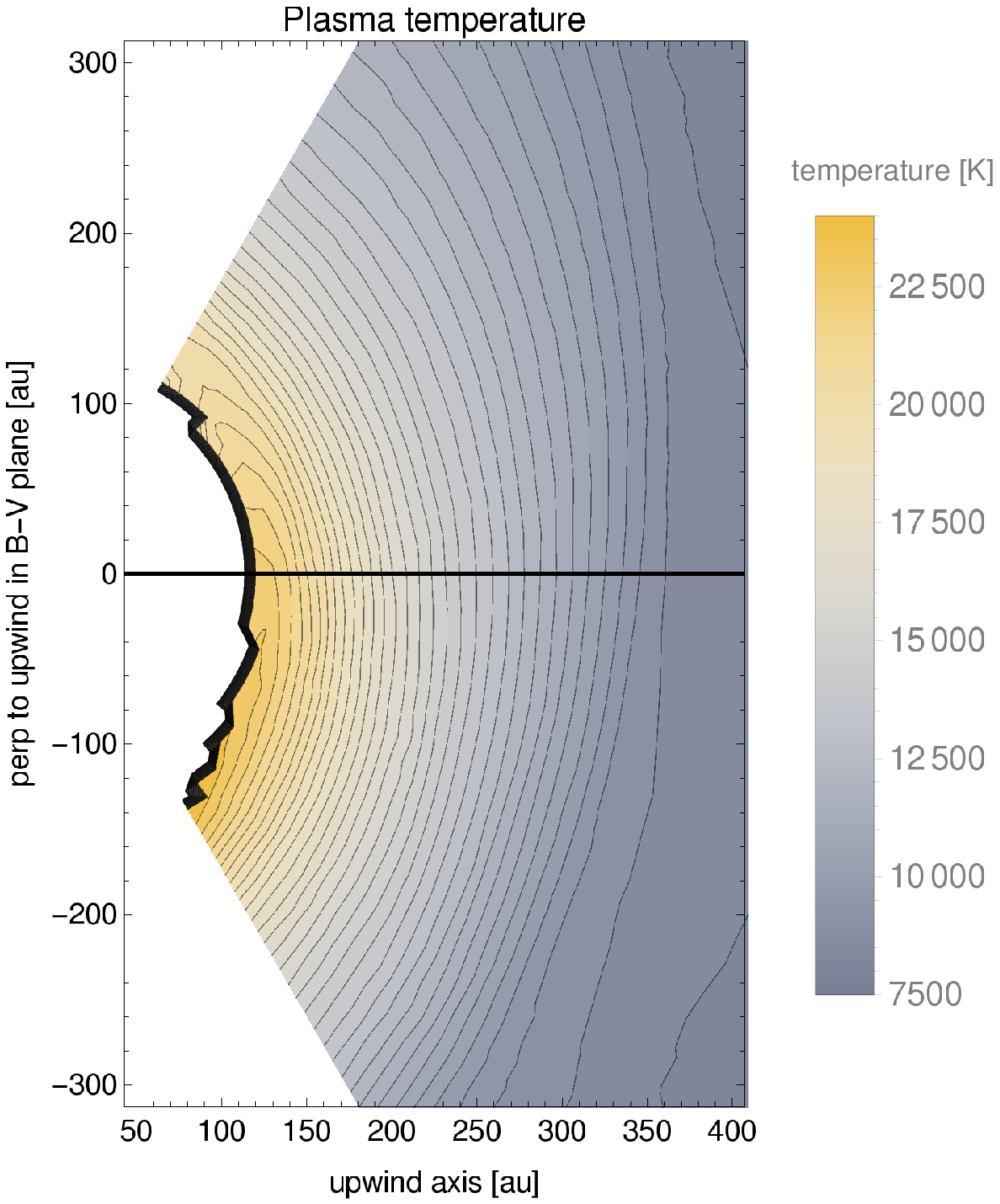}
\plottwo{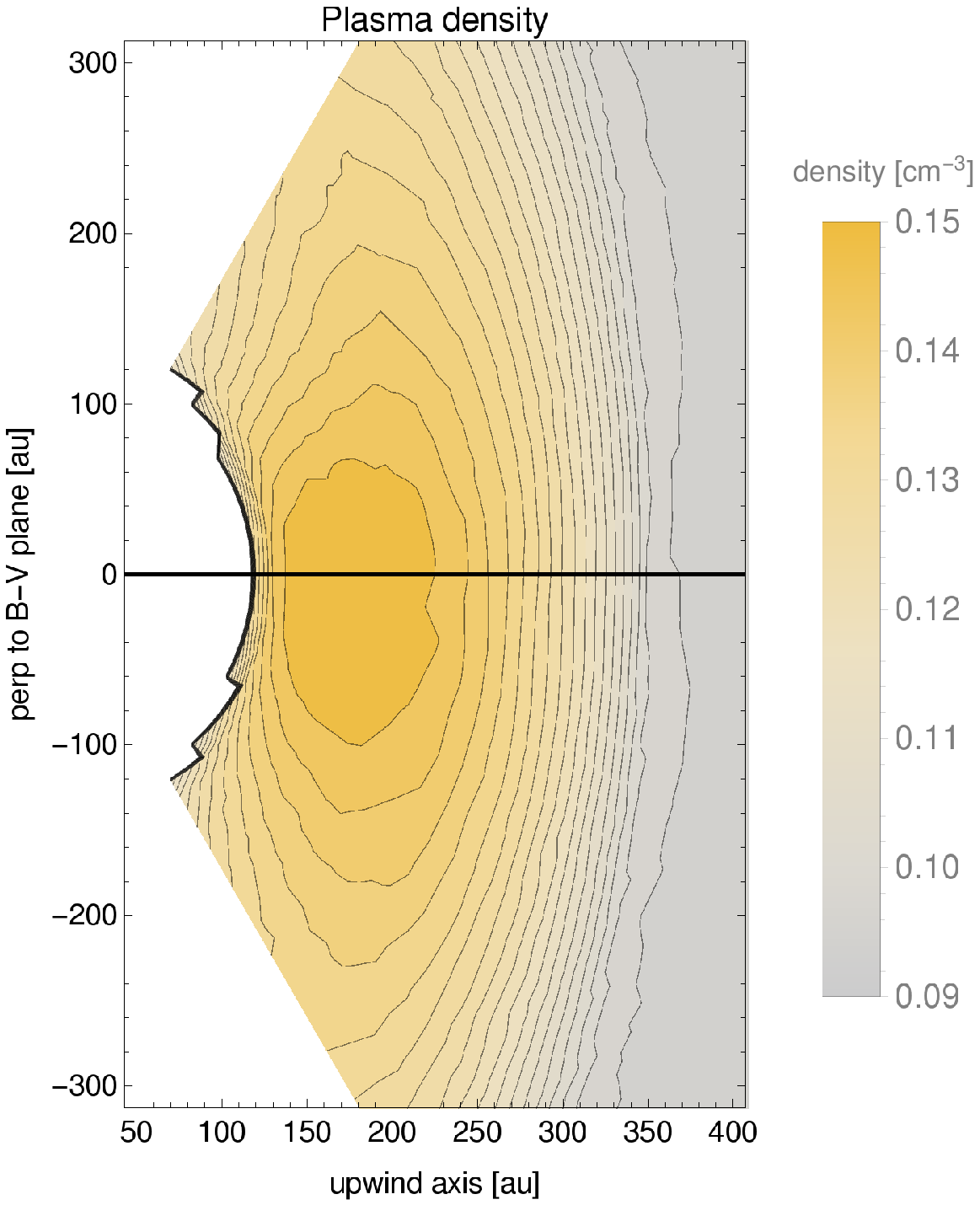}{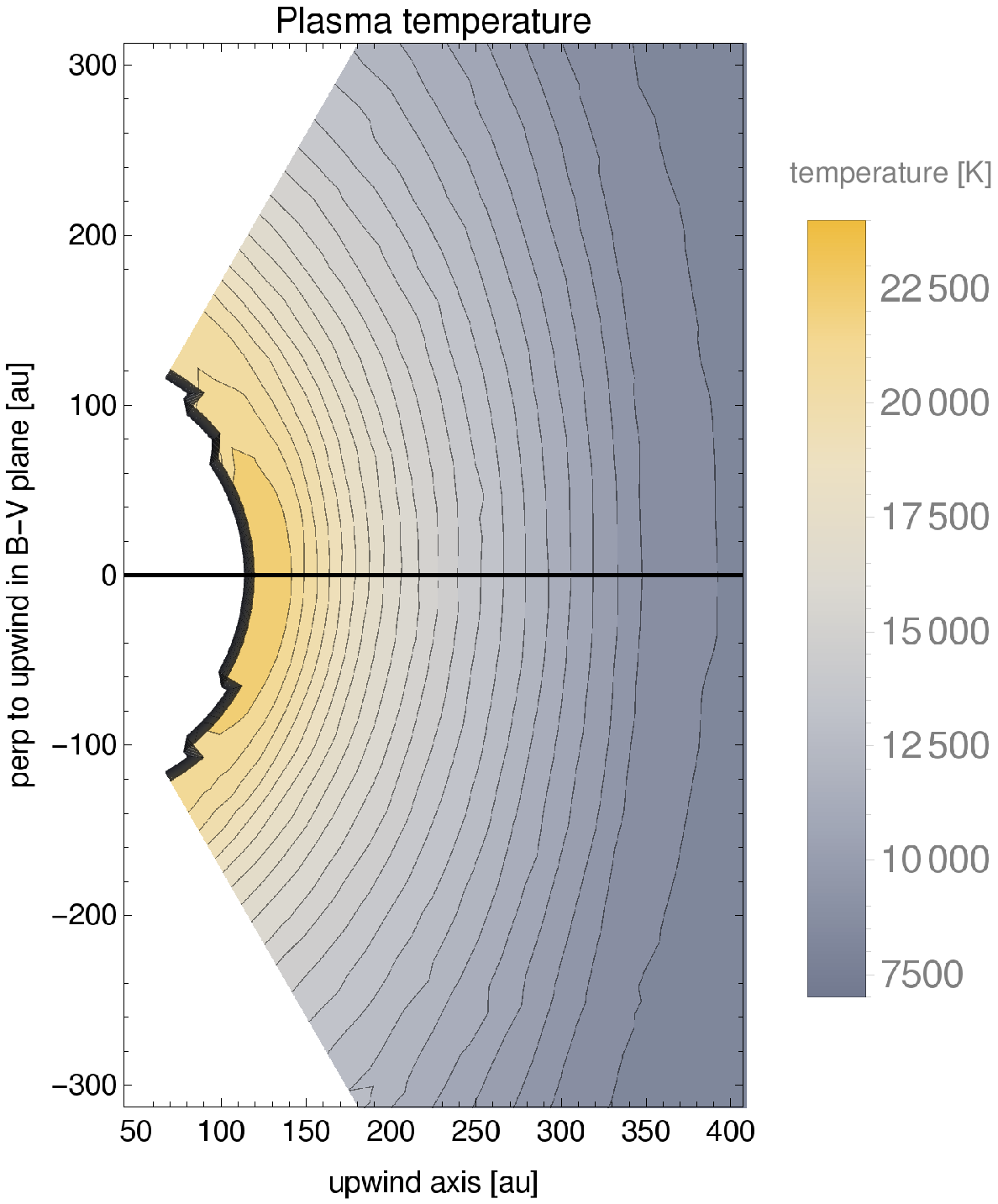}
\caption{Maps of the plasma density (left-hand column) and temperature in the OHS (right-hand column) obtained from the MHD-kinetic model of the heliosphere \citep{heerikhuisen_etal:10a, zirnstein_etal:16b}. The upper row presents the spatial variation of these parameters in the B-V plane, the lower row in a plane perpendicular to the B-V plane. The black horizontal lies mark the upwind direction. The black contour marks an approximate location of the heliopause. 
}
\label{fig:OHSPlasma}
\end{figure*}

\begin{figure*}
\includegraphics[height=0.45\textwidth]{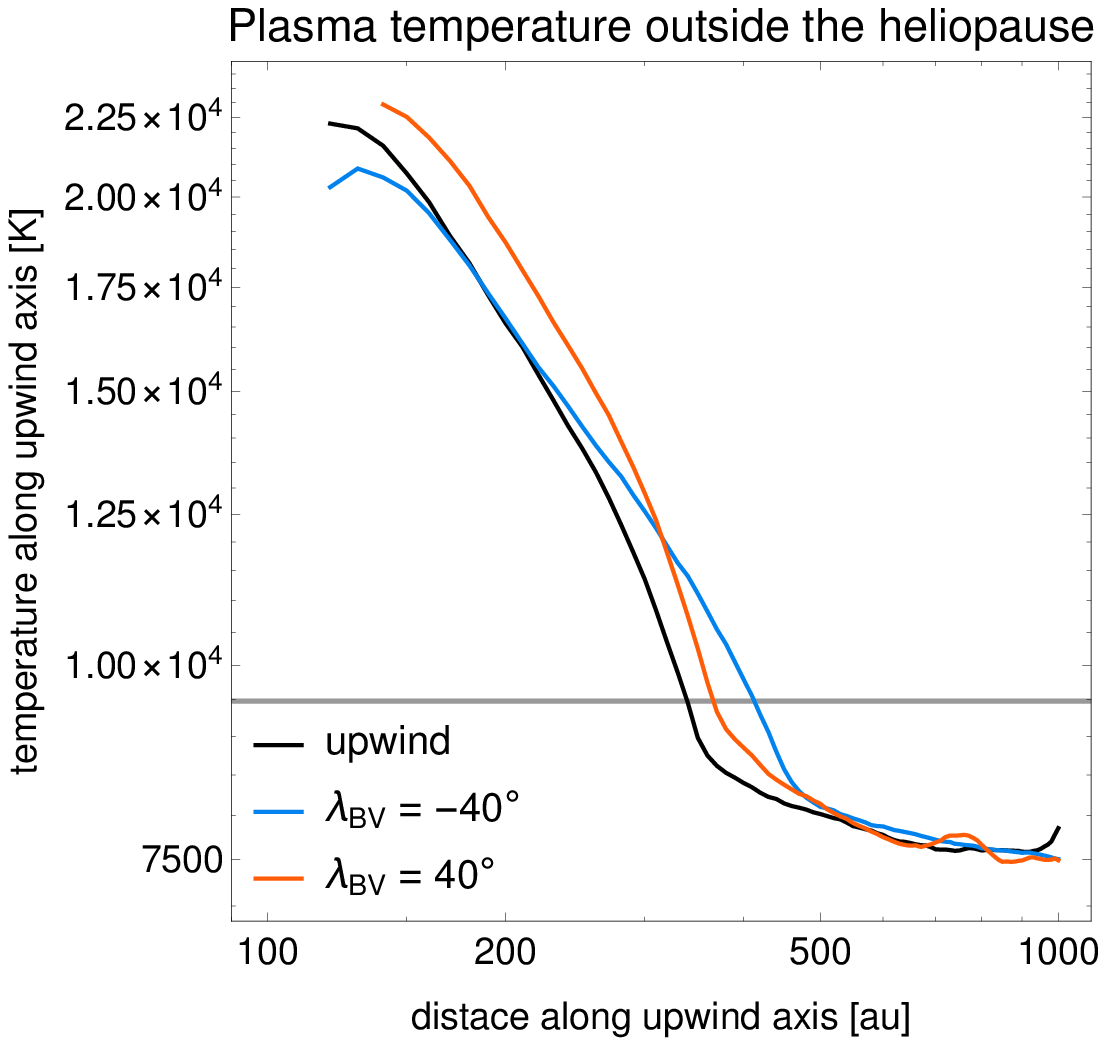}
\includegraphics[height=0.45\textwidth]{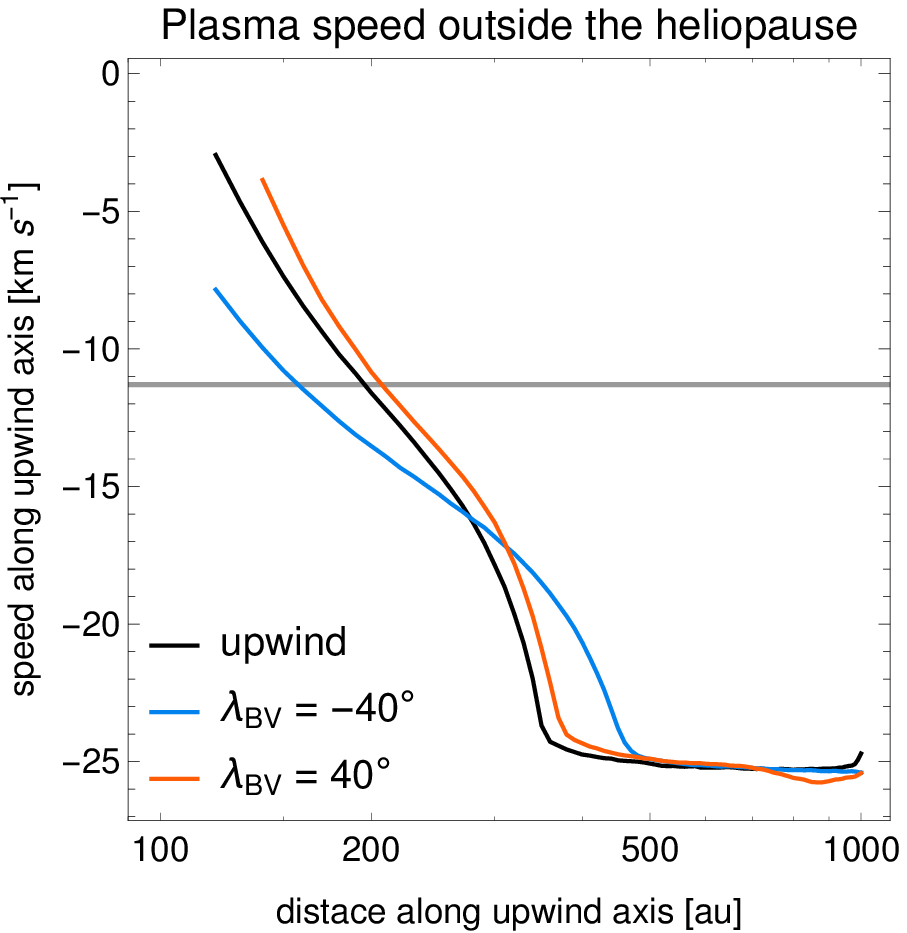}
\caption{Radial profiles of the plasma temperature (left hand panel) and velocity component along the upwind axis (right-hand panel) for three radial lines located within the B-V plane: upwind, and at $\pm 40\degr$ off the upwind direction. The horizontal lines mark the temperature and inflow speed of the Warm Breeze obtained from fitting the two-Maxwellian model to IBEX observations by \citet{kubiak_etal:16a}.}
\label{fig:plasma_along}
\end{figure*}

As discussed by \citet{bzowski_etal:19a}, we allow for the presence of He$^+$ ions in the plasma. These ions are responsible for a portion of the interstellar plasma mass density. To first approximation, the structure of the heliosphere and the plasma flow in the OHS are governed by the ram and thermal pressures of the plasma and the magnetic pressure. Neither thermal nor ram pressure of the plasma are modified directly by the fact that a portion of the plasma is He$^+$ ions. Only a posteriori, we assume that a certain portion of the plasma density is locked in He$^+$ ions.  Therefore, this simplification does not affect our conclusions.

The solar wind is assumed to be spherically symmetric and magnetized. In the absence of an interstellar magnetic field, the plasma flow in the OHS would be axially symmetric. The presence of magnetic field of 3~$\mu$G induces a deformation of the OHS from axial symmetry \citep{ratkiewicz_etal:98a}. This asymmetry is the strongest in the B-V plane, as visible in the maps of plasma density and temperature presented in Figure~\ref{fig:OHSPlasma}. In this figure we present the plasma density and temperature in a region of the OHS that is ballistically connected to IBEX at 1 au, i.e., the region penetrated by He atoms that are detected by this spacecraft \citep{bzowski_etal:19a}. Note a distortion from symmetry about the upwind line in the B-V plane (the upper row of panels), which is due to the interstellar magnetic field, and the lack of asymmetry in the perpendicular plane, shown in the lower row of panels. Also note that even though strong deviations of the plasma density from the magnitude of 0.09~\ccm{} and of the temperature from 7500~K only begin at $\sim 350$~au from the Sun, still the plasma farther away from this distance is somewhat compressed and heated. Consequently, the ISN He and the ambient plasma are no longer in thermal equilibrium in this region, which creates favorable conditions for the creation of the secondary population of ISN gas already in this region.  

\citet{kubiak_etal:16a} found the temperature of the secondary He population equal to $\sim 9500$~K and a mean speed of 11.3~\kms. Figure~\ref{fig:plasma_along} presents radial cuts of the adopted plasma model  in the OHS for the upwind direction and for the directions $\lambda_{BV} = \pm 40\degr$ in the B-V plane. It illustrates variations and asymmetries of the plasma model within the B-V plane and clearly shows that the temperature and speed of the Warm Breeze are not exactly representative for the plasma conditions in the OHS. 

\subsection{Maps of the distribution function}
\label{sec:mapDiFu}
\subsubsection{Radially towards the Sun along the upwind line}
\label{sec:mapDiFuRad}

\begin{figure}
\plottwo{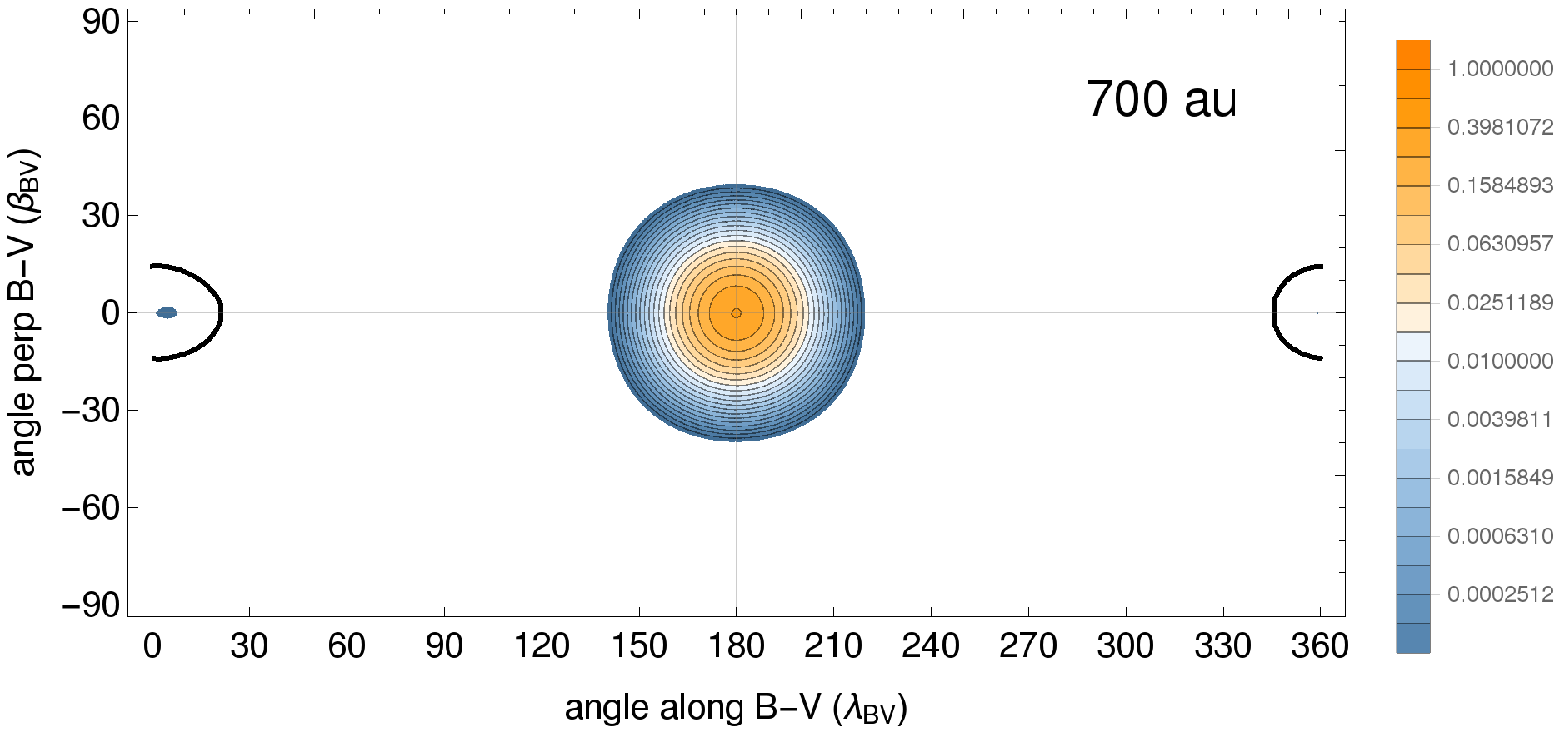}{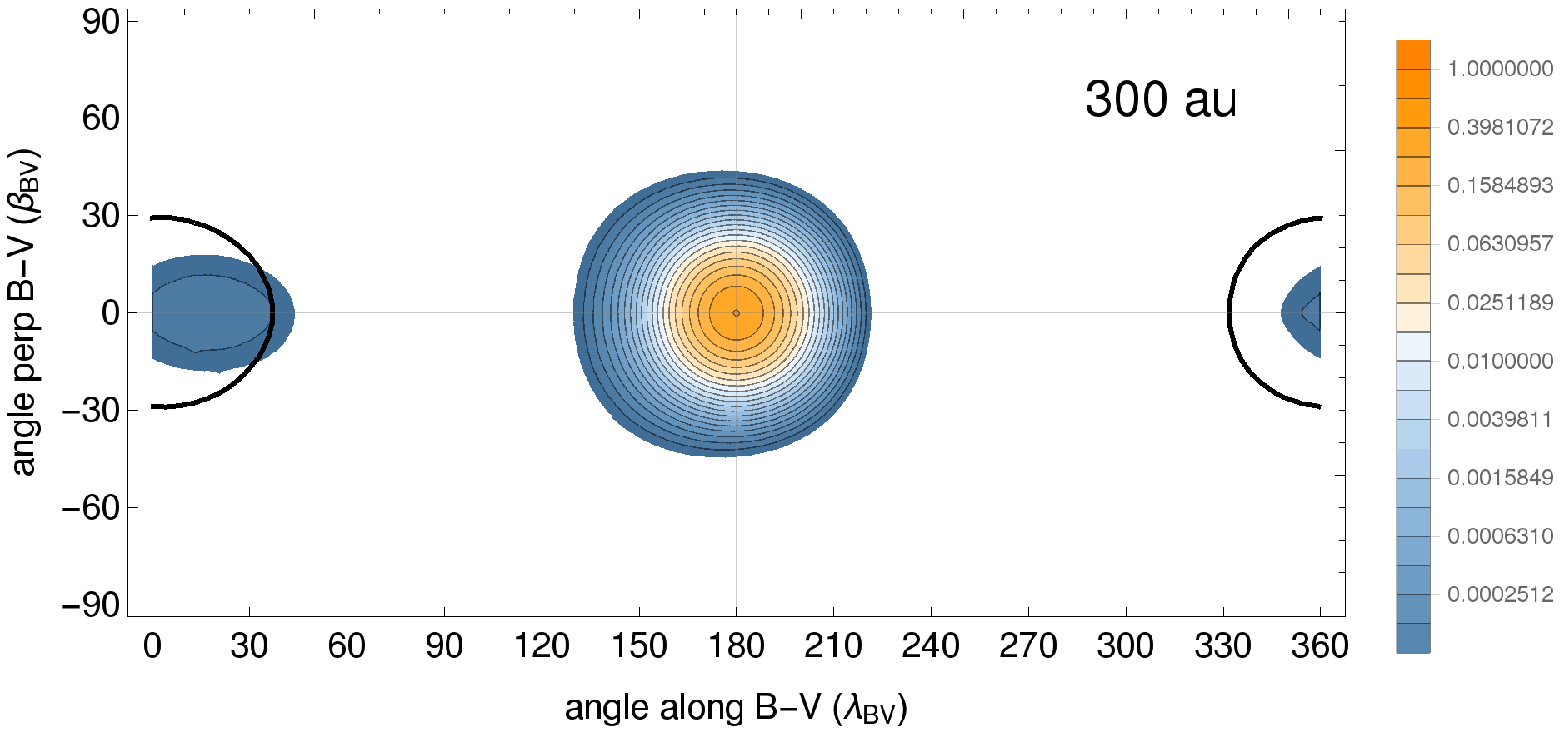}
\plottwo{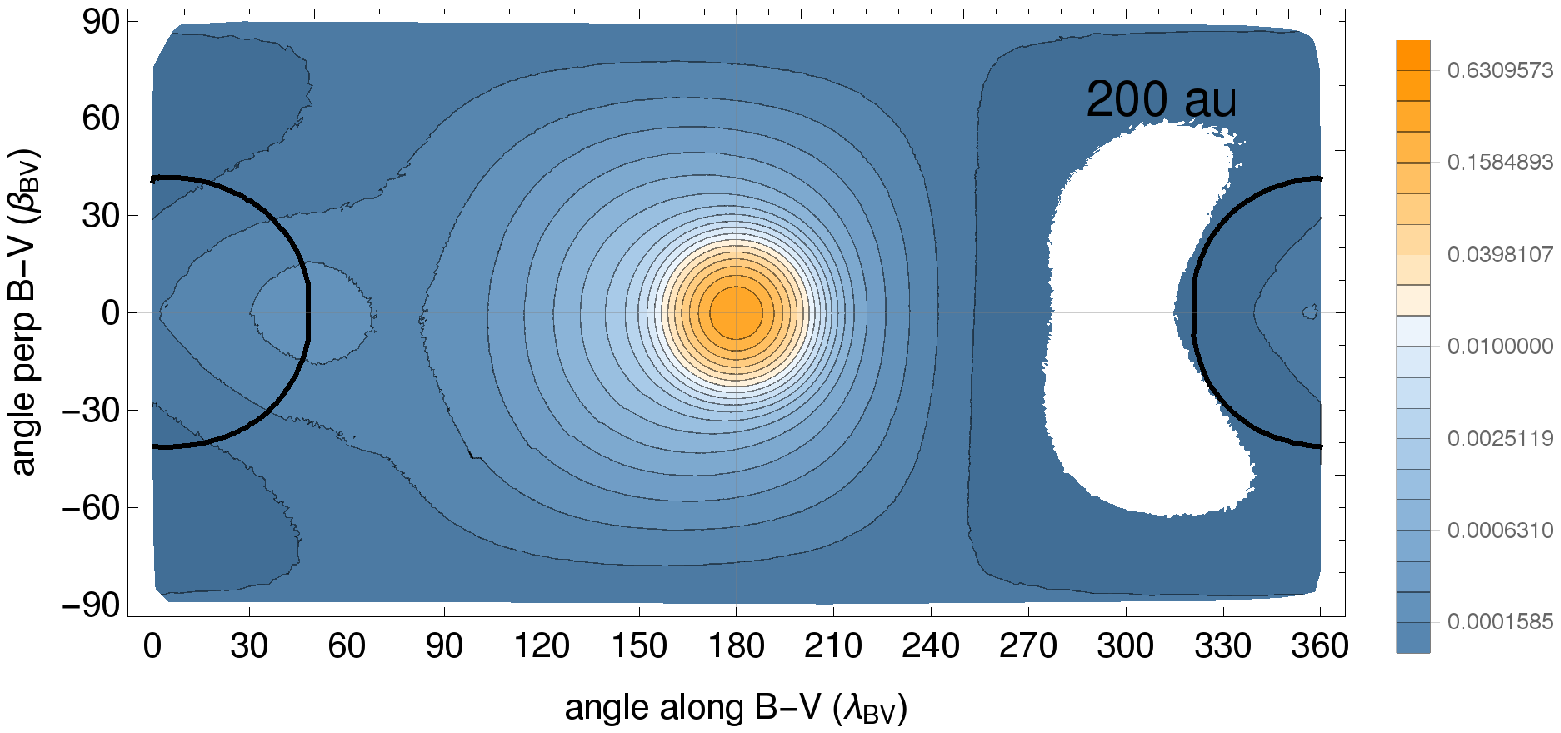}{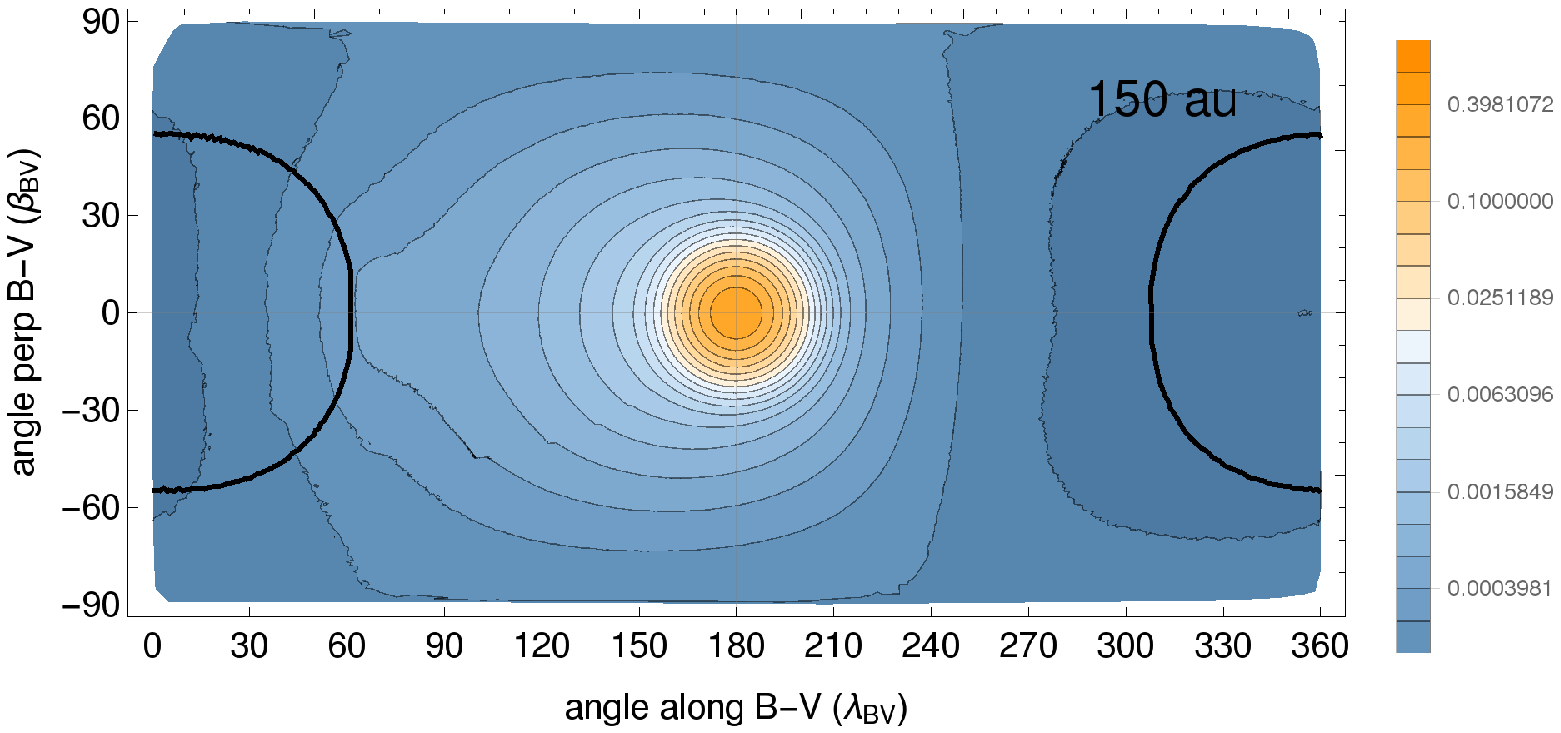}
\plottwo{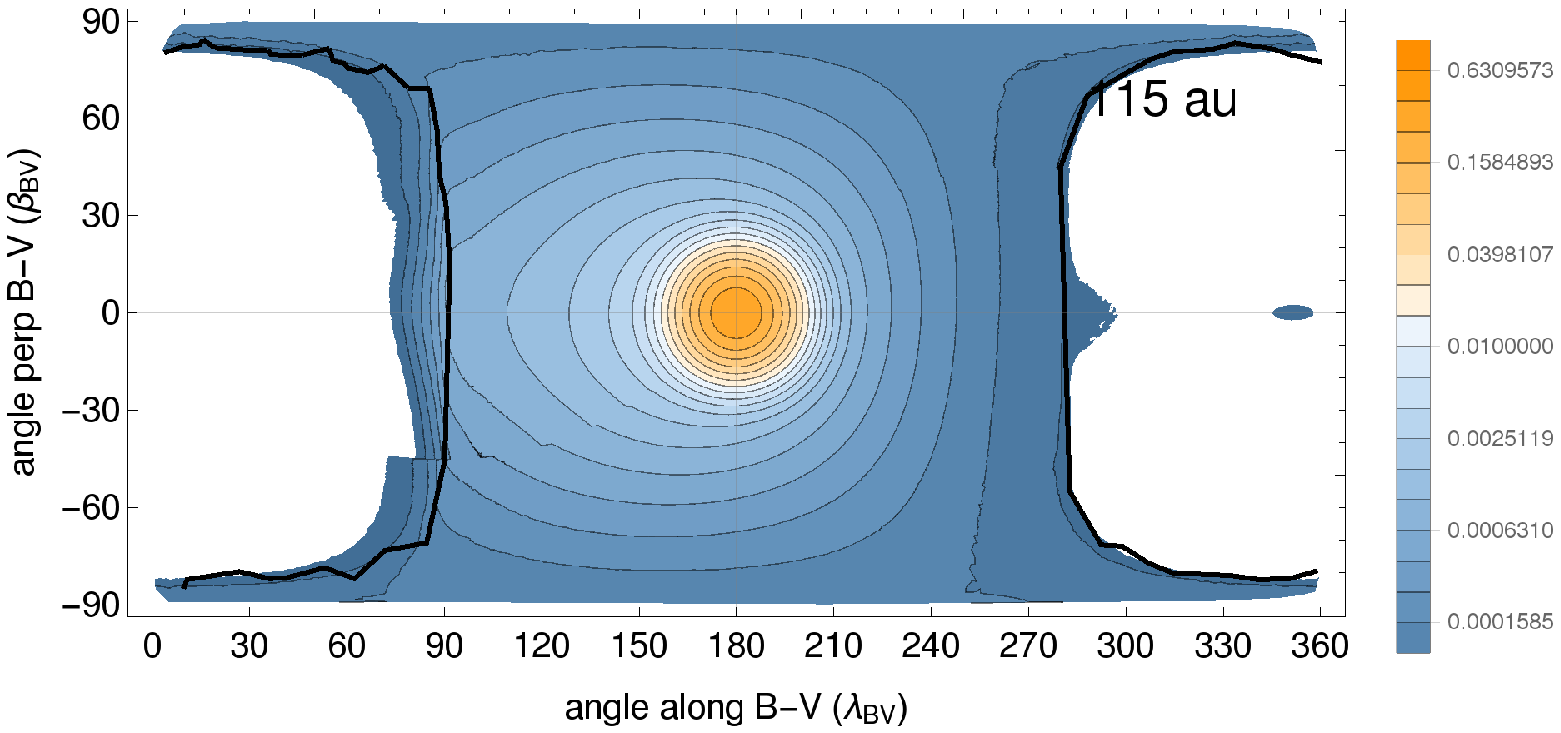}{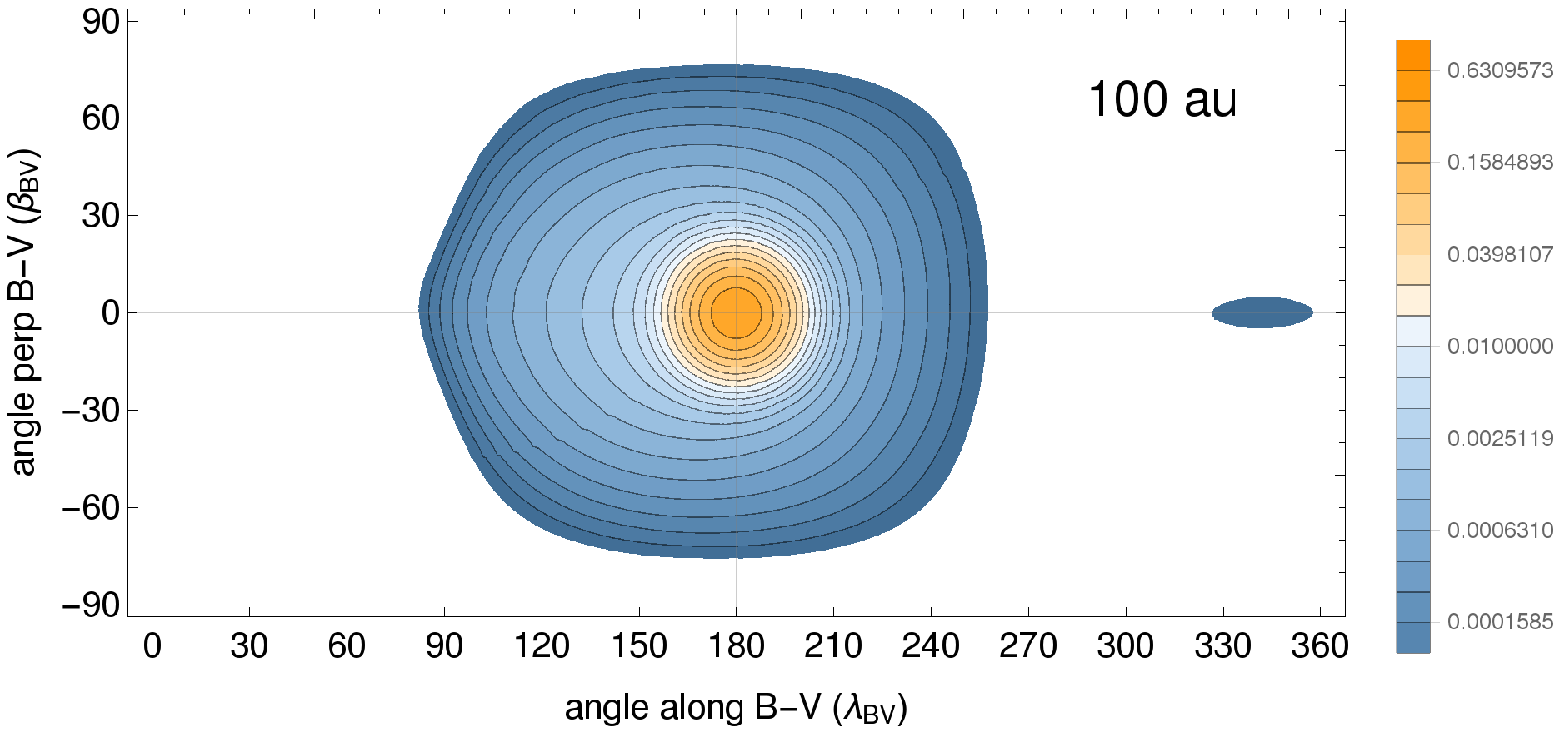}
\plottwo{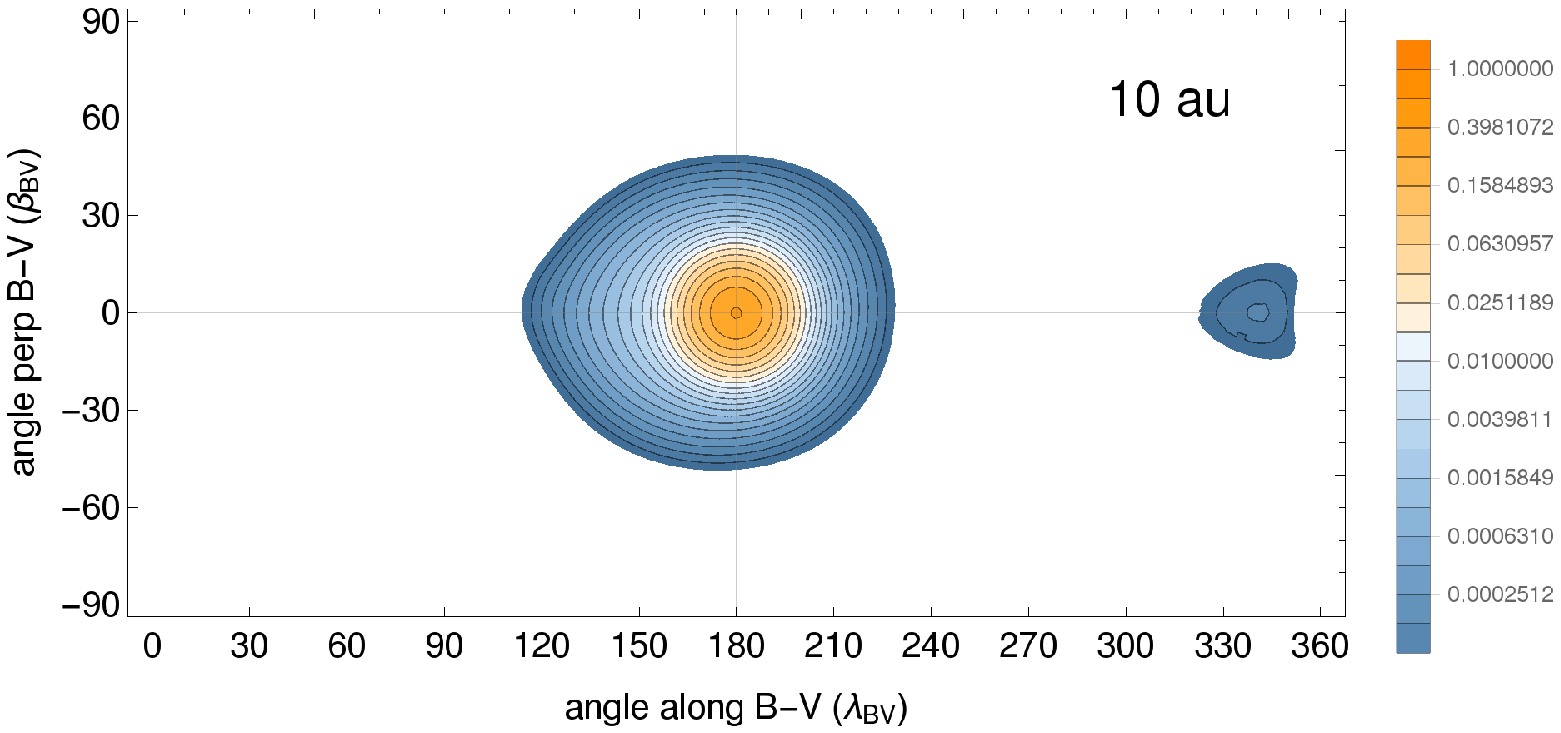}{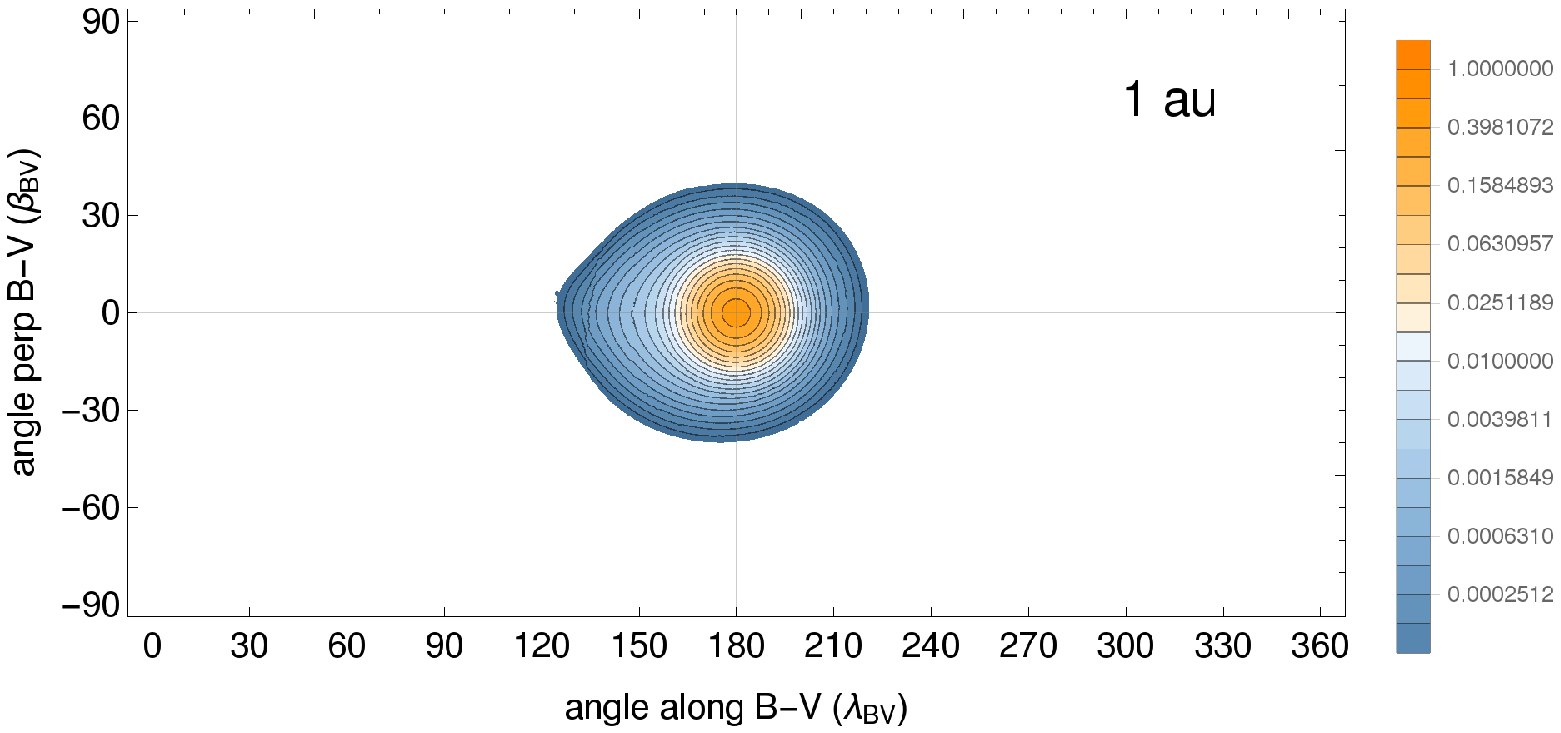}
\caption{Maps of speed-integrated distribution function of ISN He $F_{V}$, defined in Equation~\ref{eq:difuVInt}, along the upwind line for observer at selected distances from the Sun. The black contour marks the directions to the heliopause boundary as seen by the observer from those locations in space. The horizontal axis is the azimuth angle (longitude) along the B-V plane, and the vertical axis is the elevation angle (latitude) perpendicular to the B-V plane. The yellow peaks close to $\lambda_{BV} = 180\degr$ corresponds to atoms coming from the upwind direction, i.e., the primary ISN He. Note that the distribution function isocontours are logarithmically spaced.The lowermost contour traces the $10^{-4}$ isoline of the distribution function; the empty regions in the plot mark the regions where  the magnitudes of $F_{V}$ are below this threshold.}
\label{fig:radialUp2D}
\end{figure}

We begin the discussion of the distribution function of ISN He from presenting its evolution along the upwind direction, starting at 700~au, where modifications of the unperturbed function are very small, and ending up at 1~au. This evolution is shown in Figure~\ref{fig:radialUp2D}.

The speed-integrated distribution function $F_V$ at 700~au from the Sun, i.e., in the region of very little perturbed plasma, closely resembles the unperturbed function. The yellow peak is centered at 180\degr{} in B-V longitude and 0\degr{} B-V latitude, which means that the gas is flowing directly towards the Sun. This is the primary ISN He population, which persists in the maps for all distances from the Sun shown in Figure~\ref{fig:radialUp2D}. The 700~au map has a trace of a narrow secondary peak, approximately centered at the heliopause. The heliopause seen from 700~au is a small object, centered near the Sun. This tiny peak corresponds to a small subset of the secondary population atoms that were produced close to the heliopause with close-to-antisolar velocity vectors and were not subsequently lost due to charge-exchange collisions between their origin and the observer at 700~au. 

At 300~au from the Sun, the primary peak is a little widened and features a small asymmetry along the B-V plane. It is due to the secondary atoms produced locally. The secondary peak that was barely visible at 700~au becomes wider, although it still represents a very small portion of the secondary atoms. It is formed by atoms that originate from He$^+$ ions that happen to have almost purely anti-solar velocity vectors. Note that the location at 300~au upwind is close to the boundary of the region of strongly perturbed plasma in the OHS, as shown in Figure~\ref{fig:OHSPlasma}. In other words, the secondary He atoms are hardly present between 1000 and 300~au from the Sun because of the very low local production rate and strong ballistic selection effects. Due to these effects, the secondary population in these regions consists of two counterstreaming flows: one flowing radially \emph{away} from the Sun, produced closer to the Sun, and the other one flowing \emph{towards} the Sun, produced locally. However, the abundance of the secondary population in this region relative to the primary population is almost negligibly low. 

At 200~au, the ISN peak persists, and the secondary atom come in from all directions except a relatively small region near $\sim 300\degr$, which at this distance from the Sun is farther from the Sun than the region where most of the secondary He production operates. While the distribution function is symmetric with respect to the B-V plane, telltale asymmetries along the B-V plane show up due to the asymmetry of the OHS. In addition to the primary ISN He peak, a secondary peak at $\sim 45\degr$ appears. It is entirely formed by the secondary atoms. The inflow directions of these atoms point towards the region of maximum production, which is within the B-V plane but away from the upwind direction, and closer to the Sun than 200~au. 

At 150~au, there are no regions in the sky void of the incoming atoms. The structure of the distribution function does not show a clear correlation with the heliopause contour, which is the boundary surface inside which the production of secondary atoms ceases. This is because here we are located within the secondary atom production region and the He atoms arrive from all directions. The secondary peak that we noticed at 200~au, at 150~au is almost 90\degr{} away from the primary peak and is partly covered by the very broad structure that has formed around the primary peak. The magnitude of the distribution function seems to be the largest among all the cases shown. All this suggests that this location is close to the region where the production of secondary He atoms is the most intense, with the peak production rate within the B-V plane. 

At 115~au, we are within 1~au from the heliopause. A void region reappears in the map. It occupies almost the entire downwind hemisphere. This implies that the region where most of the secondary He atoms are produced are farther away from the Sun. The secondary peak (partly blended with the ISN He peak) persists but seems to be shifting towards the primary peak, which suggests that the region of maximum production rate is farther away from the Sun than 115~au. The tiny peak located at $\sim 340\degr$ is due to the indirect beam of the secondary population. The region occupied by this beam grows with a decreasing distance from the Sun and disappears only inside 10~au.

At 100~au we are well inside the heliopause and the distribution function is rapidly transforming: since the production does not operate in this region, the secondary atoms present here must have been produced somewhere away.  Their velocity vectors cluster in a broad region around the primary peak. The global distribution occupies a large portion of the hemisphere and bears traces of the asymmetry in B-V-longitude created due to the offset of the region of maximum production from the upwind axis. The atoms forming the distribution function in this location in space are filtered by ballistic selection effect: only a portion of secondary atoms is able to reach it, namely those that have perihelion distances less than 100~au and an appropriate angular momentum vector. 

Going further towards the Sun (see the 10~au and 1~au maps), the distribution increasingly narrows due to the aforementioned selection effect and the acceleration by Sun's gravity. The asymmetry is still present, but increasingly difficult to notice because of the low abundance of secondary He in general on one hand, and the geometric effect on the other hand: the angular separation of the upwind direction and the location in space of the maximum production region of secondary He is small because of the large distance.

\subsubsection{Laterally along B-V plane close to maximum production distance}
\label{sec:mapDiFuLateral}
\begin{figure}
\plottwo{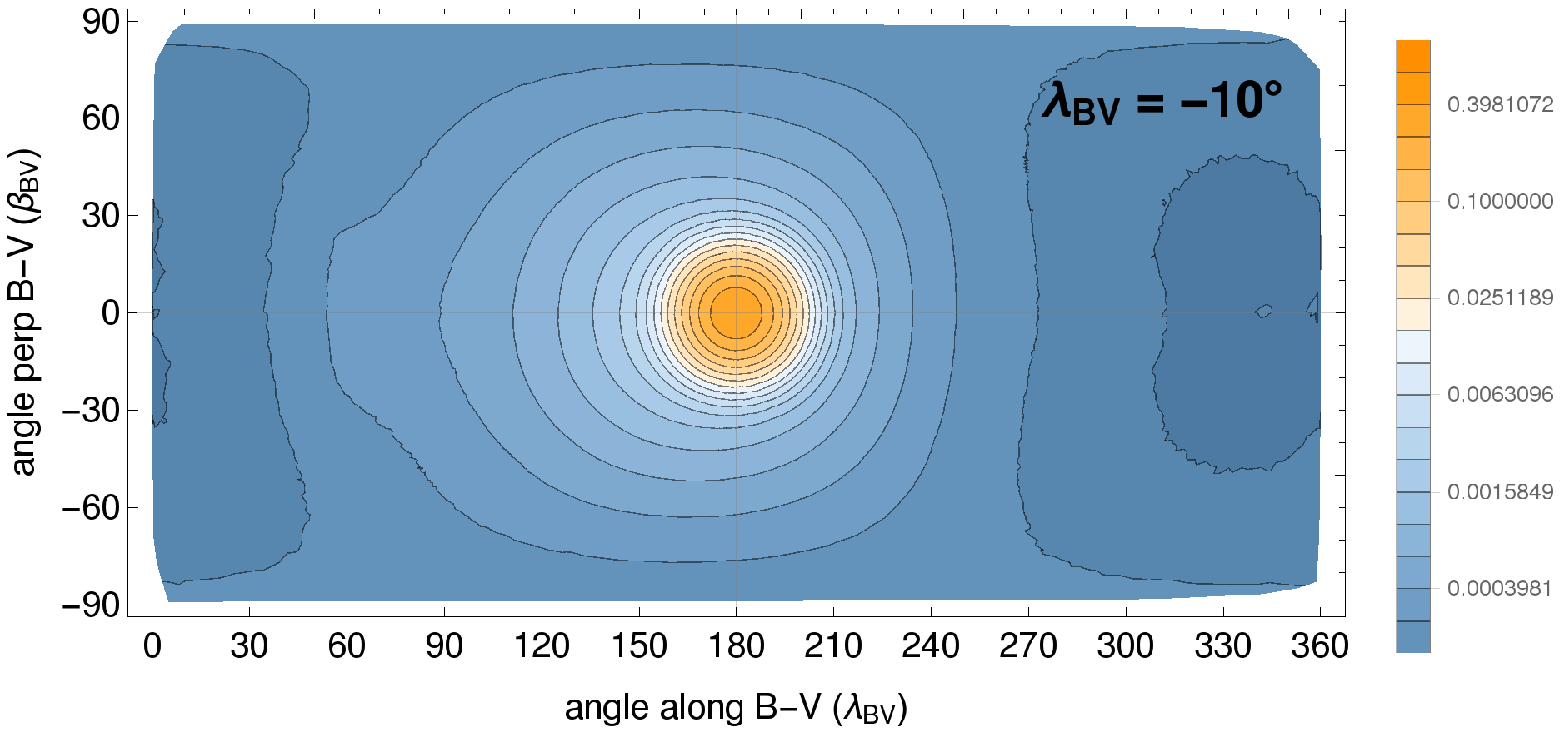}{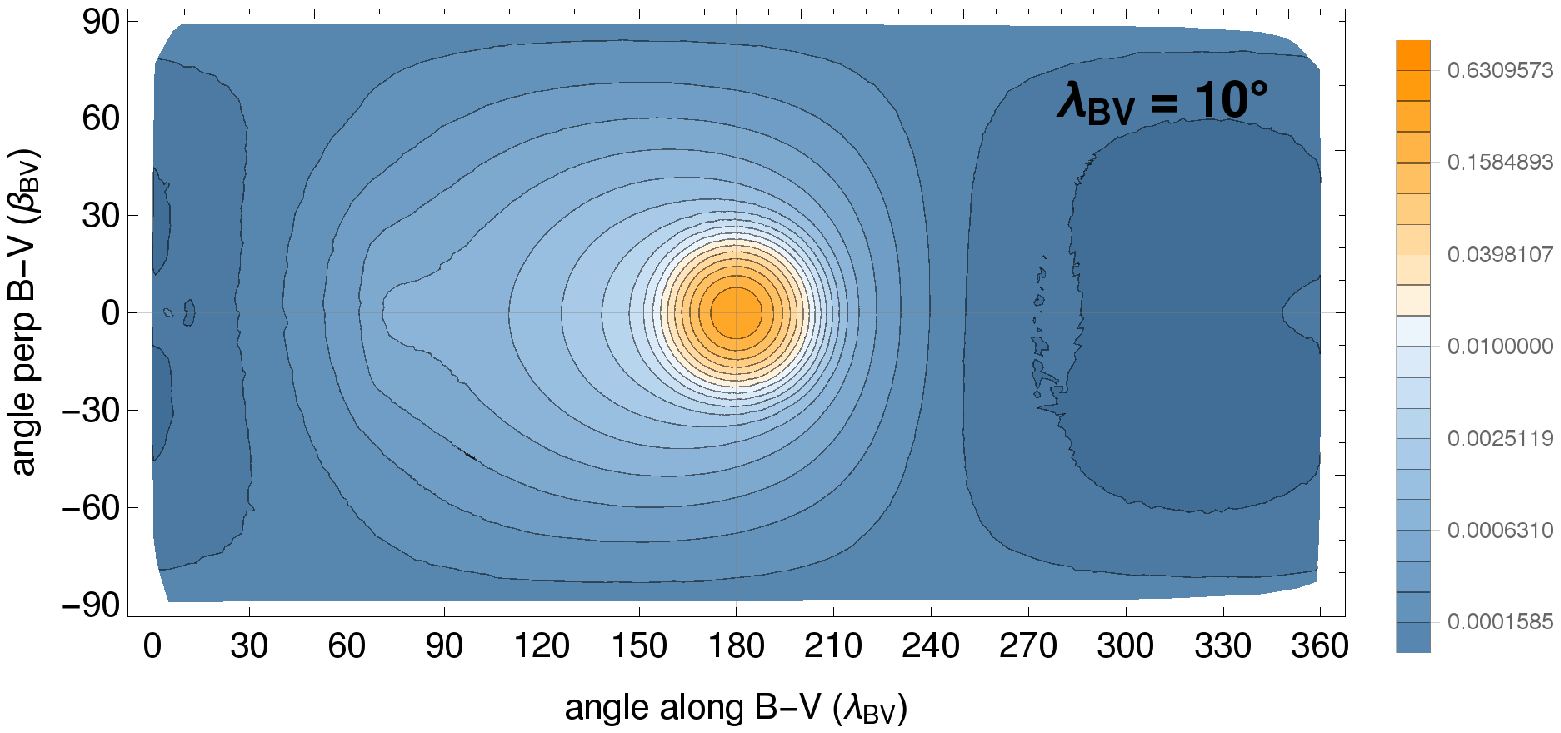}
\plottwo{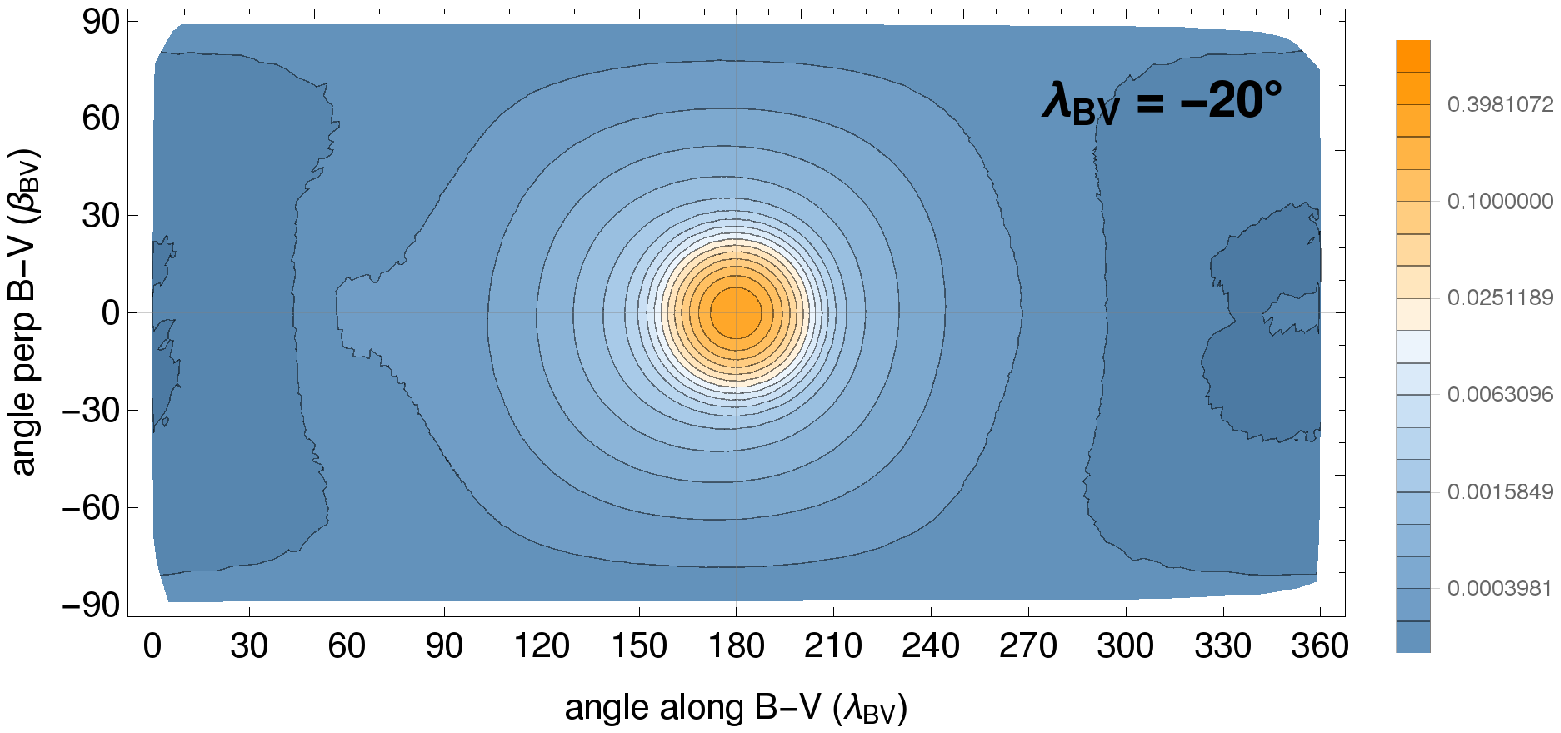}{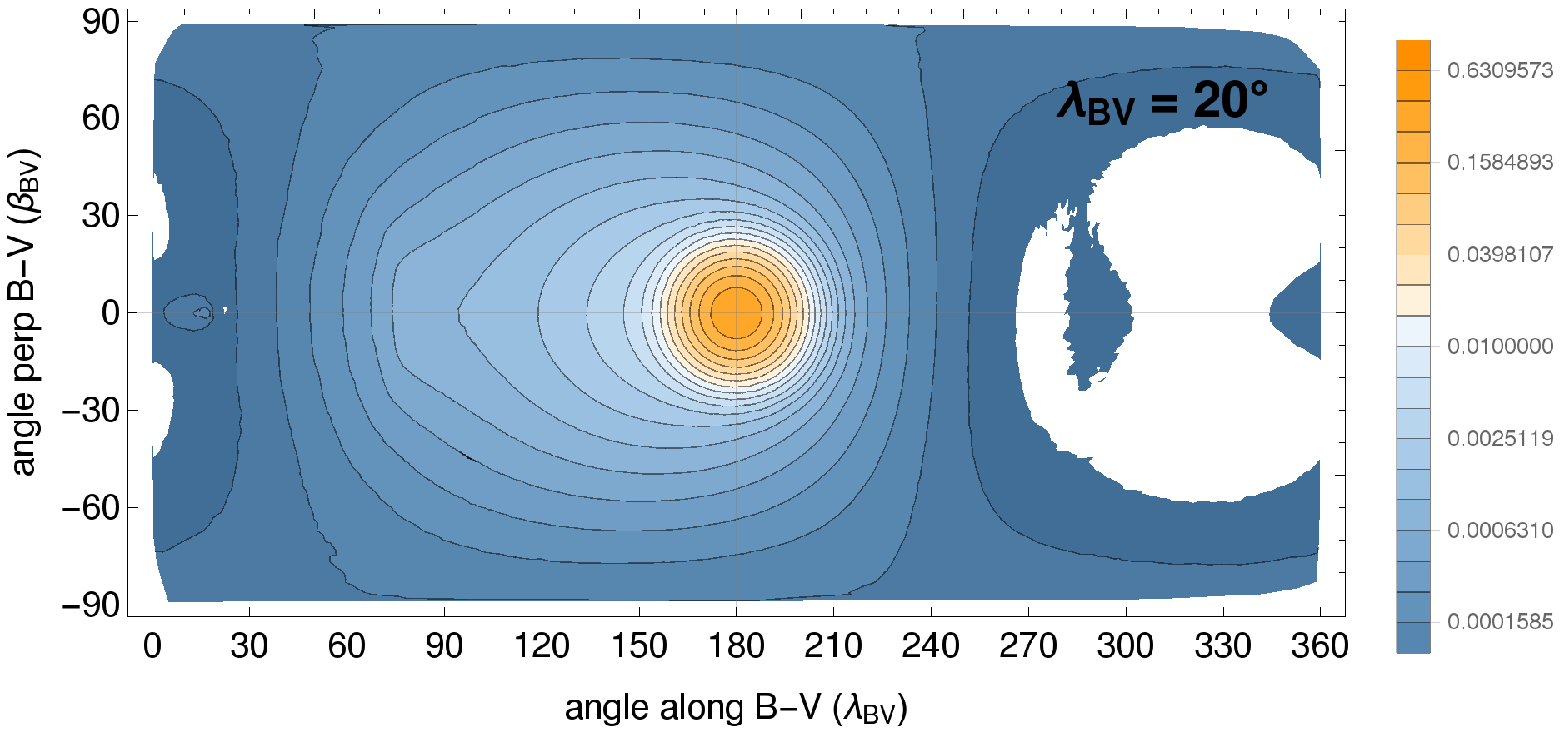}
\plottwo{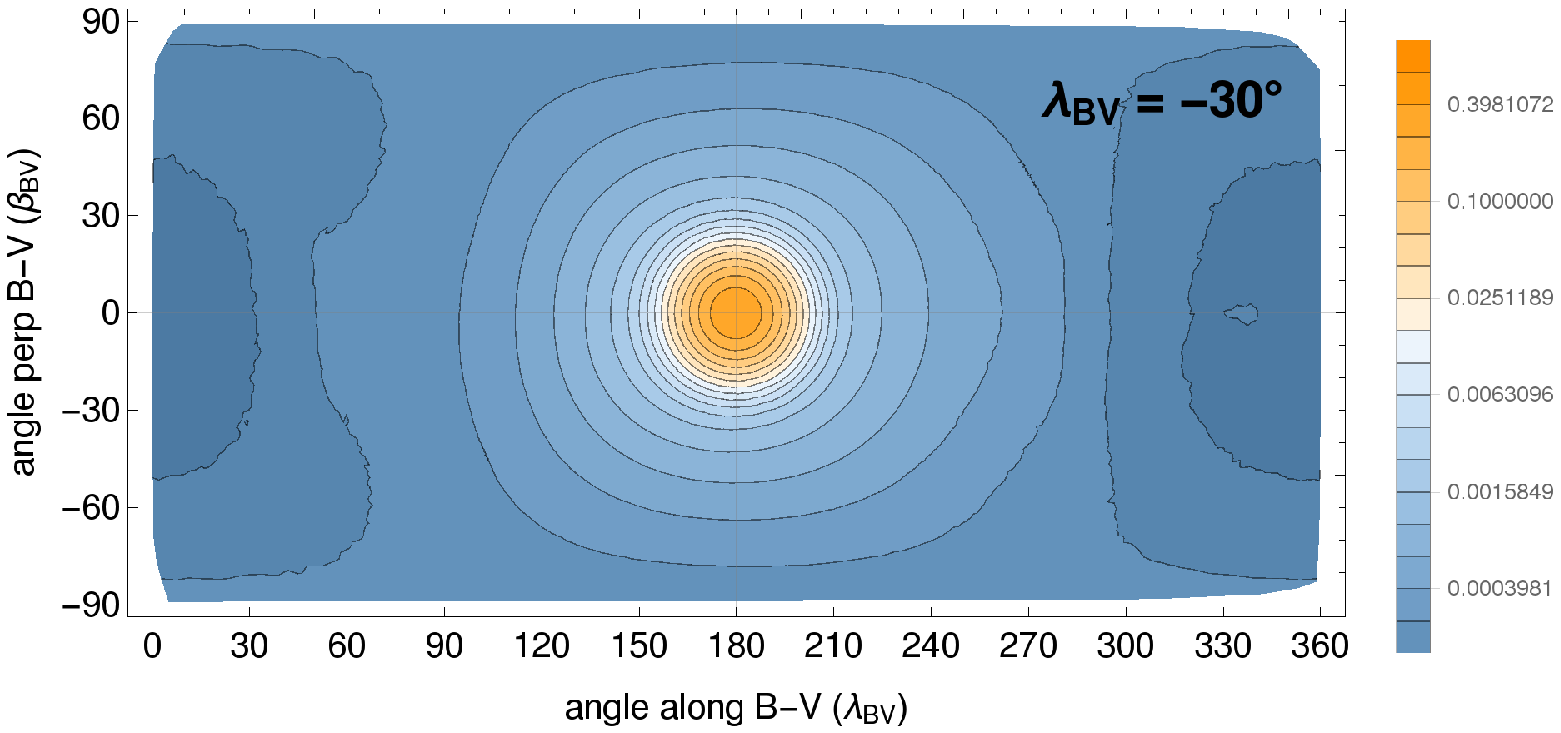}{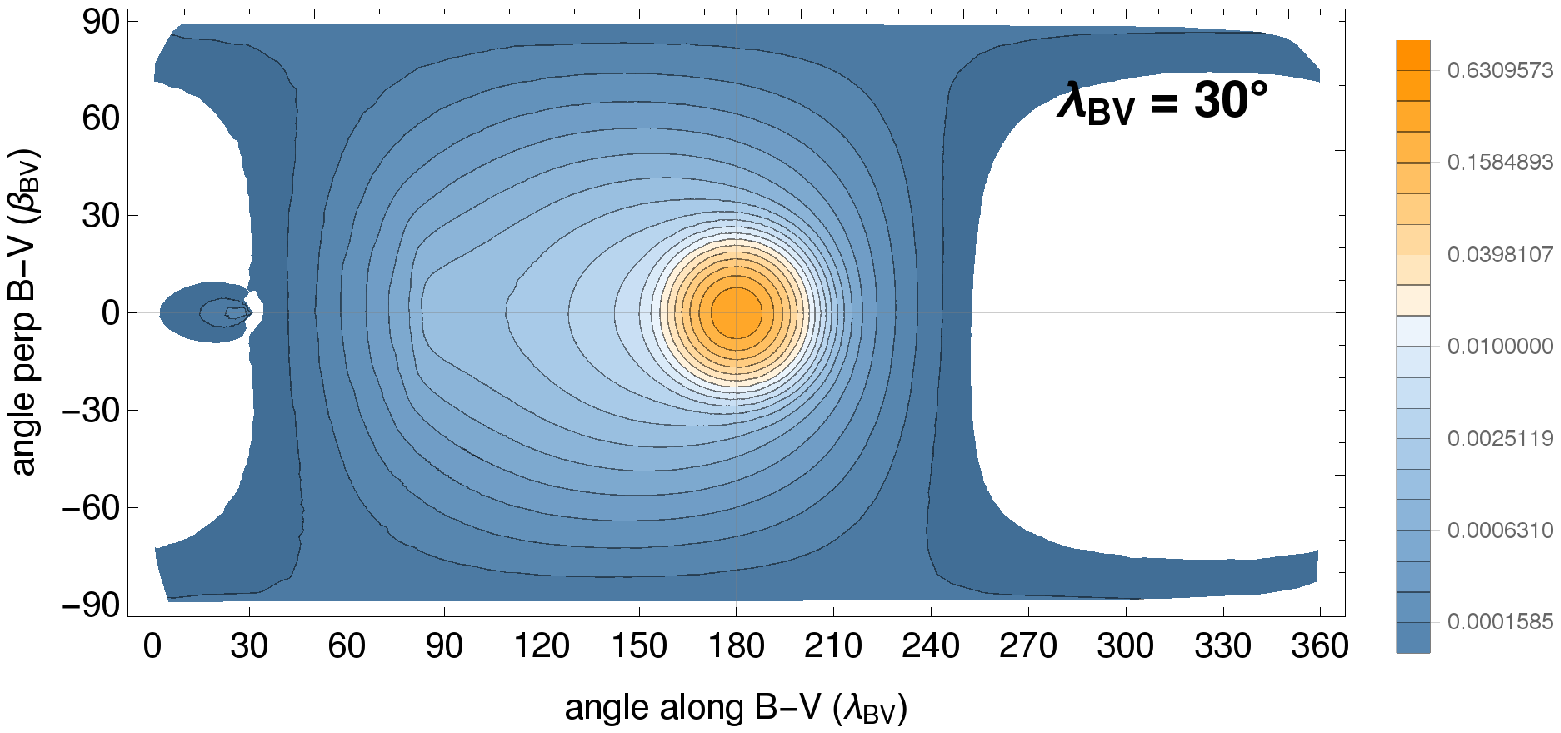}
\plottwo{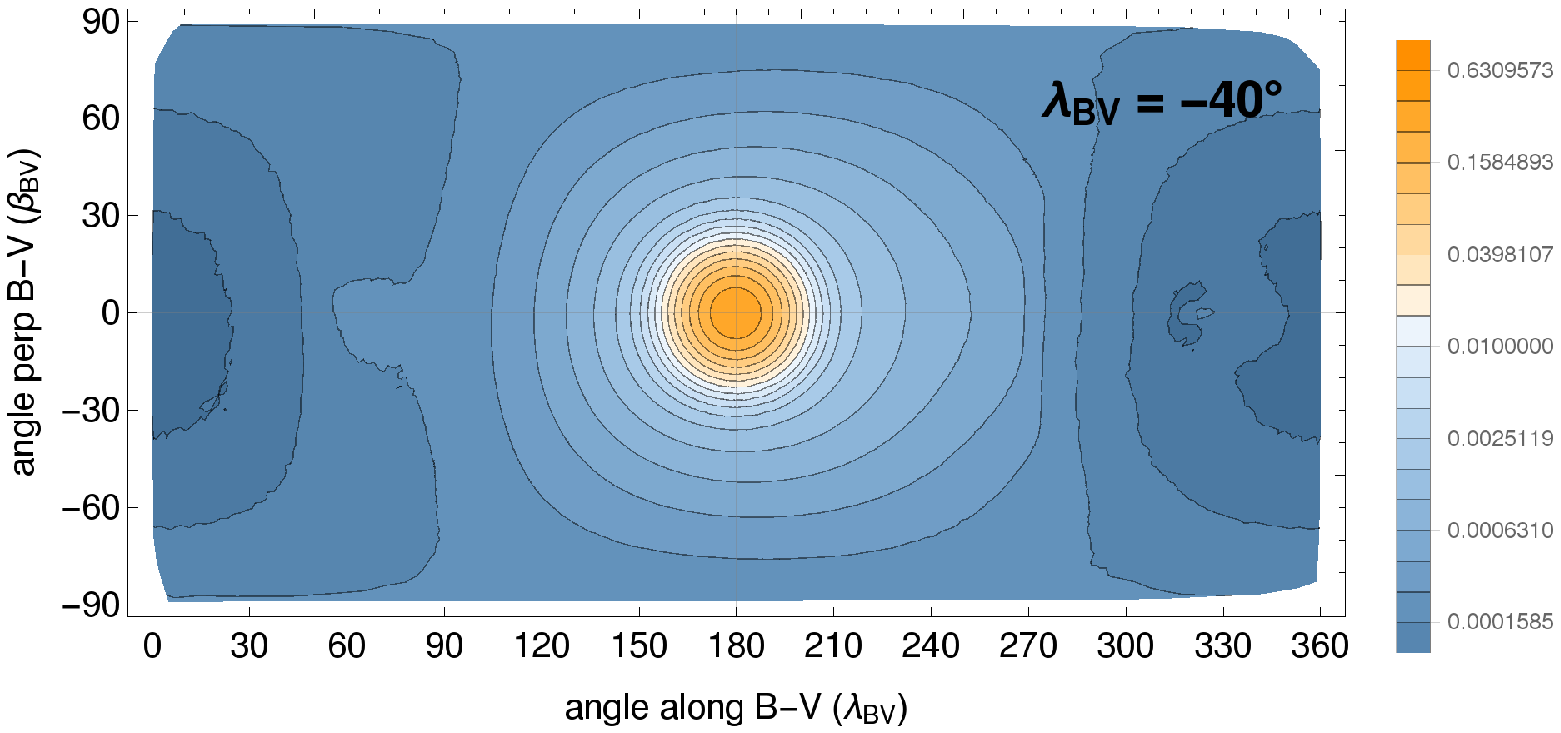}{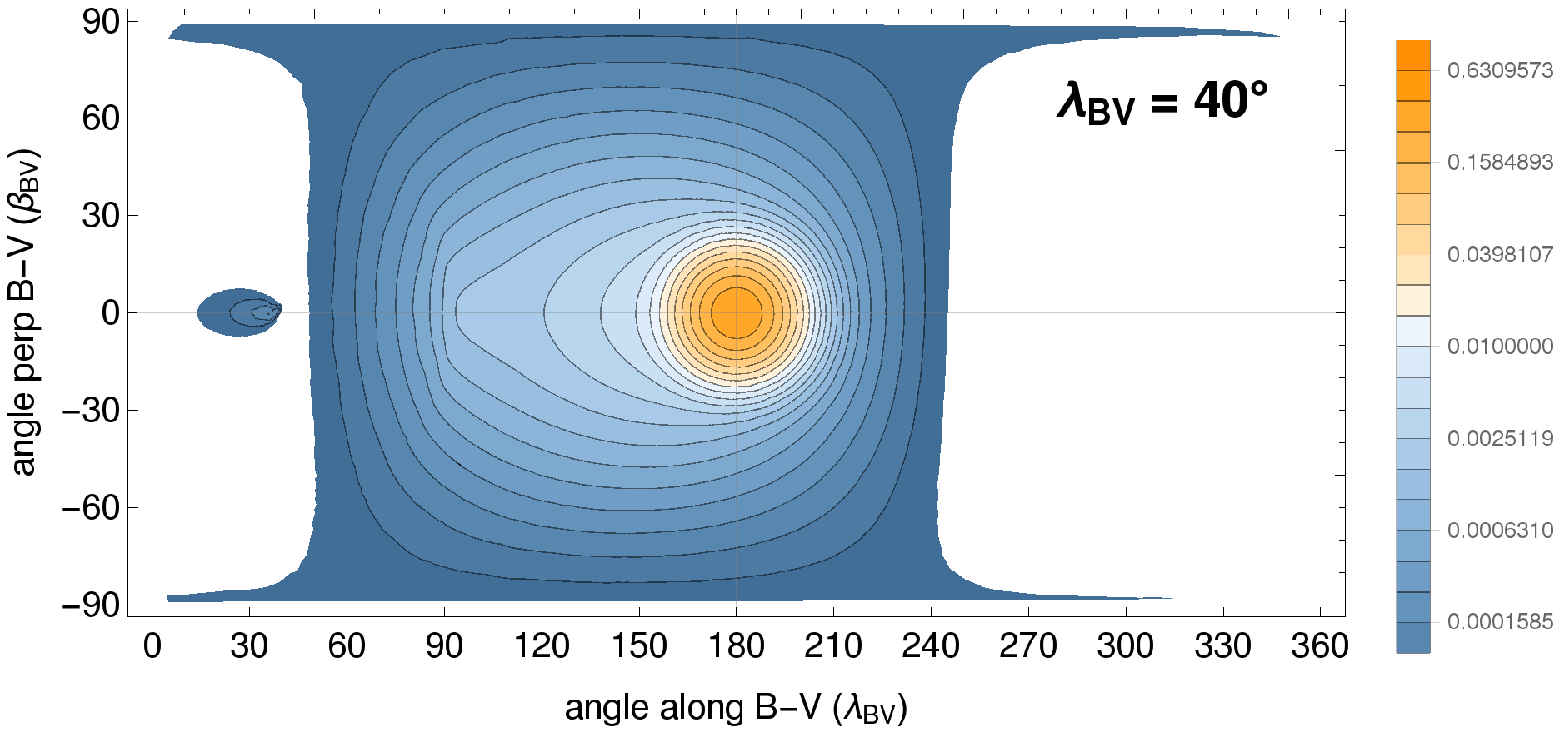}
\caption{Maps of speed-integrated distribution function of ISN He $F_V$, defined in Equation~\ref{eq:difuVInt}, along the B-V plane at 150~au from the Sun. The BV-longitudes of the locations are marked in the panels. The function is normalized to the peak value of $F_V$ at 1000~au upwind. The locations shown are distributed in the B-V plane symmetrically relative to the upwind direction. The flow direction of the Warm Breeze, used in the two-Maxwellian approximation is (172.1\degr, -1.1\degr) while the flow direction of the primary population in these coordinates is (180\degr, 0\degr). 
}
\label{fig:alongBV}
\end{figure}

To further investigate the distribution function in the OHS, we now study maps for locations along the B-V plane, symmetrically distributed at $\pm 10\degr, \pm 20\degr$, $\pm 30\degr$, and $\pm 40\degr$ from the upwind direction at 150~au from the Sun, as shown in Figure~\ref{fig:alongBV}. In all of the selected locations, the ISN He peak does not change its position: the gas flows towards the Sun. However, the secondary population, which occupies almost the entire sky (except for the location at 30\degr and 40\degr) features characteristic asymmetries. While at $\lambda_{BV} \ge 0\degr$ (right-hand column in Figure~\ref{fig:alongBV}) the flow of the secondary atoms is consistently directed towards BV-longitudes between 0\degr{} and 180\degr, in the locations at the other side of the upwind direction the asymmetry of the flow changes the direction somewhere between $\lambda_{BV} = -30\degr$ and $-20\degr$. For heliocentric distance 200~au (not shown), this asymmetry change is also present, for slightly different directions: between $-20\degr${} and $-10\degr${}. This suggests that the behavior of the macroscopic features of the secondary population is related to the direction of interstellar magnetic field, which is deflected from the upwind direction by 39.5\degr. Most likely, the maximum of the production rate of the secondary population is at a distance of about 150--200~au from the Sun, and the asymmetry change occurs at this distance at $\lambda_{BV}$ close to the inflow direction of the Warm Breeze found by \citet{kubiak_etal:16a}. This may imply that the direction of inflow of the secondary population obtained from analysis of ISN He atom observations using the two-Maxwellian approximation points  towards the direction of maximum production of the secondary population in the OHS. This topic requires more in-depth studies, which will be subject of a future paper. 

\subsection{Profiles of B-V longitude and latitude integrated distribution function}
\label{sec:longiLatiProfiles}
\subsubsection{Radially along the upwind direction}
\label{sec:longLatiRadial}
\begin{figure}
\plottwo{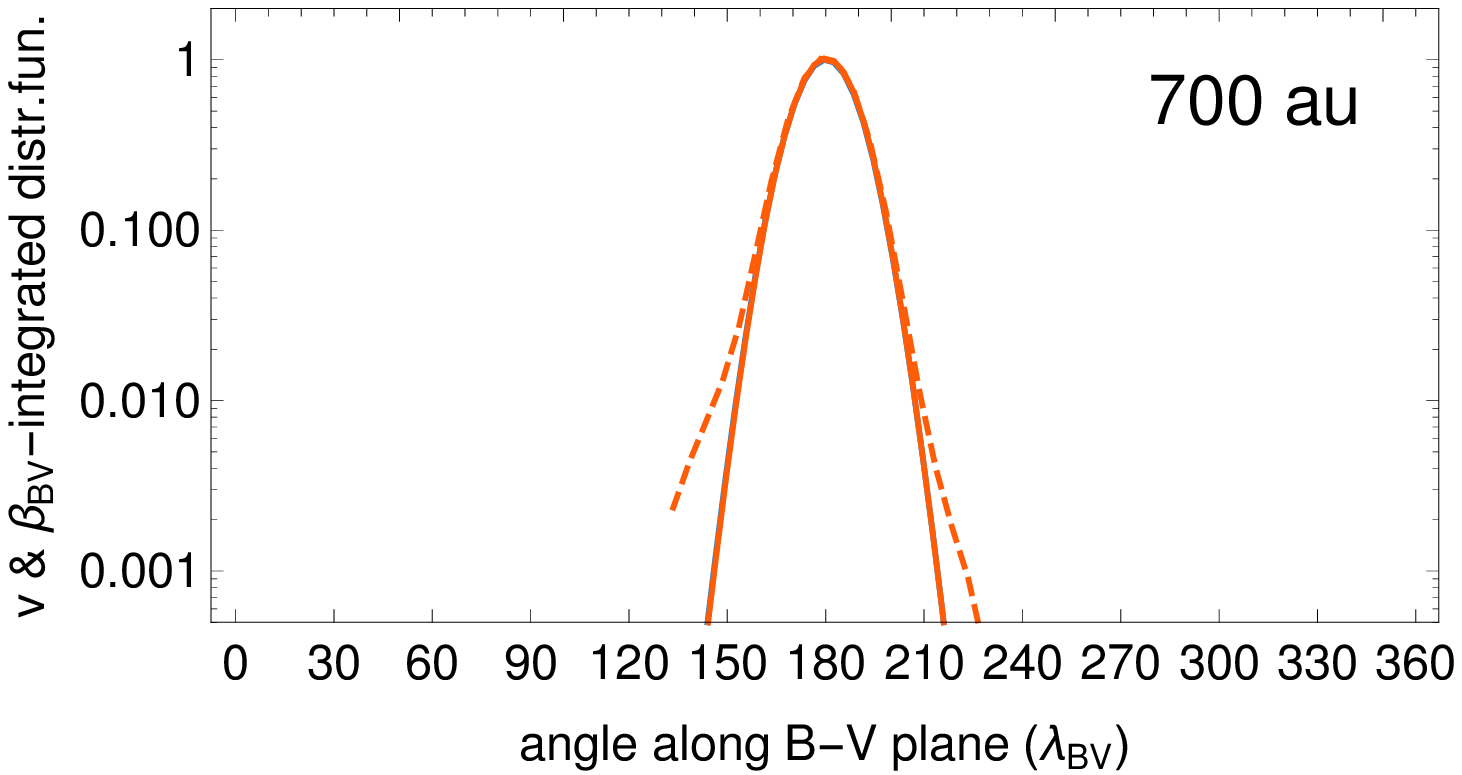}{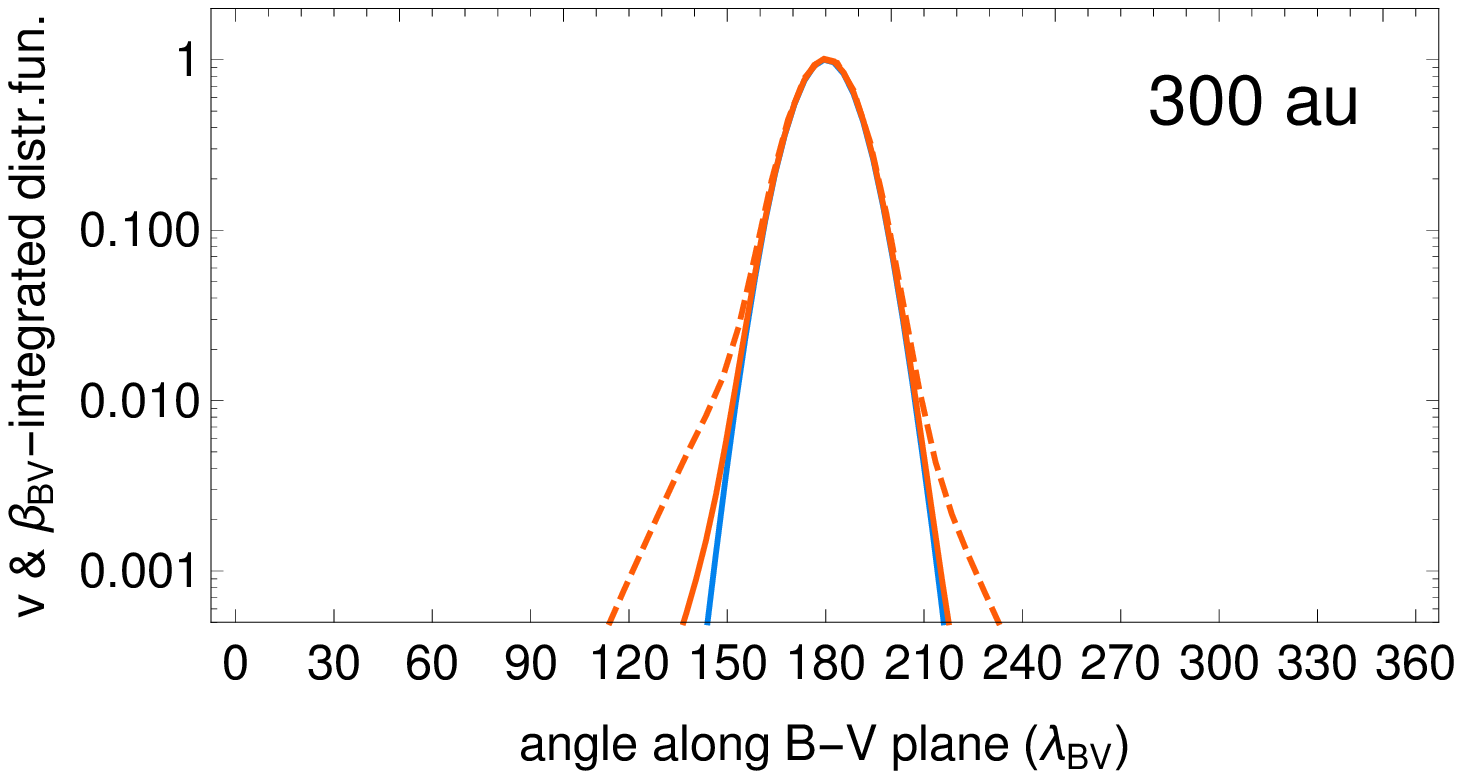}
\plottwo{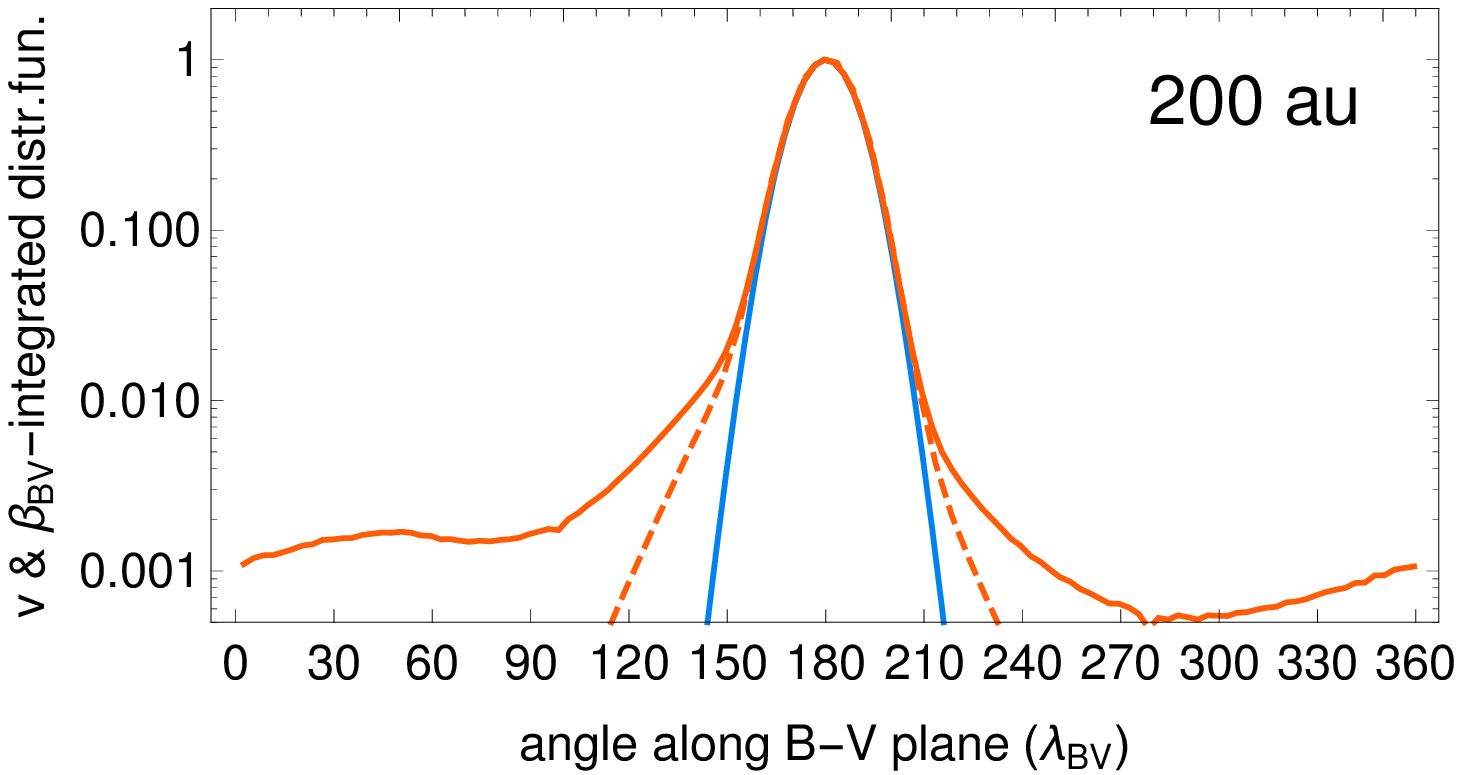}{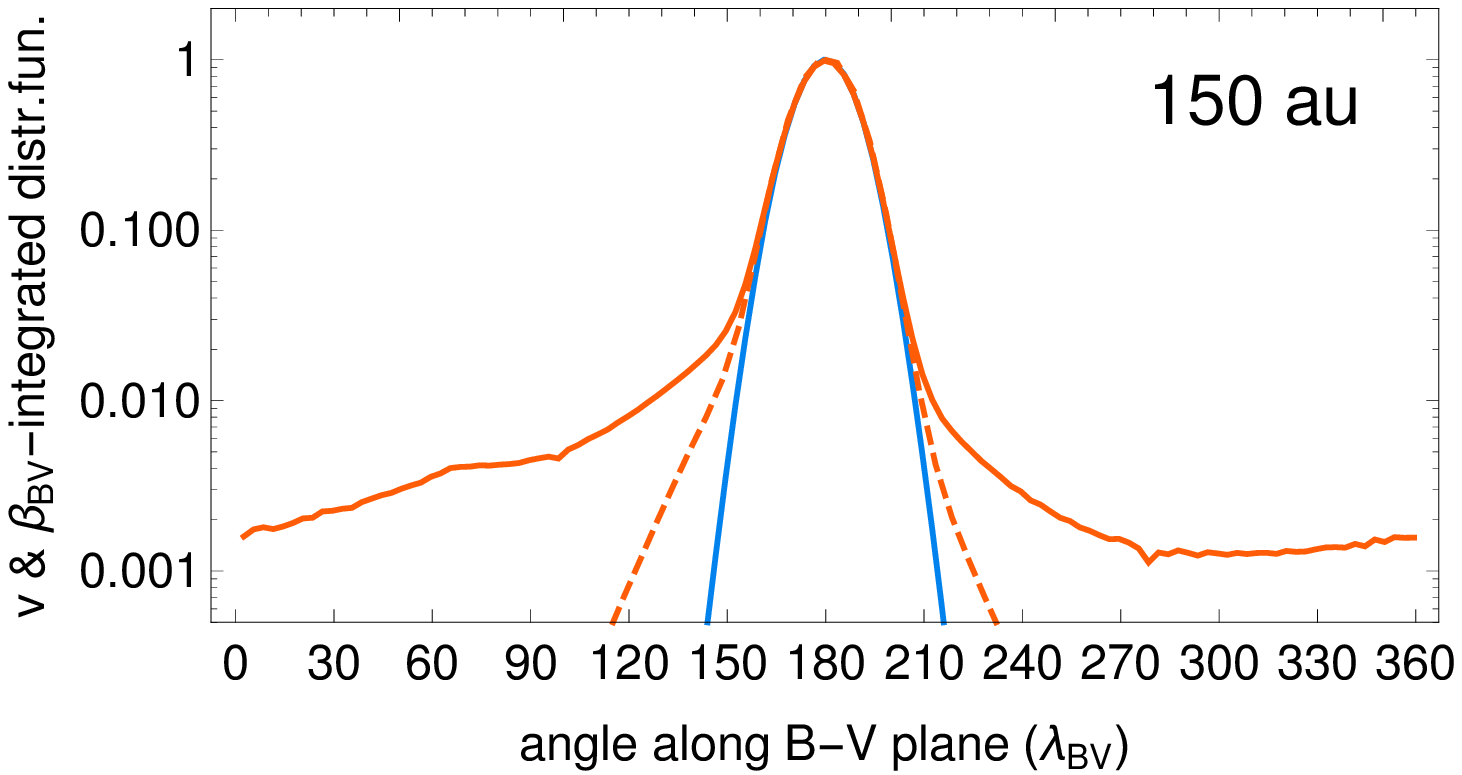}
\plottwo{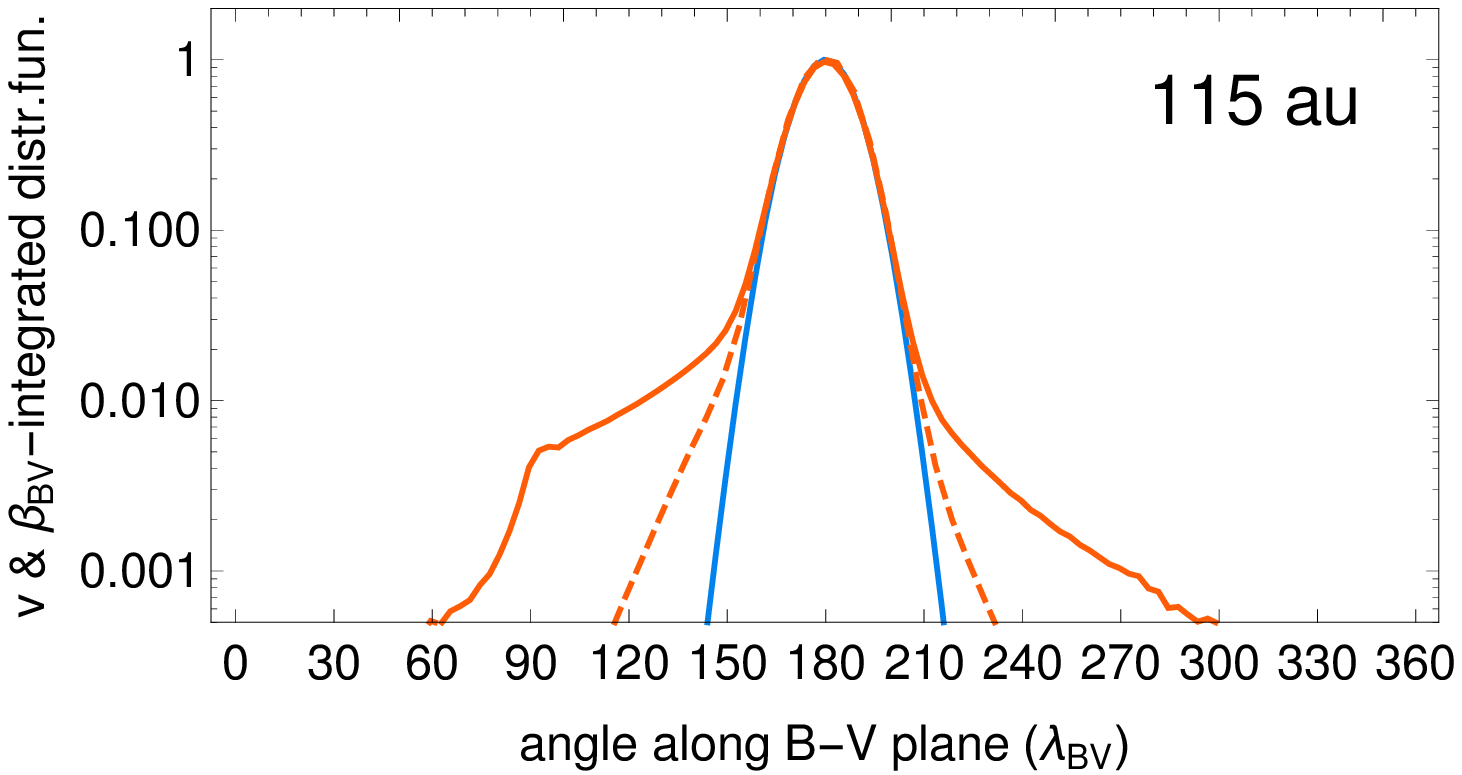}{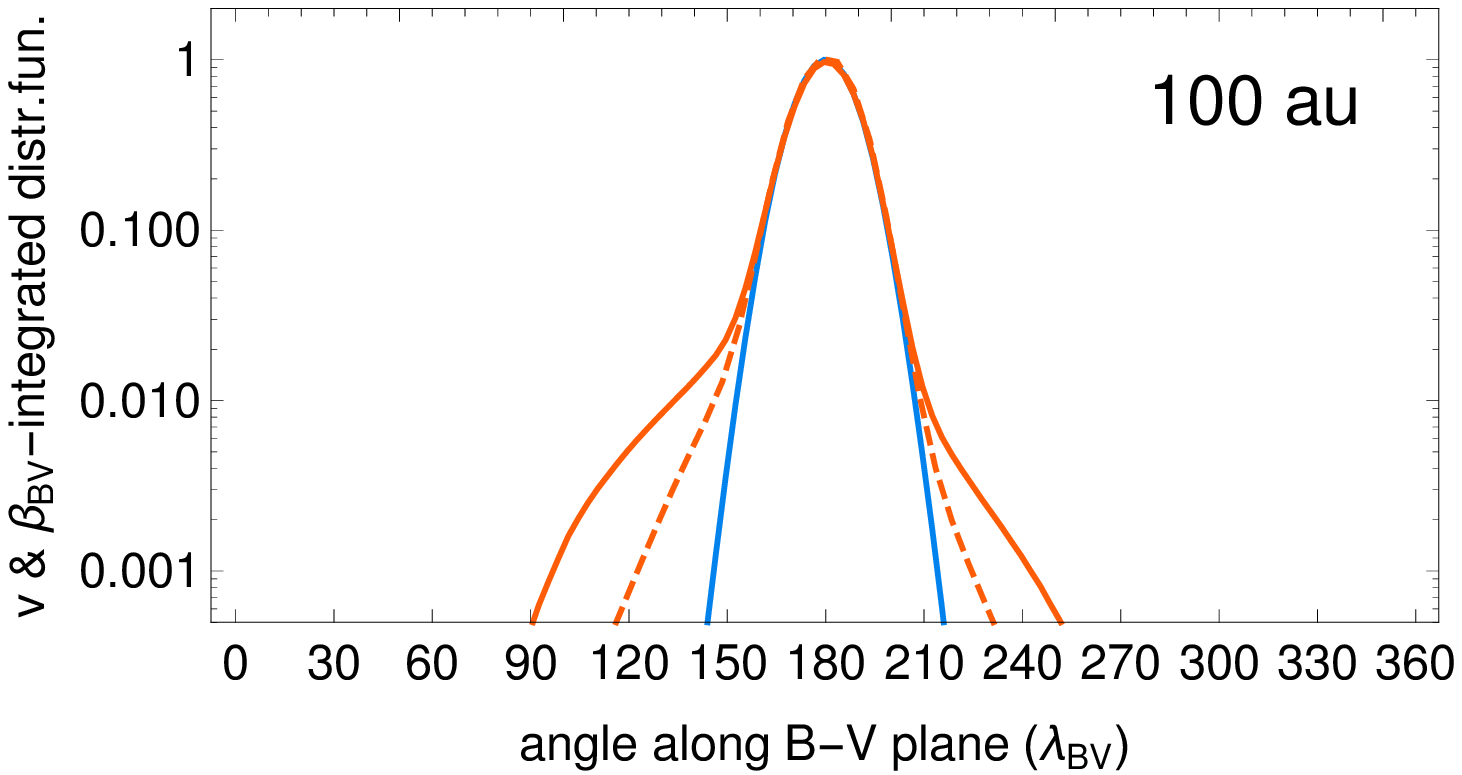}
\plottwo{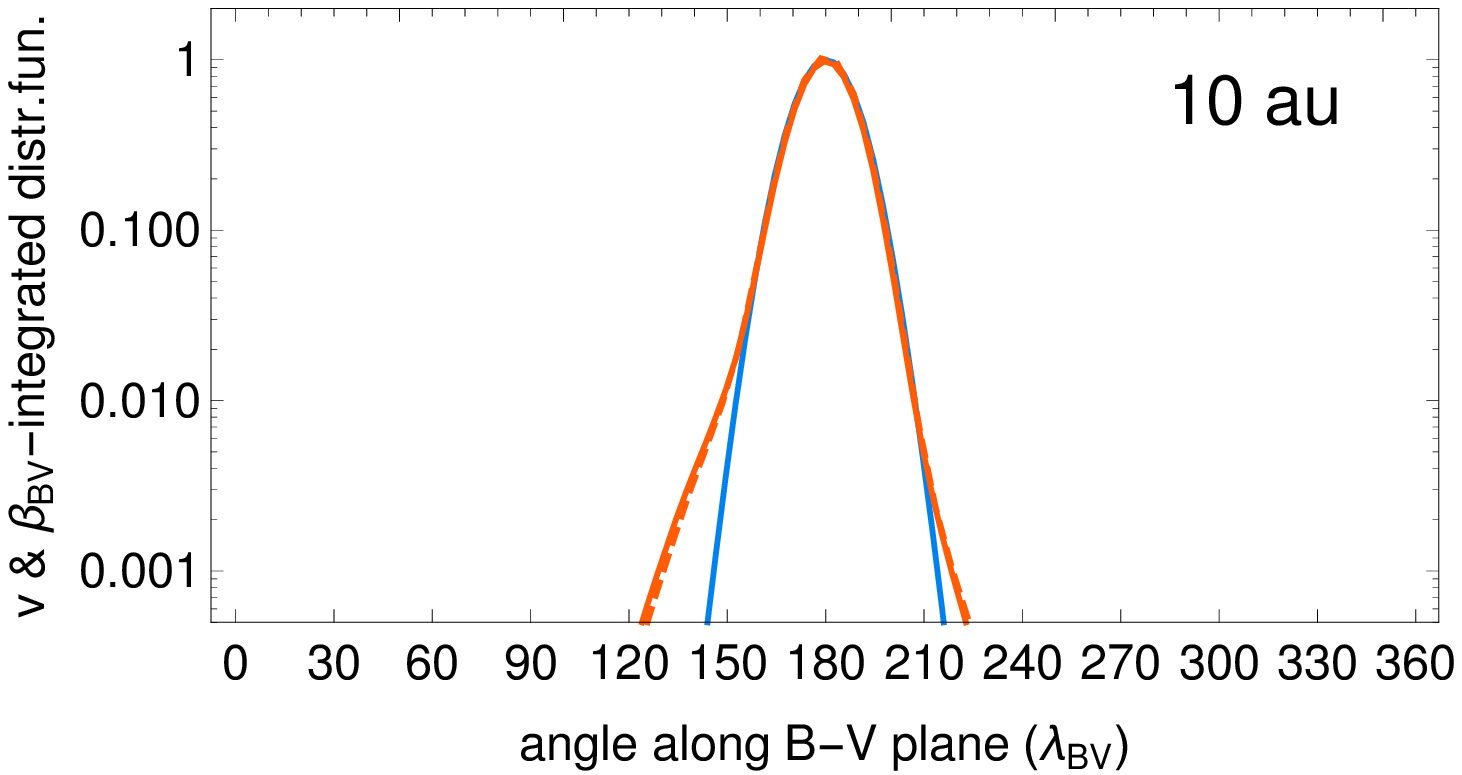}{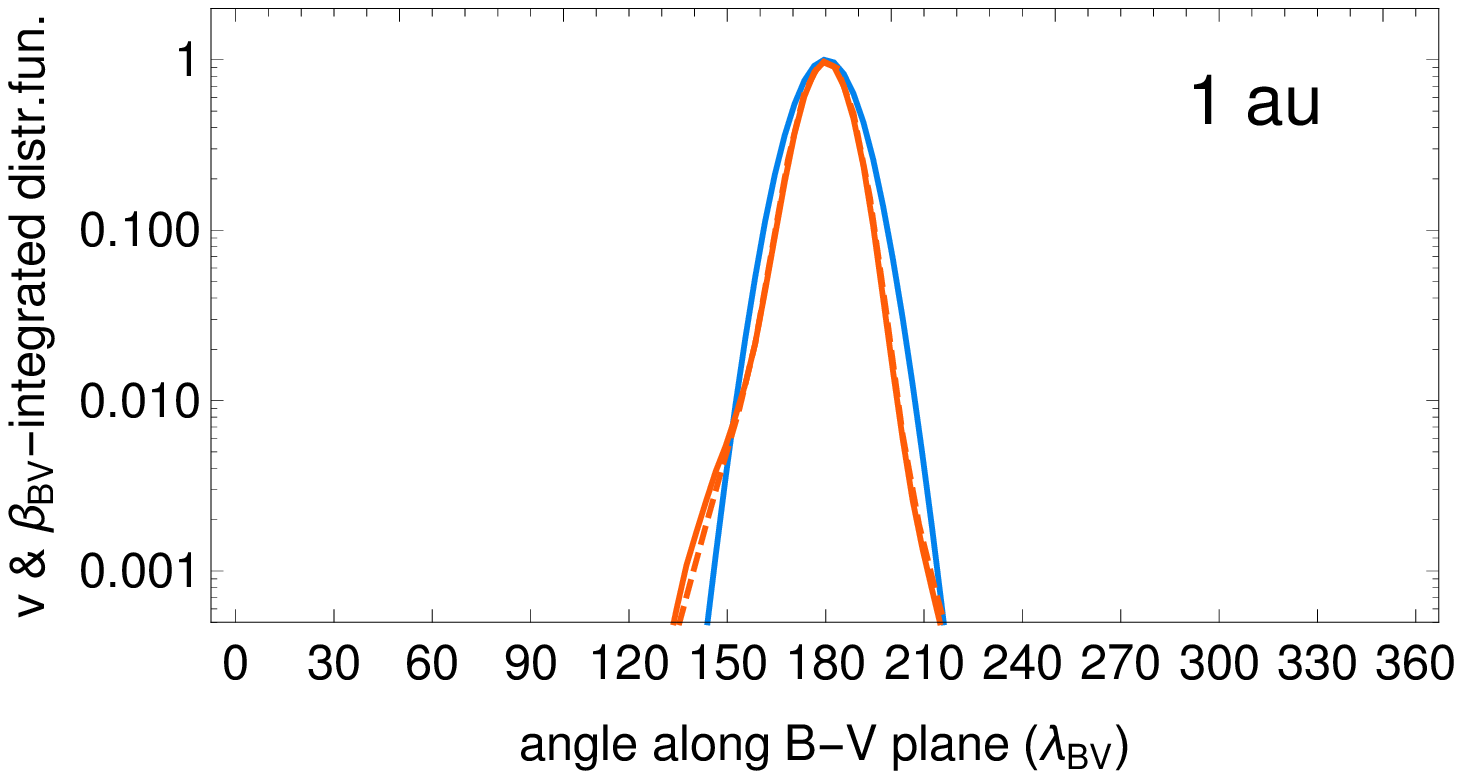}
\caption{Speed- and $\beta_{BV}$-integrated distribution function $F_{V\lambda}$ of ISN He defined in Equation~\ref{eq:difuBetaInt}, shown radially along the upwind direction for selected distances from the Sun, both in the OHS and inside the heliopause. The functions are normalized to the peak values of $F_{V\lambda}$ at 1000~au upwind. The blue line presents the normalized $F_{V\lambda}$ for 1000~au upwind. It is repeated in all panels for reference. The orange solid lines represent the synthesized distribution function obtained in this paper (Equation~\ref{eq:locDiFu}), and the broken line the two-Maxwellian approximation (Equation~\ref{eq:2MaxwDF}) with the parameters adopted from \citet{kubiak_etal:16a} for the secondary population and \citet{bzowski_etal:15a} for the primary population.}
\label{fig:upBVLongiInt}
\end{figure}

\begin{figure}
\plottwo{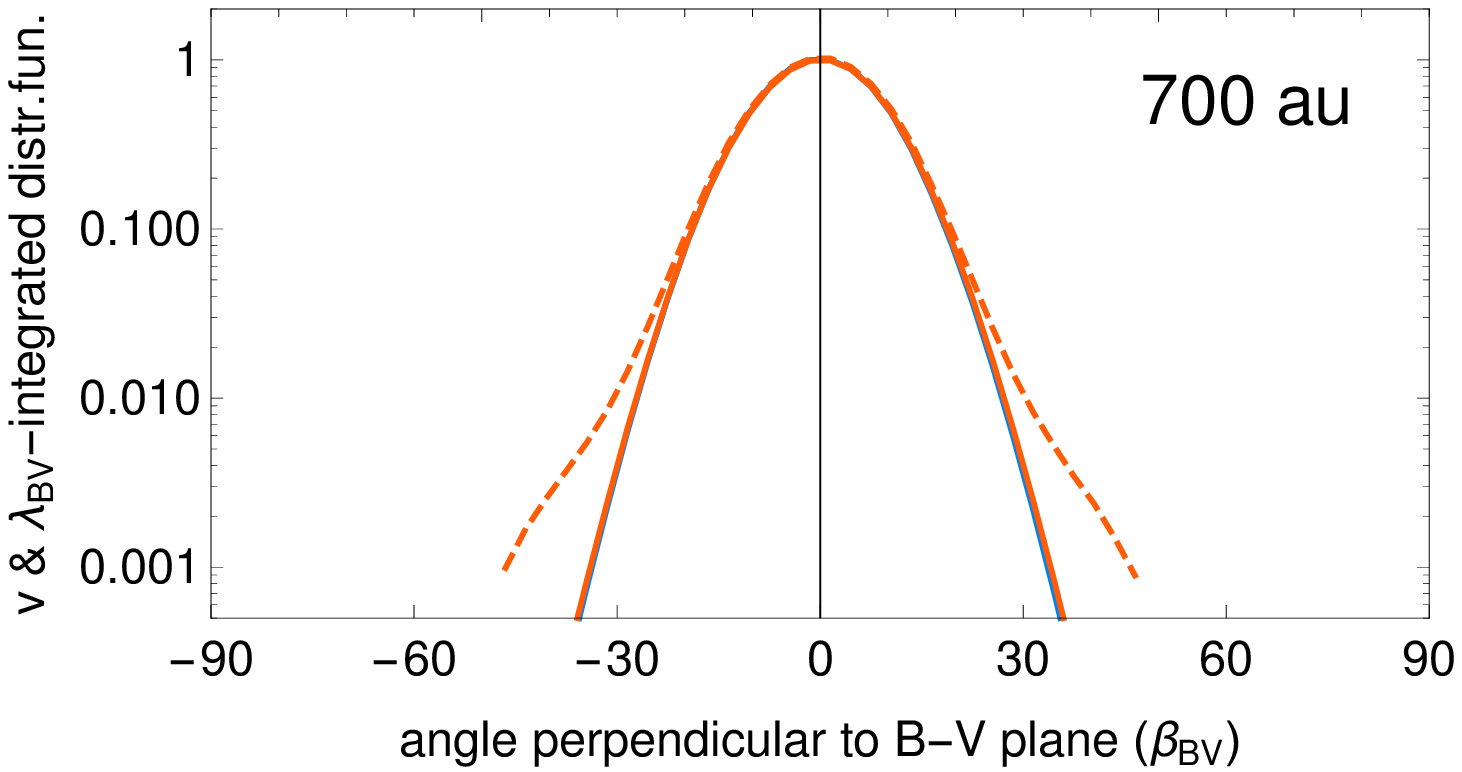}{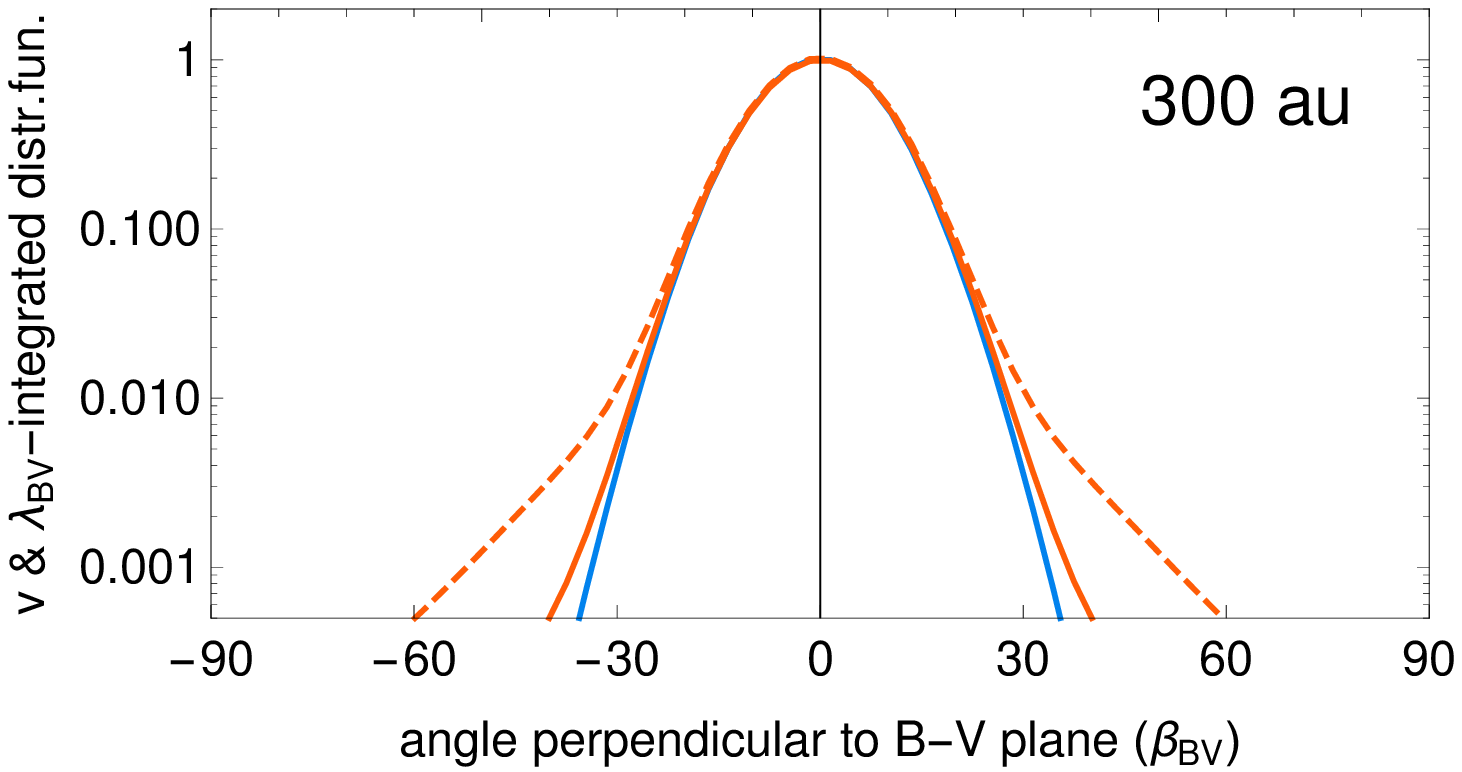}
\plottwo{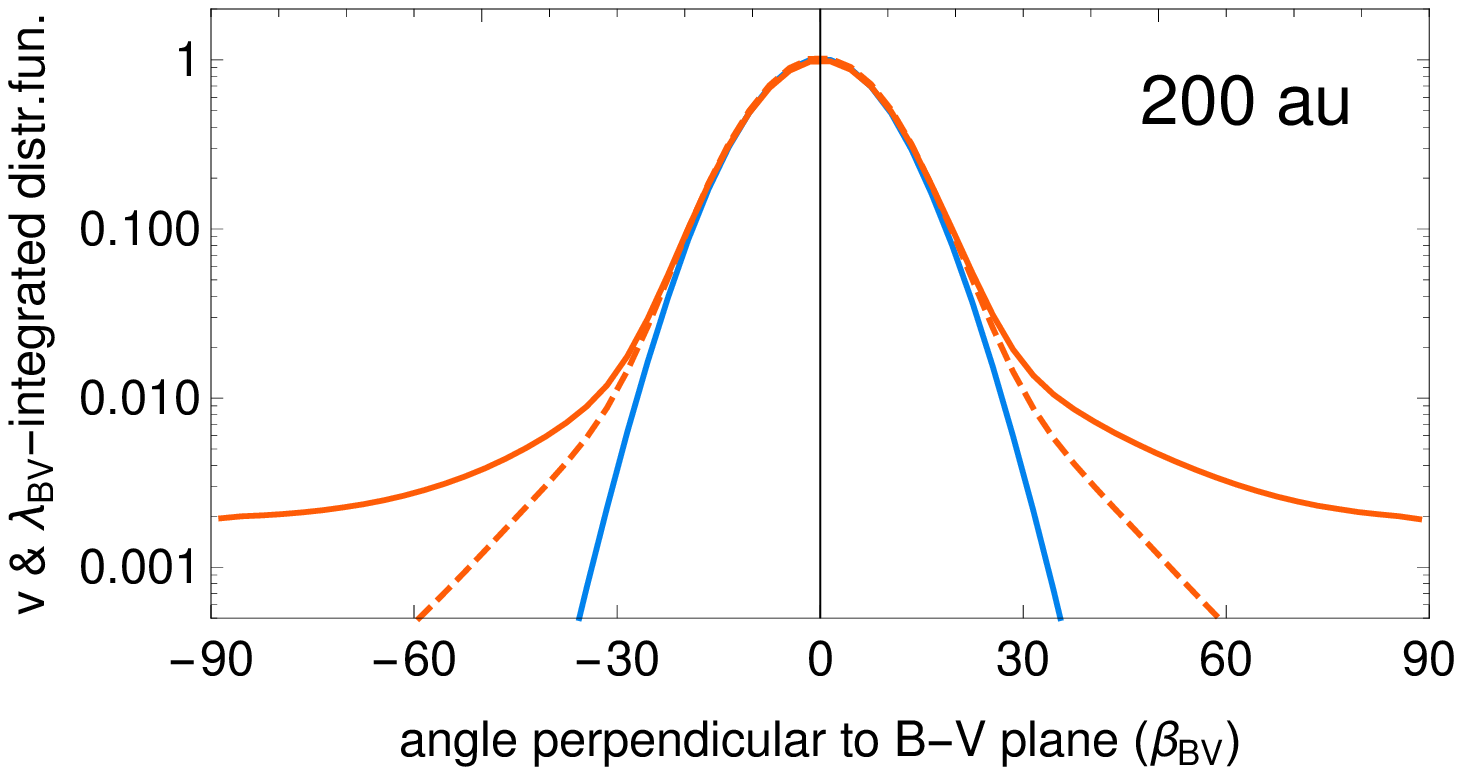}{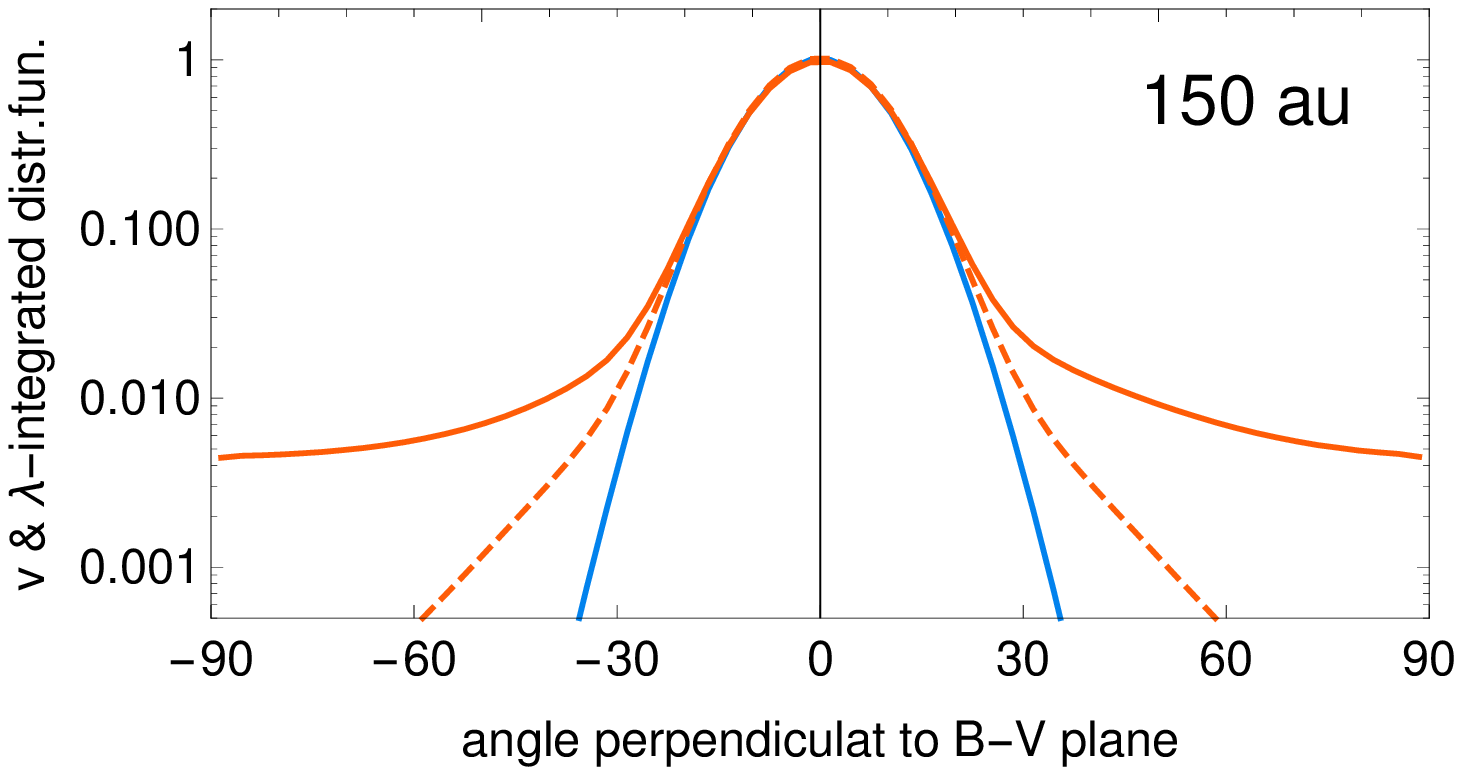}
\plottwo{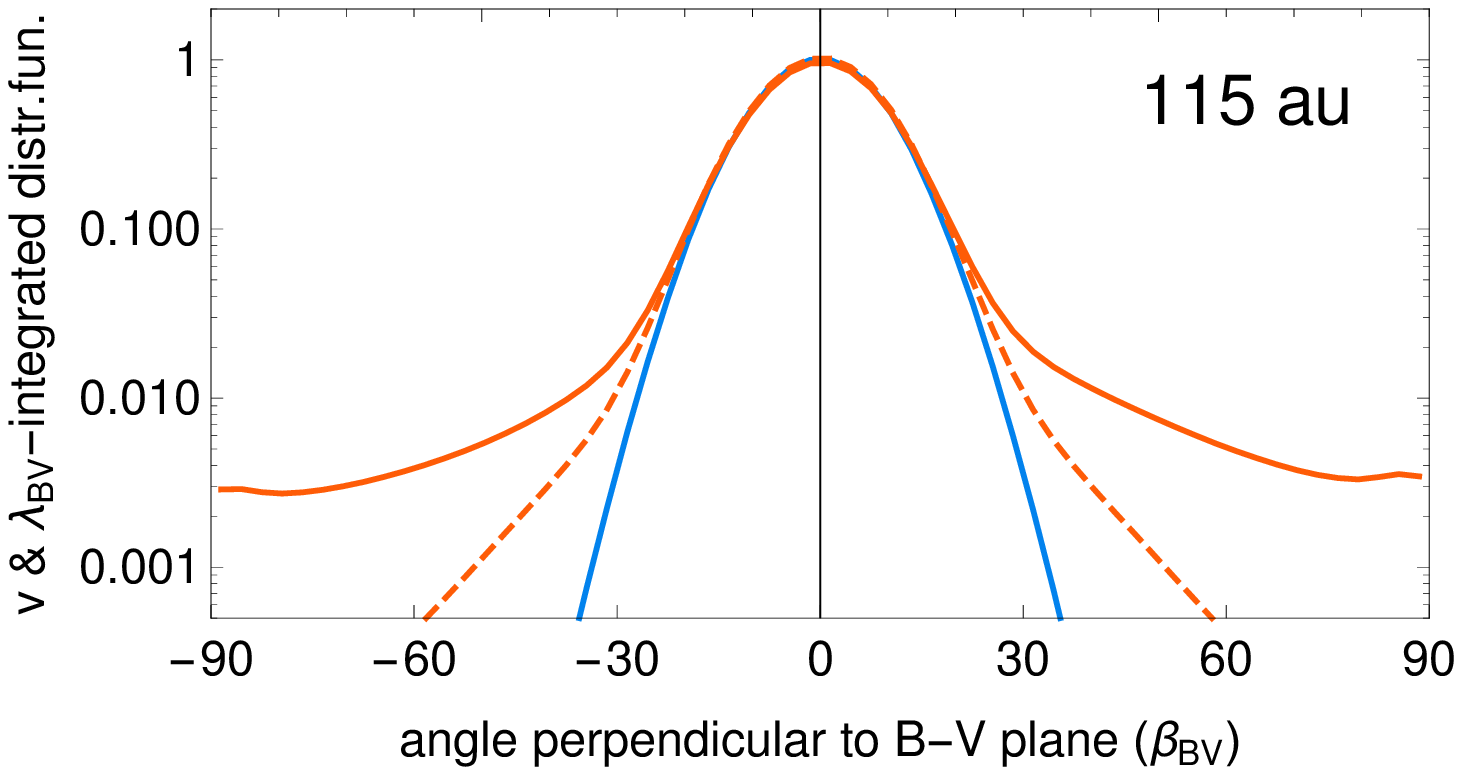}{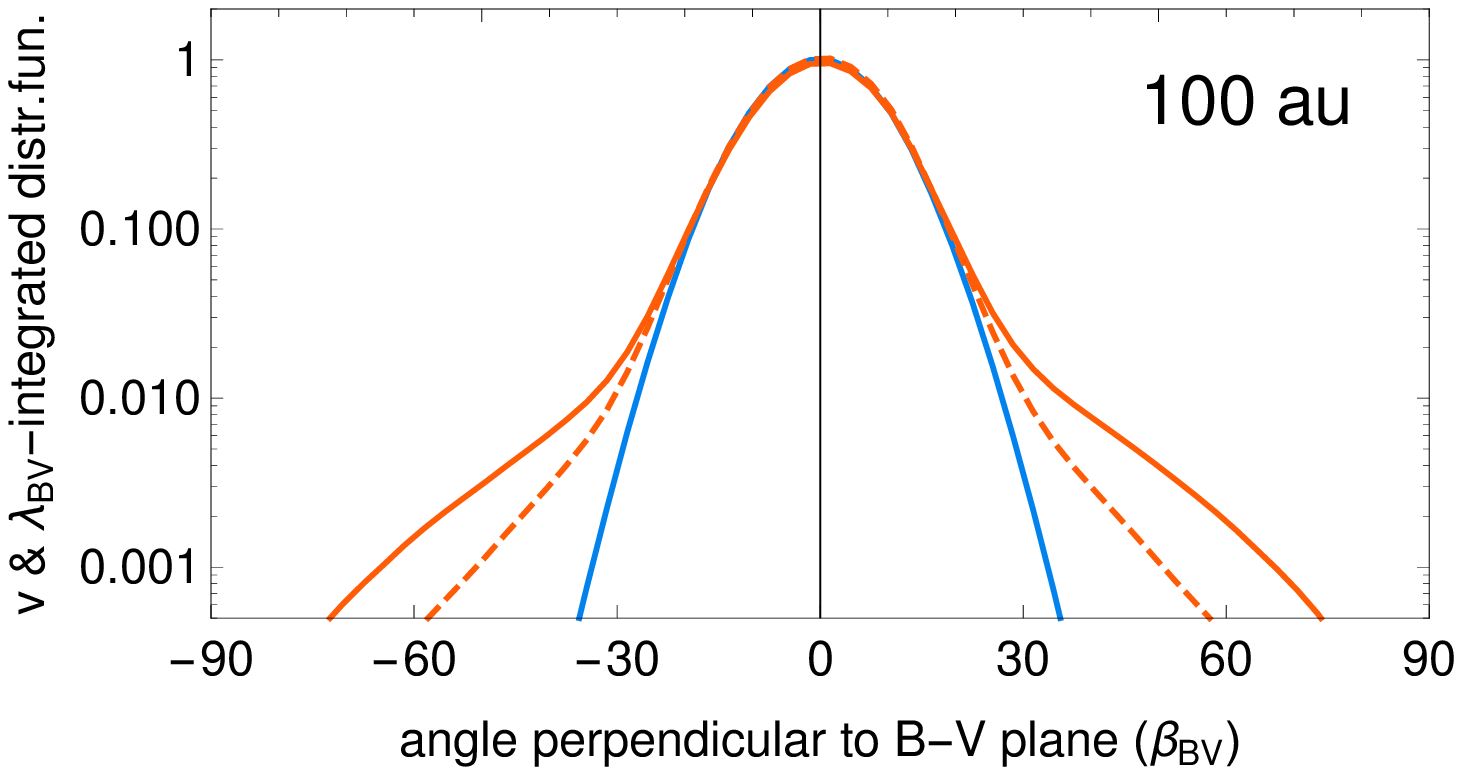}
\plottwo{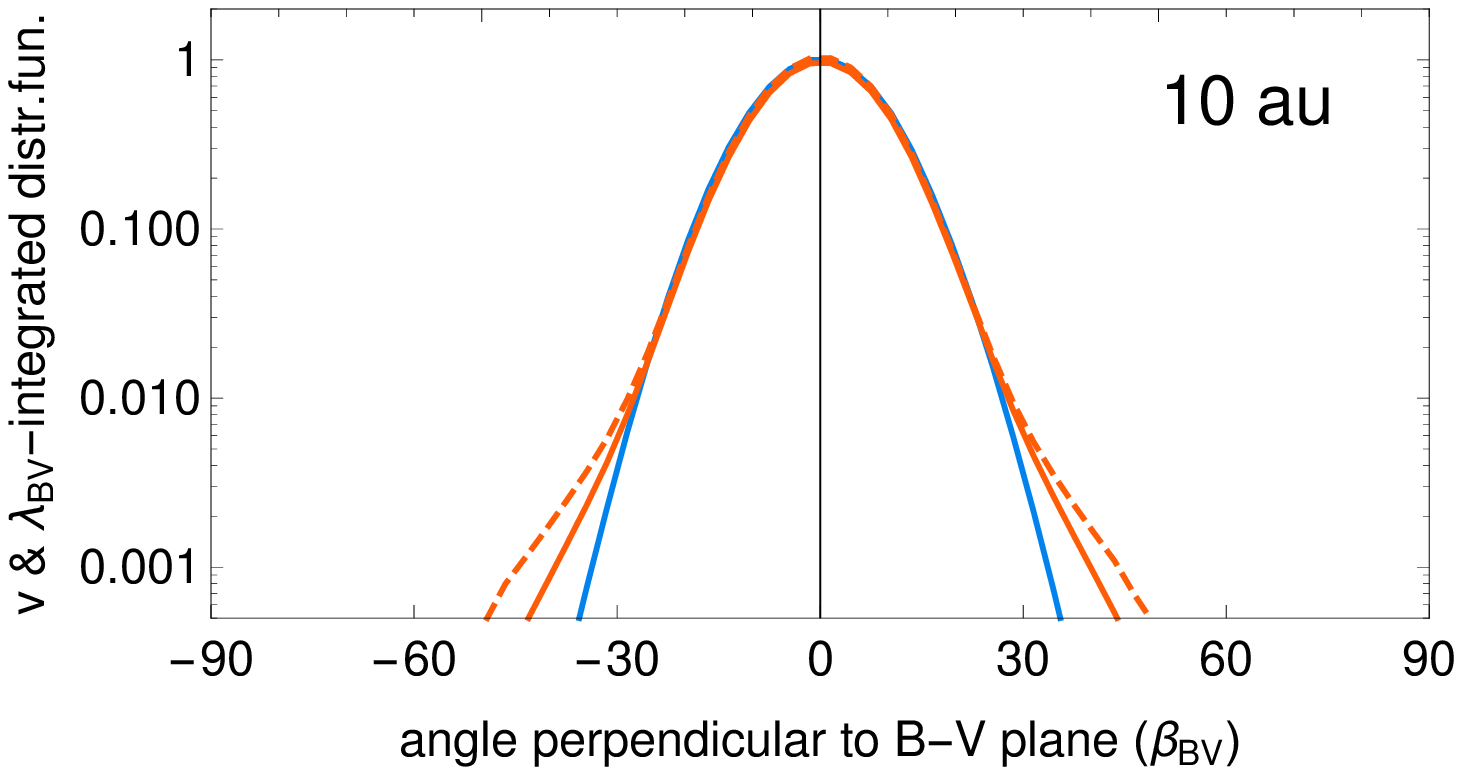}{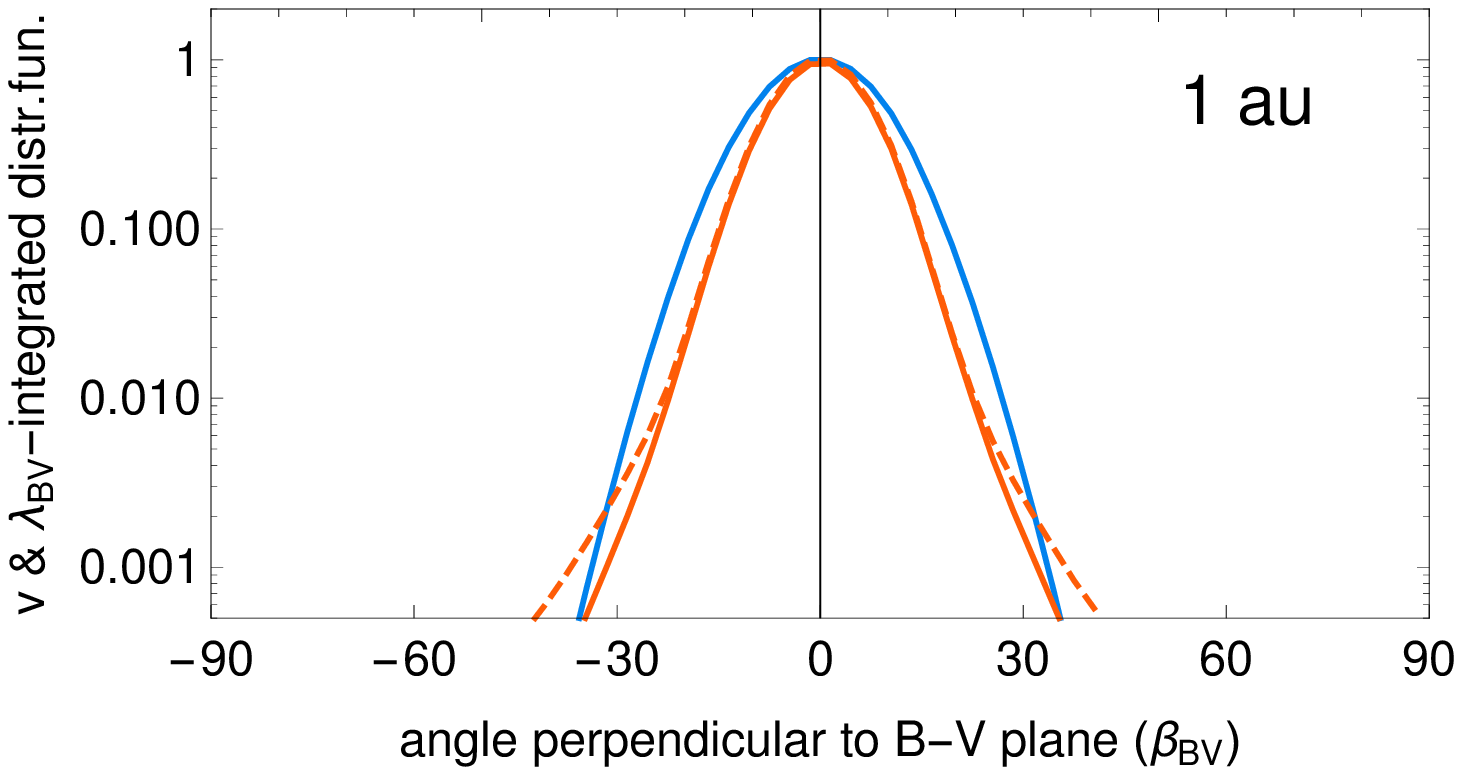}
\caption{Speed- $\lambda_{BV}$-integrated distribution function of ISN He $F_{V\beta}$, defined in Equation~\ref{eq:difuLambdaInt}, shown radially along the upwind direction for selected distances from the Sun, both in the OHS and inside the heliopause. The line style code is identical as in Figure~\ref{fig:upBVLongiInt}. The function is normalized to the peak value of $F_{V\beta}$ at 1000~au upwind.}
\label{fig:upBVLatiInt}
\end{figure}

To further illustrate the evolution of essential features of the distribution function with distance from the Sun, we consider the distribution function $F_{V \lambda}$ integrated over atom speed and the elevation angle relative to the B-V plane, as defined in Equation~\ref{eq:difuBetaInt}, and over the azimuth angle along the B-V plane, as defined in Equation~\ref{eq:difuLambdaInt}. They are presented in Figures~\ref{fig:upBVLongiInt} and \ref{fig:upBVLatiInt}.

As evident from these figures, the two-Maxwellian model of the distribution function is an inadequate approximation of the distribution function obtained using our synthesis method. In the OHS and down to $\sim 10$~au from the Sun, the two-Maxwellian approximation predicts a central core and a halo, offset by $\sim 8\degr$ towards the direction of interstellar magnetic field. The synthesized function does not resemble this approximation at all, especially in the OHS close to the region of maximum production of secondary He (at $\sim 150 - 200$~au). Closer to the Sun, however, the two-Maxwellian approximation becomes increasingly close to the synthesized function due to selection effects. At 1~au upwind, the two-Maxwellian approximation becomes almost exact along the B-V plane (cf. the lower-right panel in Figure~\ref{fig:upBVLongiInt}), while in the plane perpendicular to the B-V plane (Figure~\ref{fig:upBVLatiInt}), discrepancies appear at $\sim 0.015$ of the maximum value. While this may seem to be small, one needs to remember that a large majority of atoms are the primary ISN He. 

\subsubsection{Laterally along the B-V plane}
\label{sec:longLatiLateral}
\begin{figure}
\plottwo{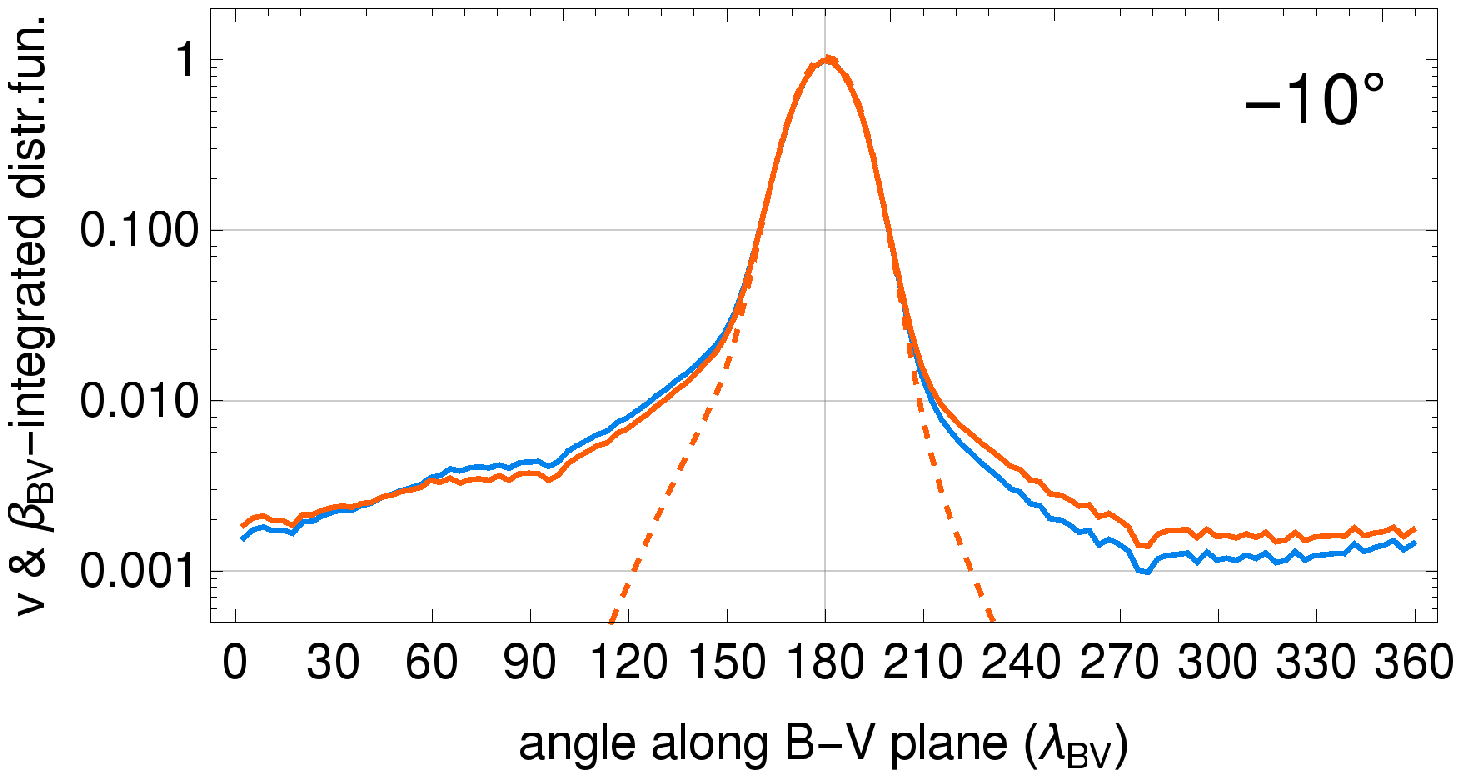}{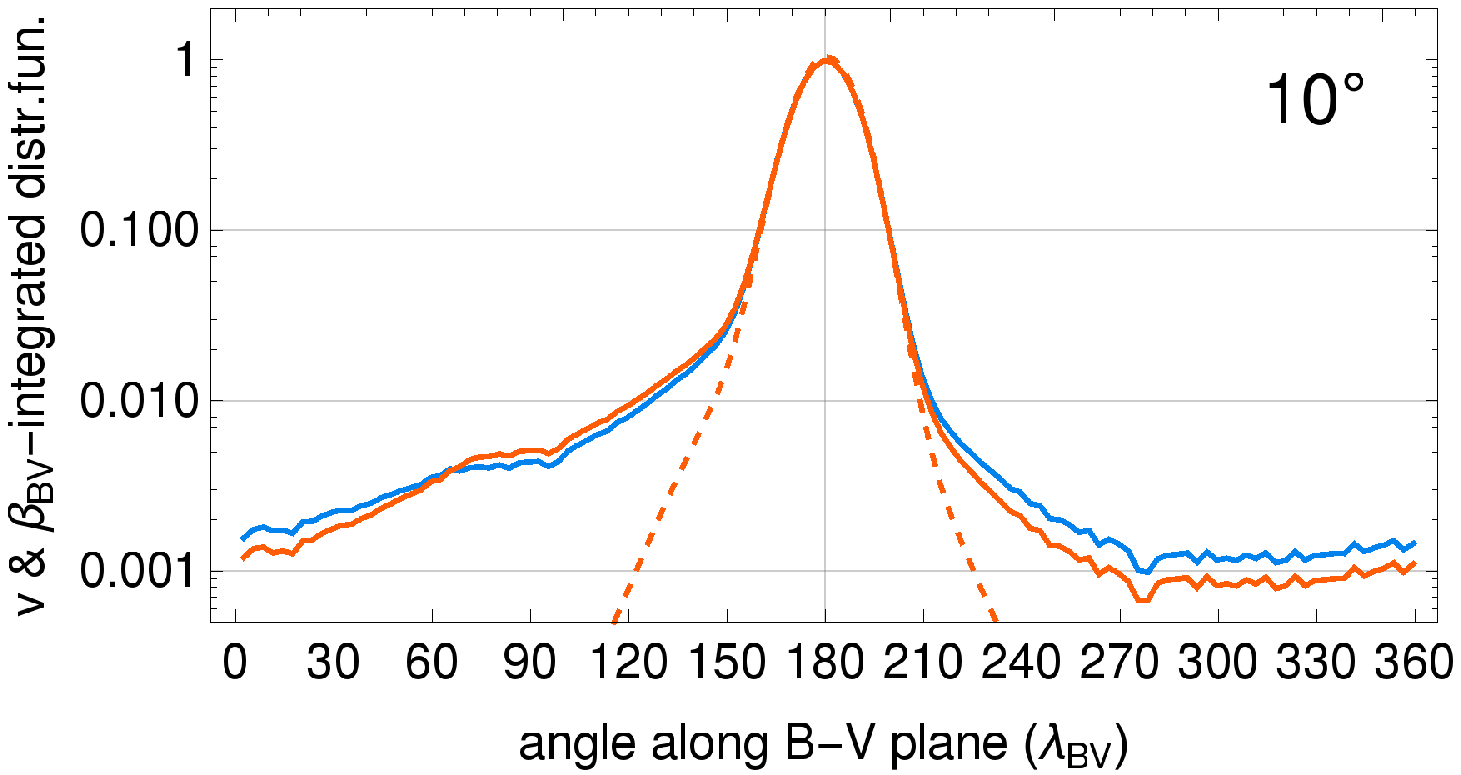}
\plottwo{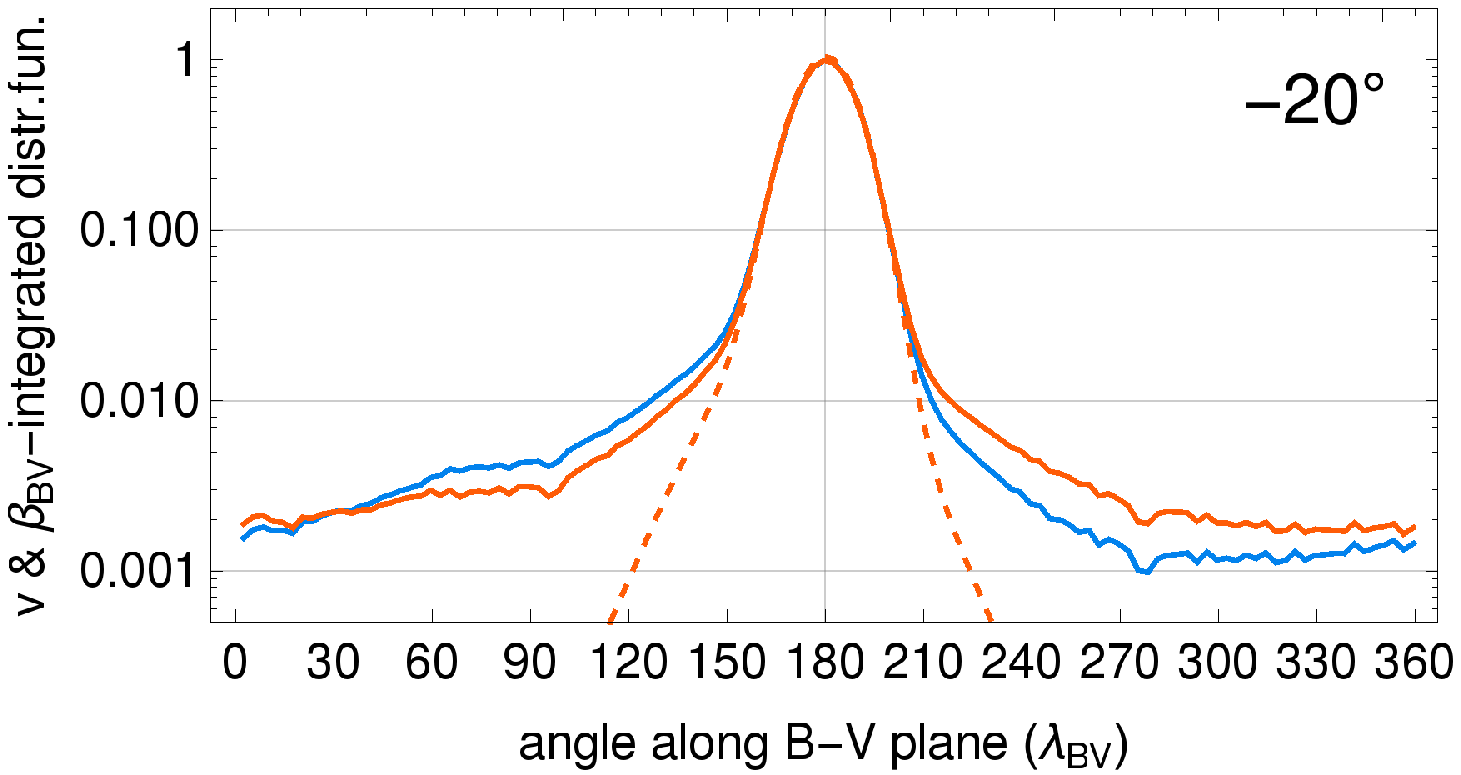}{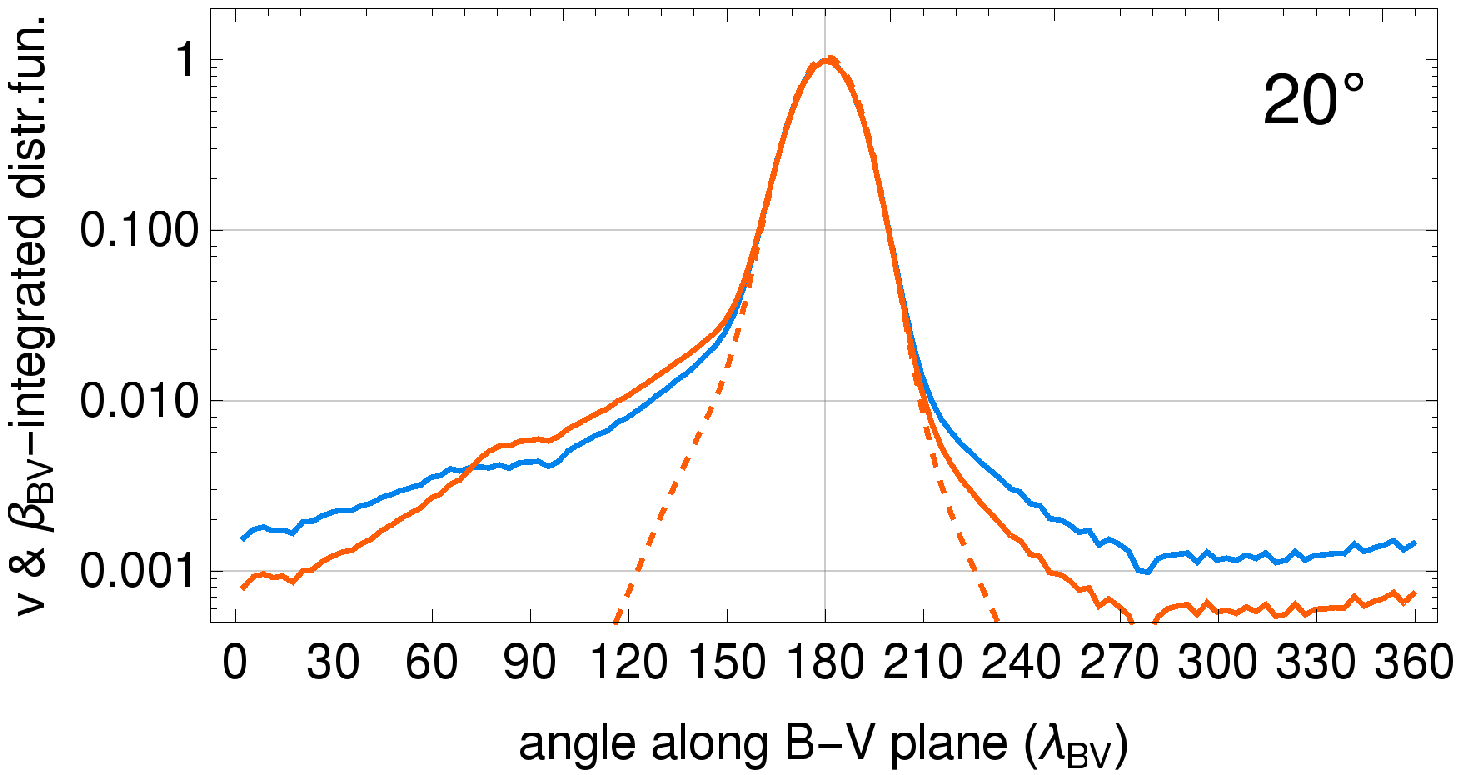}
\plottwo{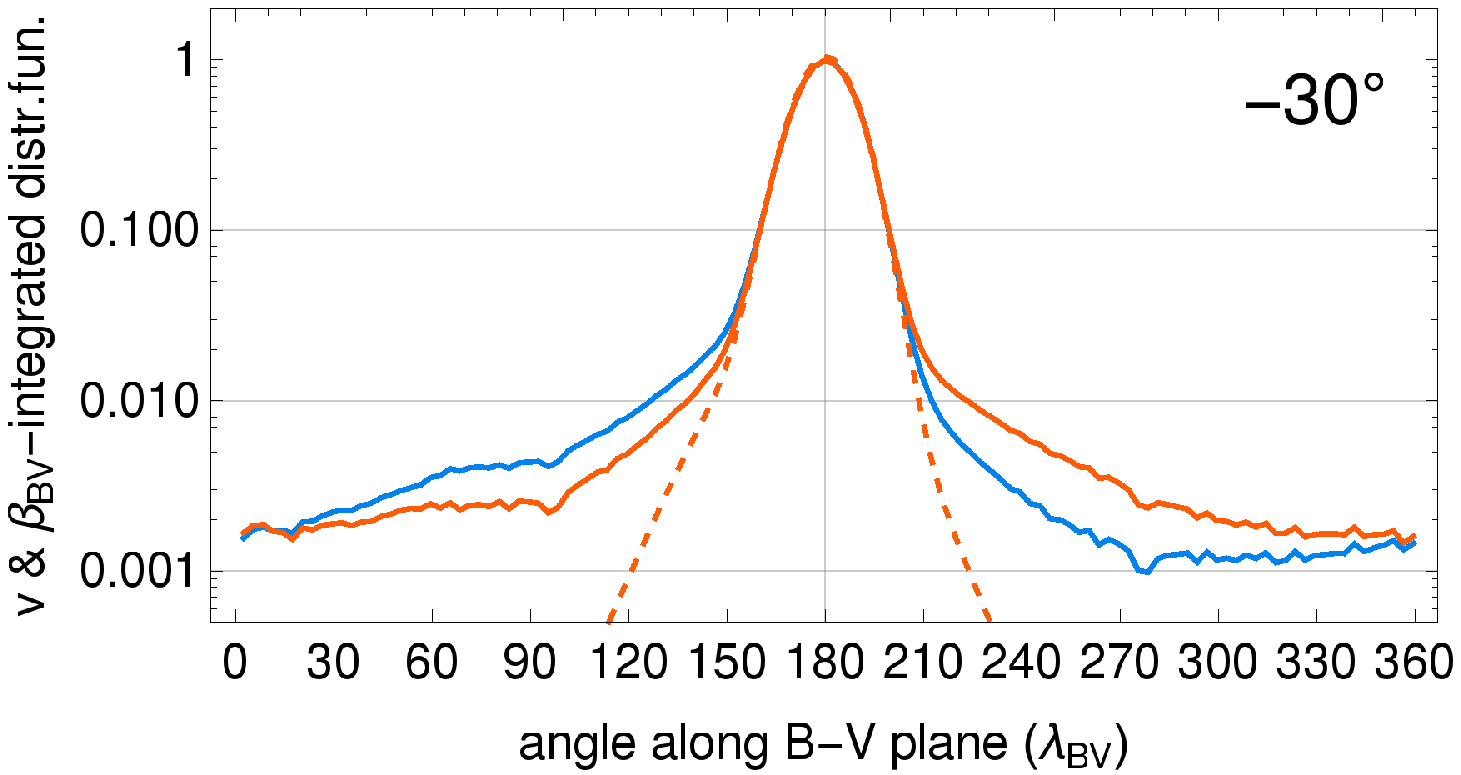}{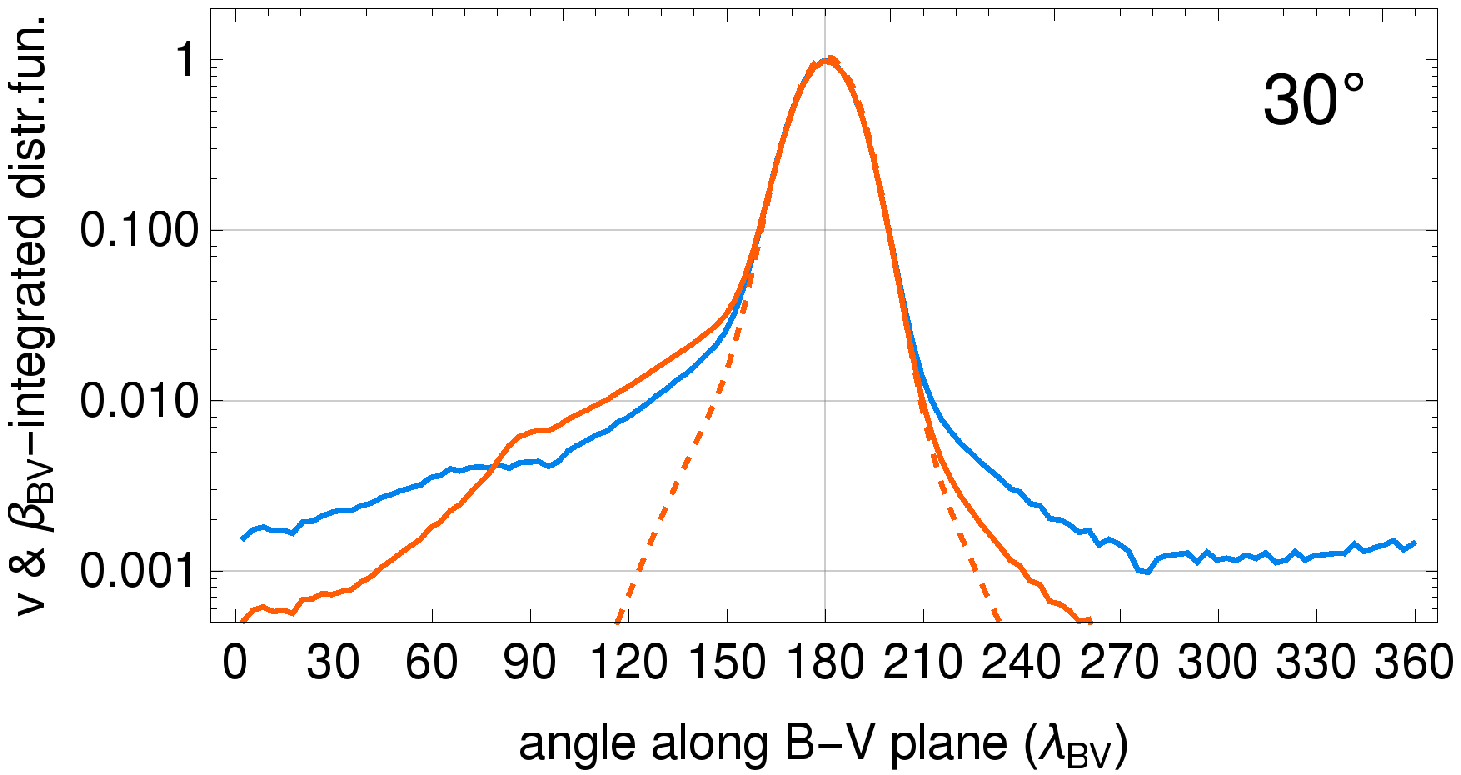}
\plottwo{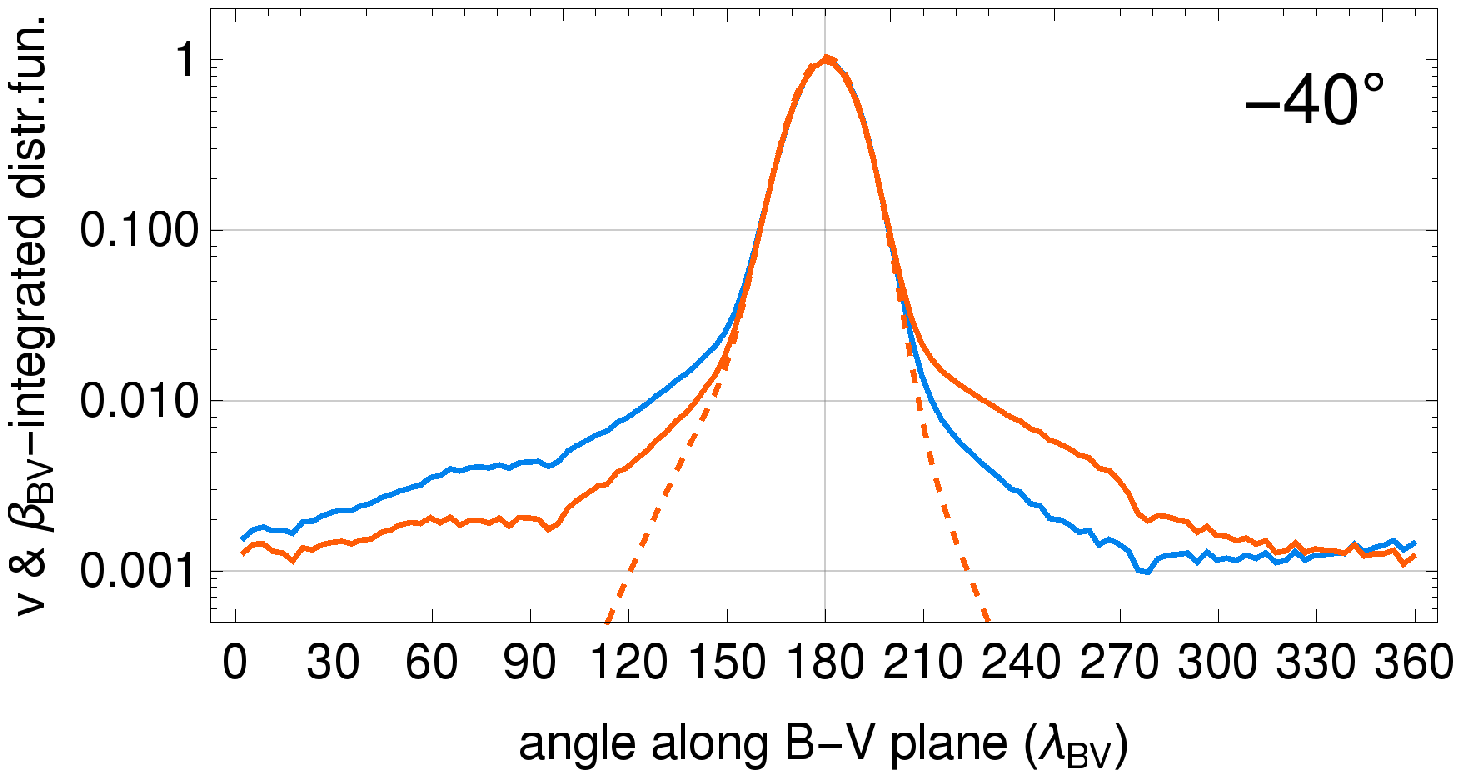}{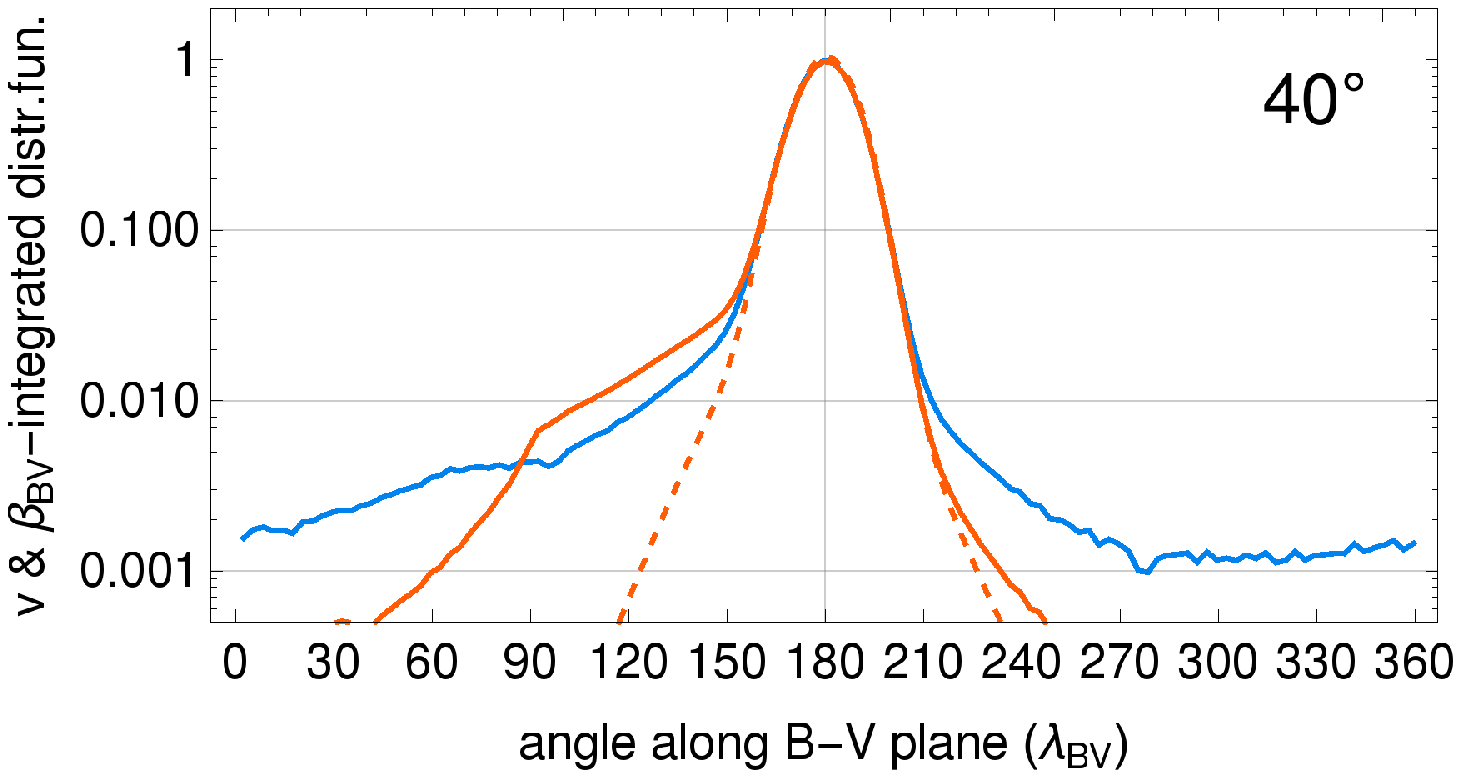}
\caption{Speed- and $\beta_{BV}$-integrated distribution functions $F_{V\lambda}$ for the production and loss processes in the OHS taken into account (solid orange lines), compared with the two-Maxwellian model (broken orange lines), shown for 150~au from the Sun along the B-V plane at offset angles from the upwind direction equal to $\pm 10\degr$ through $\pm 40\degr$. The blue line marks this function for the upwind direction. The functions are normalized to the peak value $F_{V\lambda}$ at 1000~au upwind.} 
\label{fig:df_lateral_BVLongiInt_r150}
\end{figure}

The speed- and BV-latitude integrated distribution function $F_{V\lambda}$ varies with the location along the B-V plane at 150~au, as illustrated in Figure~\ref{fig:df_lateral_BVLongiInt_r150}. Unlike within the two-Maxwellian approximation, the proportions between the left- and right-hand wings considerably vary with the angle along the B-V plane  and the distribution function does not resemble a superposition of two Maxwell-Boltzmann functions at all. The left-right asymmetry of the function varies with the location along the B-V plane and does not seem to disappear for any $\lambda_{BV}$ angle. 

\section{Distribution function at IBEX locations}
\label{sec:difuAtIBEX}
\begin{figure}
\plottwo{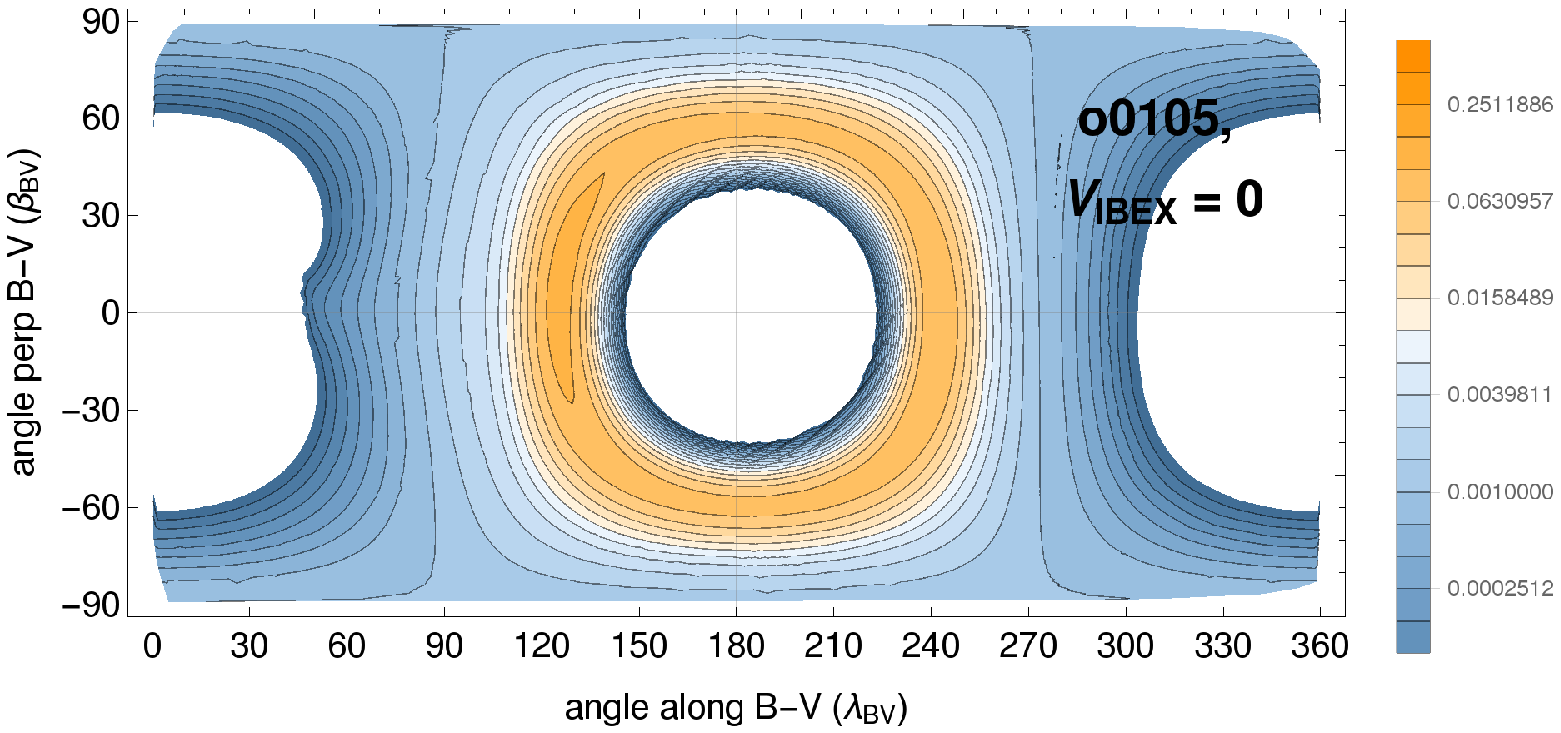}{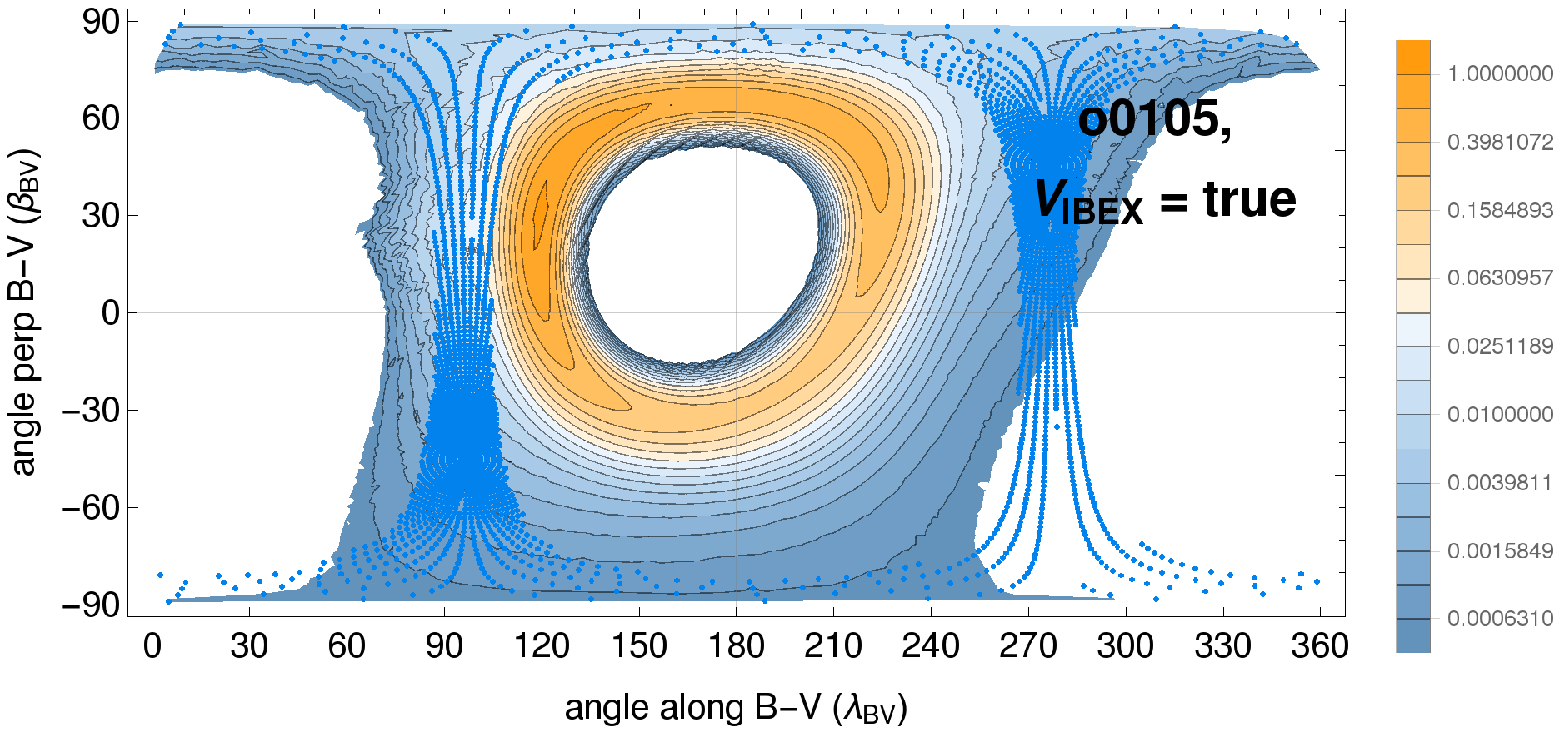}
\plottwo{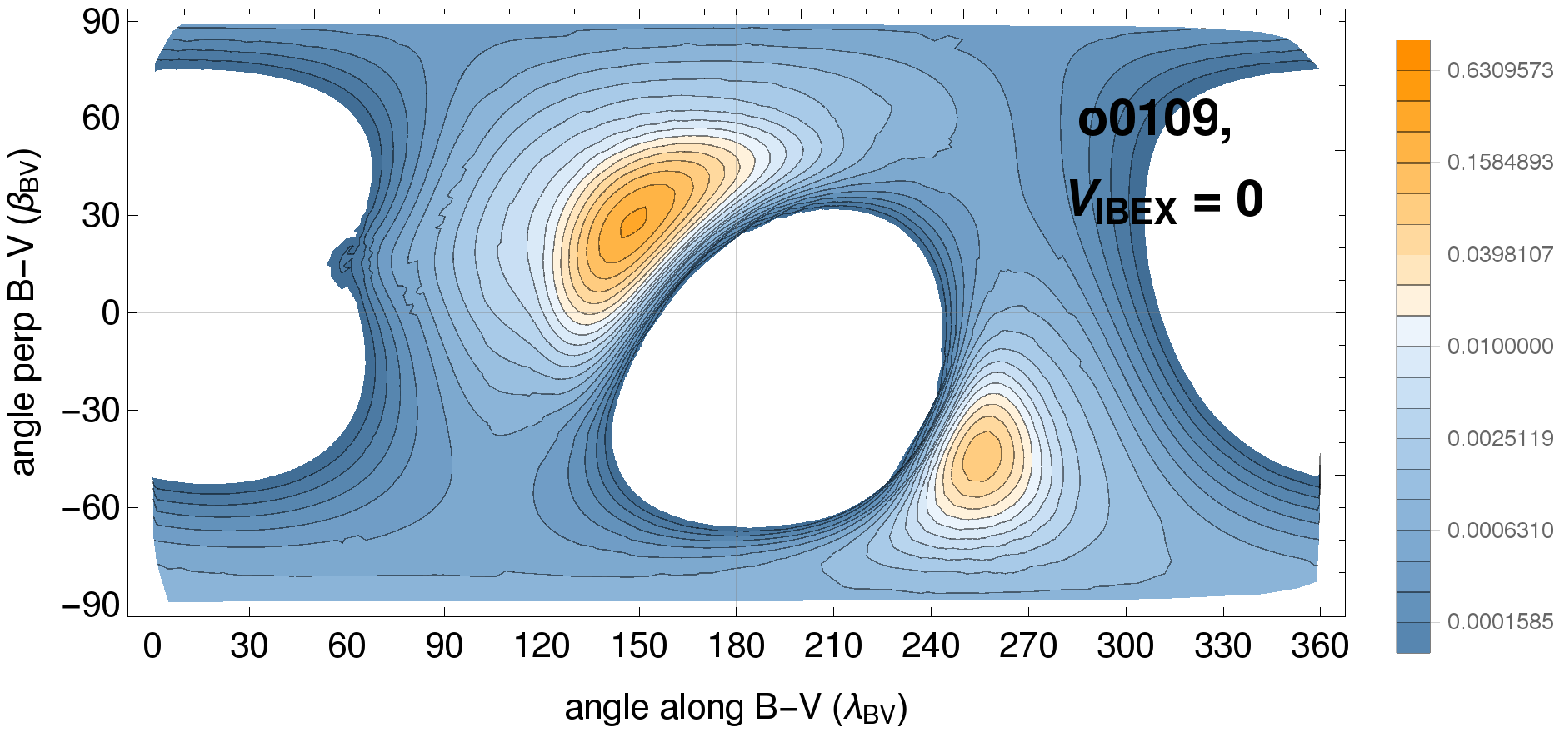}{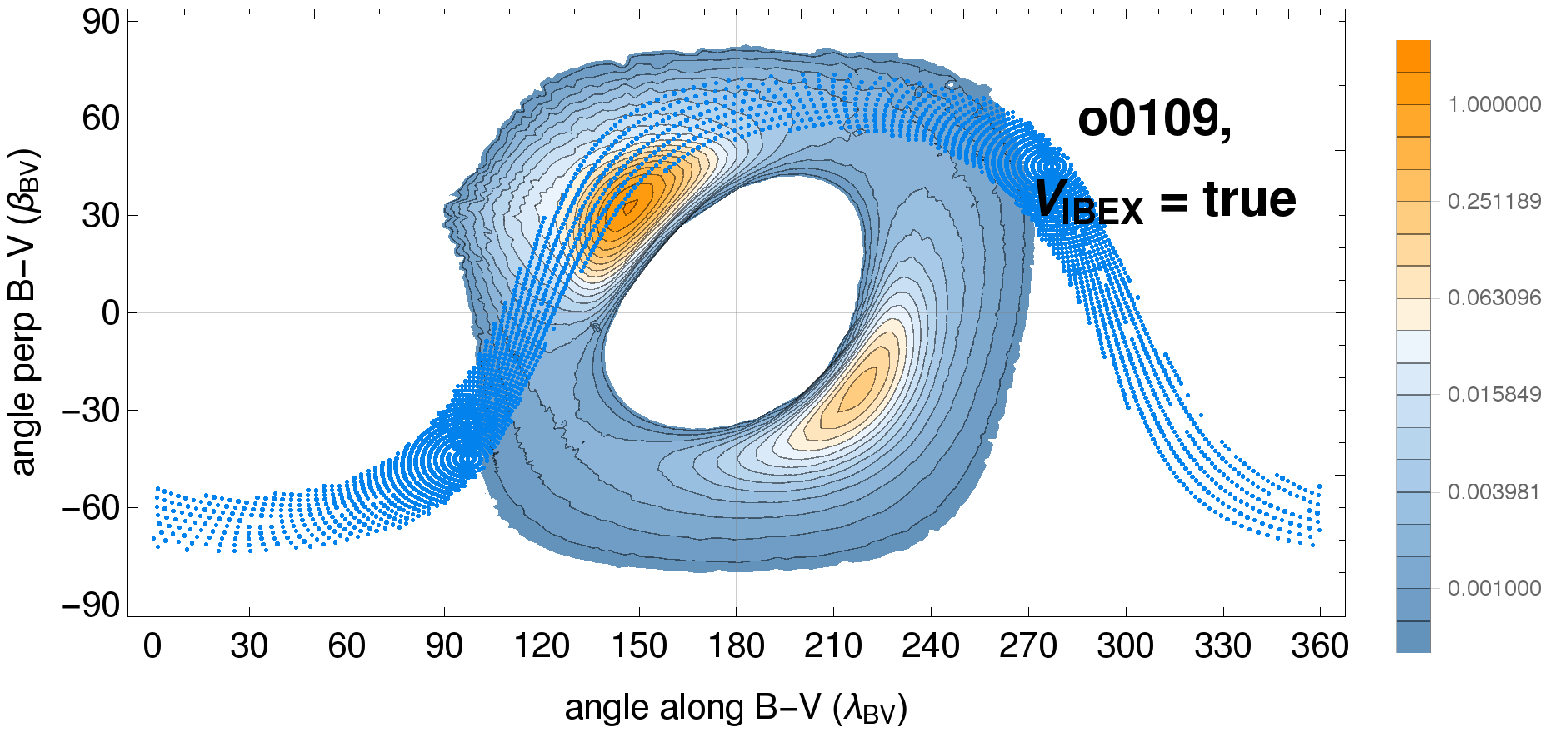}
\plottwo{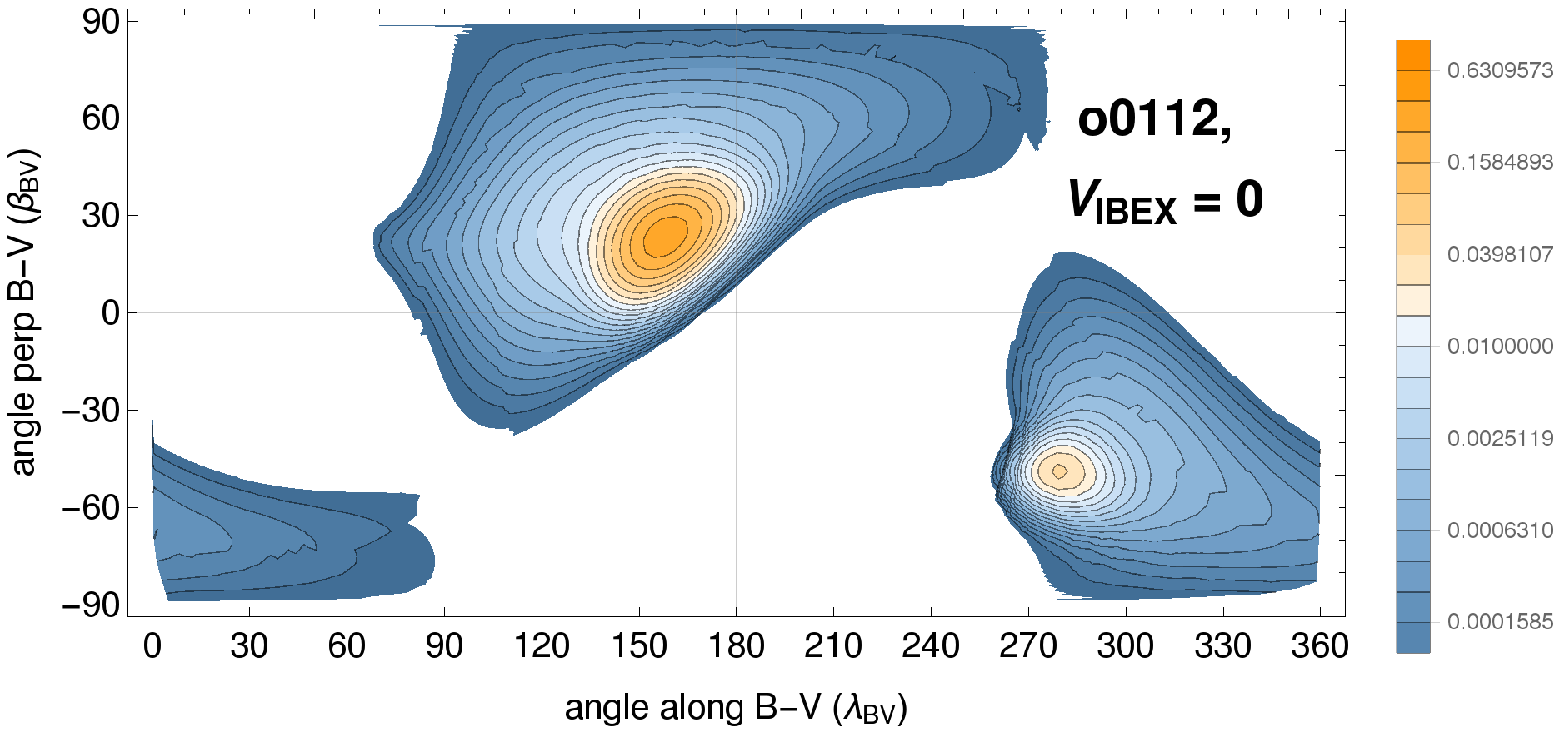}{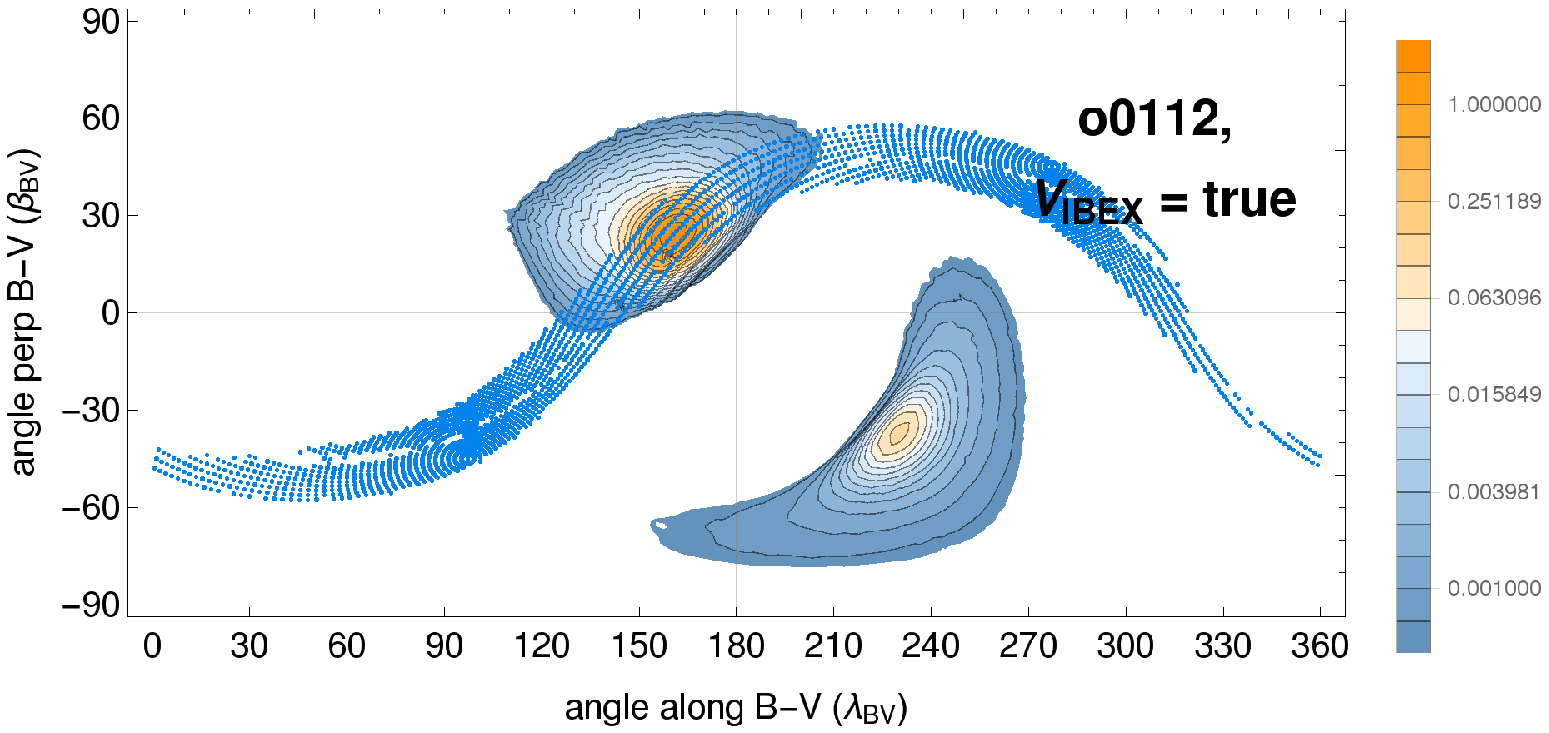}
\plottwo{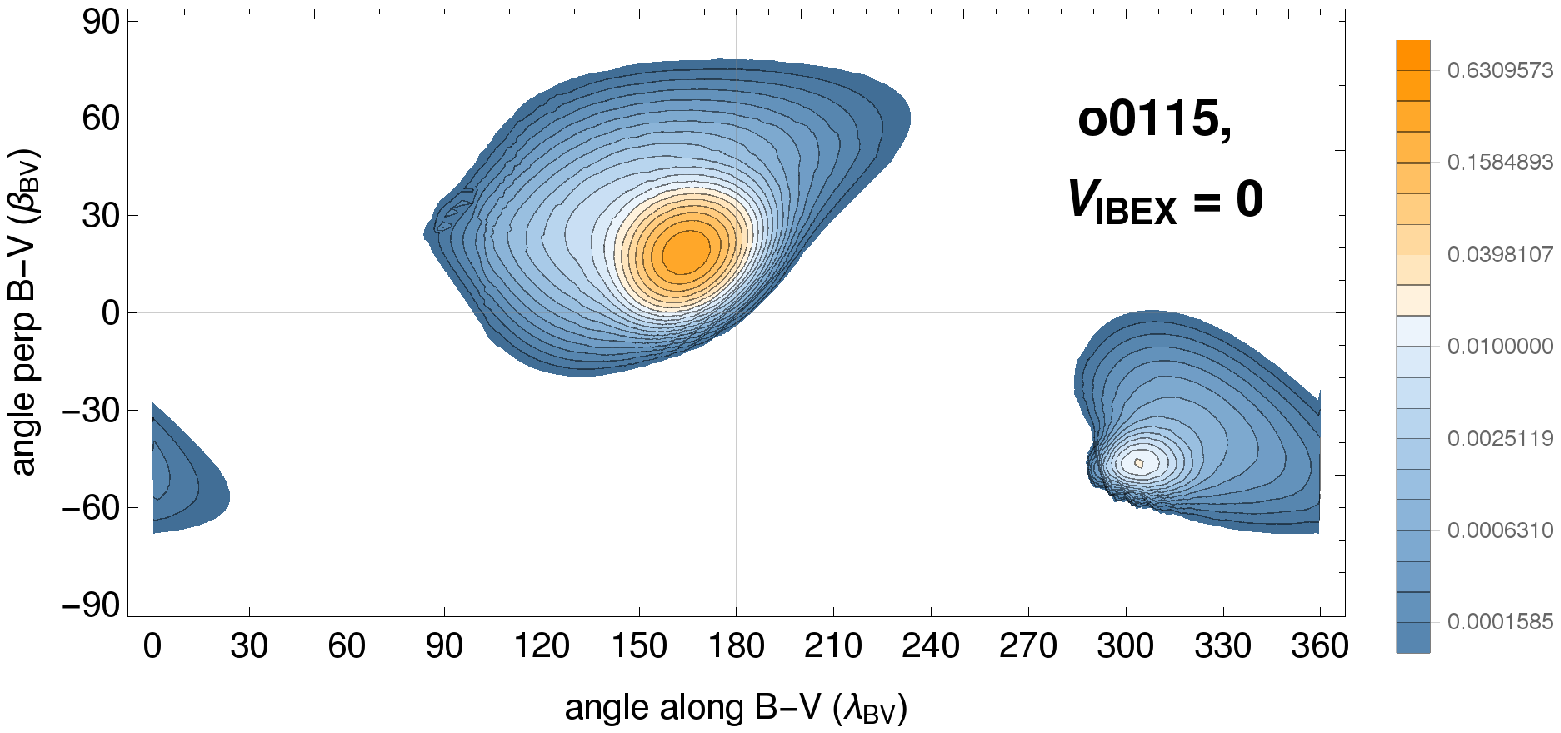}{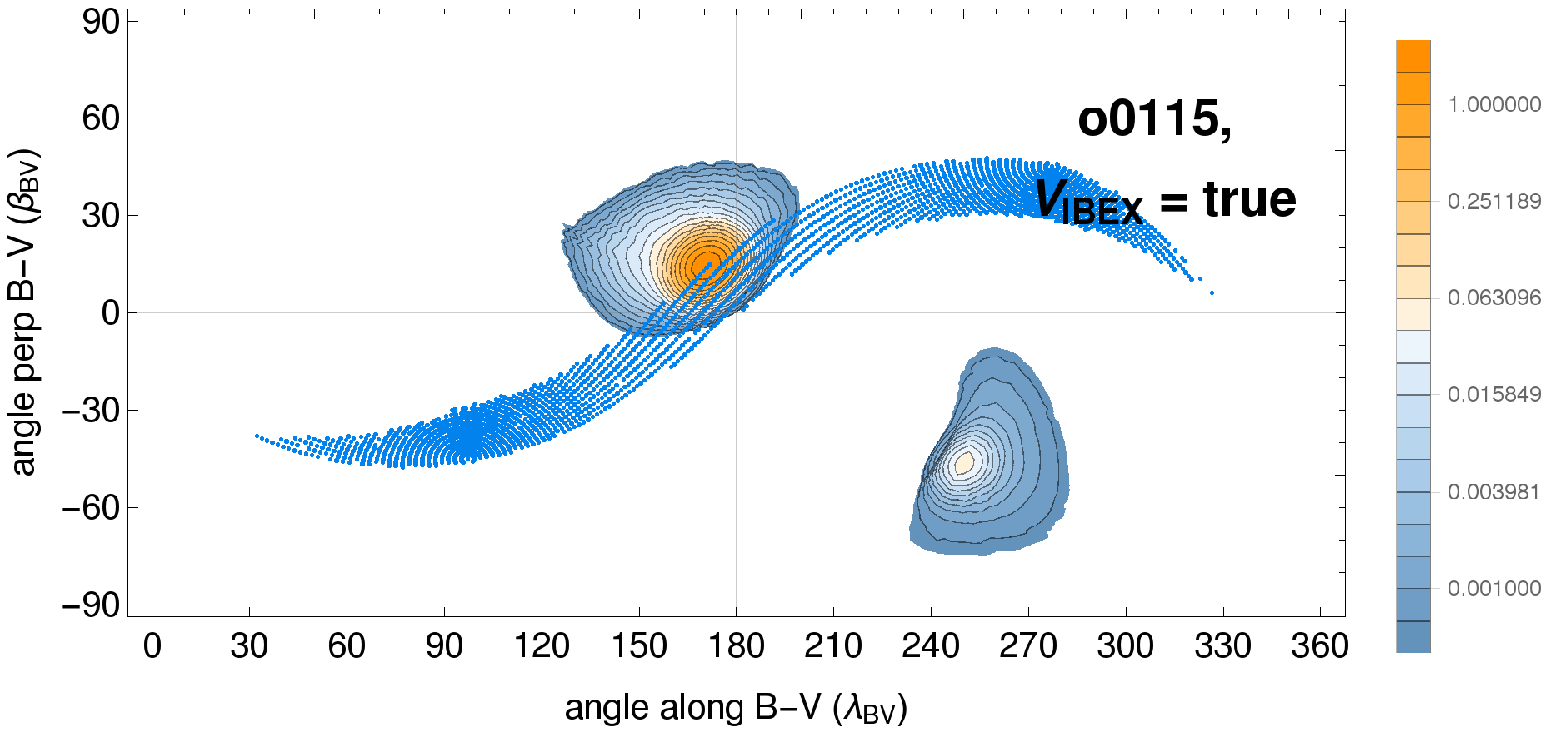}
\caption{Maps of the speed-integrated distribution function $F_V$ of ISN He at IBEX locations for orbits 105, 109, 112, and 115, presented in the B-V coordinates. The left column presents the distribution function in the solar-inertial frame. The right column corresponds to the IBEX-inertial frame, where the function, already distorted by solar gravity, is additionally modified due to spacecraft (and Earth) motion around the Sun. Blue shades in the right column trace the regions in the sky observed by IBEX-Lo in the respective orbits. The direction towards the Sun is $\lambda_{BV} = 0\degr$ throughout.}
\label{fig:difu2DMapsIBEX}
\end{figure}
In this section, we discuss the distribution function along the Earth's orbit in selected locations where IBEX observes ISN He flux. Following \citet{bzowski_etal:17a}, we selected the locations corresponding to IBEX orbits 105, 109, 112, and 115. They are located at ecliptic longitudes 83\degr, 114\degr, 137\degr, and 161\degr, respectively, i.e., in the downwind hemisphere. They correspond to time intervals when IBEX only observes the Warm Breeze (105), when the Warm Breeze and primary ISN He are in comparable proportions (109), when IBEX observes the peak of ISN He (112), and an interval after the peak ISN He, where the Warm Breeze is also detected (115). The location for orbit 105 is offset by only $\sim 8\degr$ after the downwind longitude. We start the discussion with maps of the distribution function in the solar-inertial reference frame and proceed to present the influence of the spacecraft motion.
\begin{figure}
\includegraphics[width=0.32\textwidth]{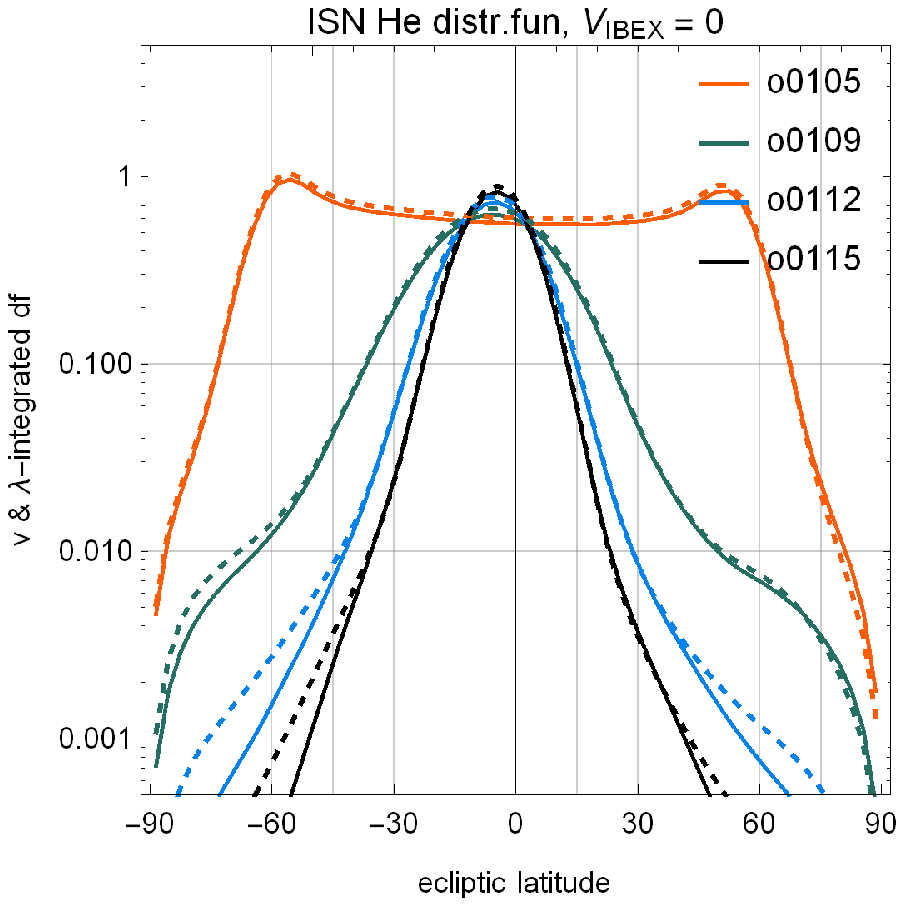}
\includegraphics[width=0.32\textwidth]{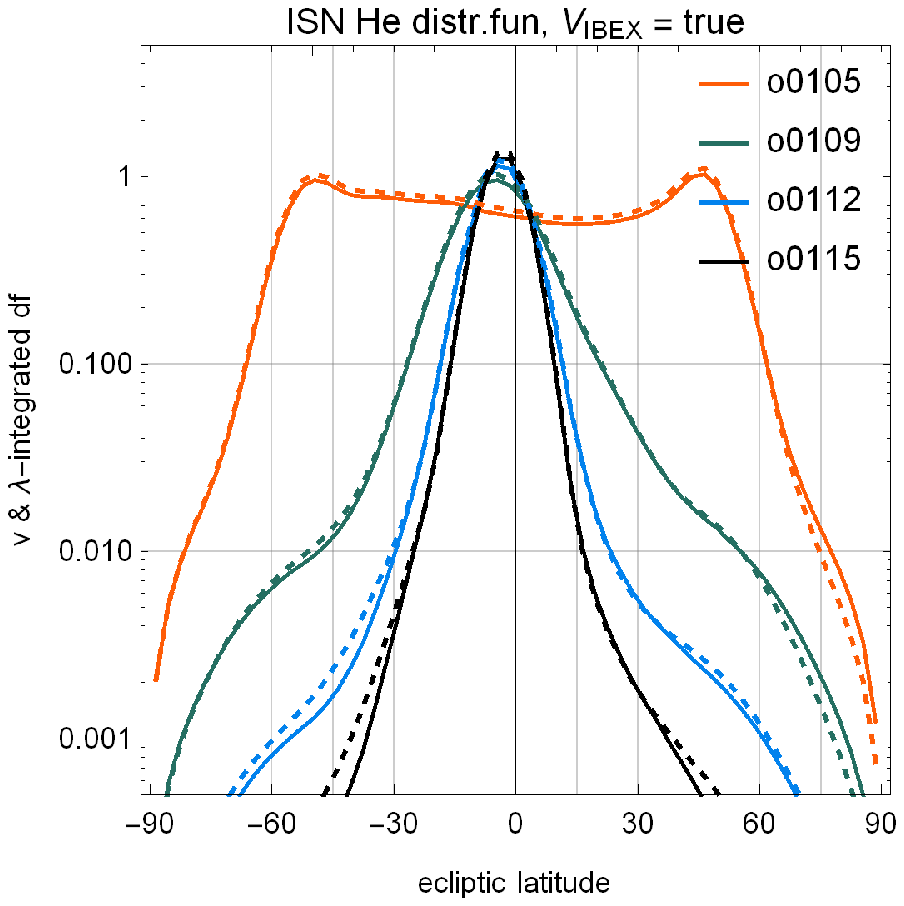}
\includegraphics[width=0.33\textwidth]{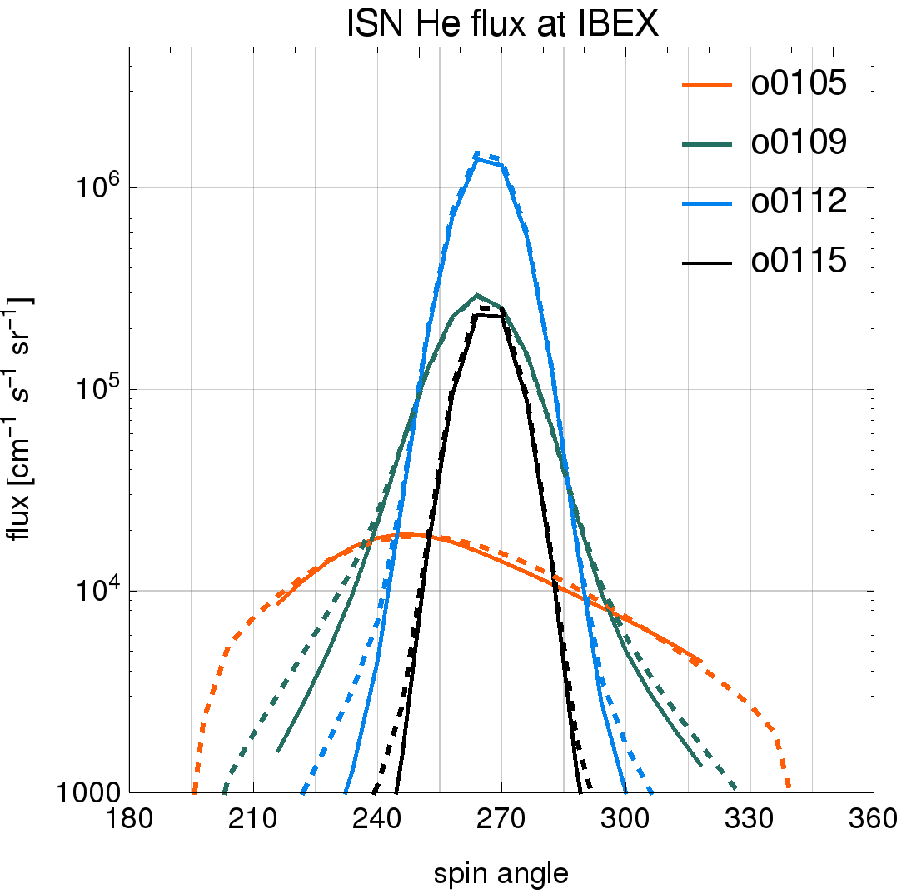}
\caption{Speed- and ecliptic longitude-integrated distribution function $F_{V \beta}$ of ISN He, normalized to the peak value at 1000~au upwind, shown for the locations of IBEX orbits 105, 109, 112, 115. The left-hand panel shows this quantity in the solar-inertial frame and the middle panel in the IBEX-inertial frame. The right-hand panel presents the model flux observed at IBEX, with the good times, binning, and collimator function taken into account. Solid lines represent the model with production and losses in the OHS taken into account, and the broken lines the two-Maxwellian approximation. Note that the horizontal axis in the right-hand panel corresponds to IBEX spin angle, with 270\degr{} corresponding to ecliptic plane. The solid lines in the right-hand panel are cut off at the spin angles corresponding to the data selection in \citet{kubiak_etal:16a}.}
\label{fig:difuLongiIntIBEX}
\end{figure}

\subsection{Solar-inertial frame}
\label{sec:SolInetrialFrame}
In the solar-inertial frame (i.e., in the frame used in the preceding parts of the paper), the geometrically regular form of the distribution function prevailing at 1~au upwind (Figure~\ref{fig:radialUp2D}, the lowest panel in the right-hand column) is strongly distorted (``lensed'') by solar gravity for locations in the downwind hemisphere. In the hot-model paradigm \citep{fahr:78} with one Maxwell-Boltzmann population of ISN He, the flow of the gas is split by solar gravity into two intersecting beams, the direct and the indirect. They are illustrated in \citet{mueller_cohen:12a}. In the downwind region, these beams effectively merge (as shown in the upper row in Figure~\ref{fig:difu2DMapsIBEX}), and a two-dimensional image of the distribution function in this region is similar to volcanic caldera, with a close-to-circular rim. We present maps for the model with production and loss processes in the OHS taken into account. The picture with a dominant direct beam and the indirect beam persists. The strength of the indirect beam is gradually reduced with the decreasing angular distance from the upwind direction. Note that to facilitate assessment of the effects of gravitational distortion of the distribution function, we show the maps in B-V coordinates, as in Figures~\ref{fig:radialUp2D} and \ref{fig:alongBV}. 

Comparison of the full and the two-Maxwellian models at IBEX locations is facilitated with the distribution function integrated over ecliptic longitude (Figure~\ref{fig:difuLongiIntIBEX}, left panel). The differences are relatively small in the core and increase towards the wings, asymmetrically towards the north and south ecliptic latitudes. Generally, the two models are easily discerned only away from ecliptic latitude 0. 

\subsection{IBEX-inertial frame}
\label{sec:IBEXInertialFrame}
IBEX \citep{mccomas_etal:09a} is a spin-stabilized spacecraft in an elongated Earth's orbit. The rotation axis of the spacecraft is periodically adjusted to point within several degree off the Sun and the ecliptic plane. ISN He is observed using the IBEX-Lo neutral-atom camera \citep{fuselier_etal:09b}, with the boresight perpendicular to the spin axis and the field of view defined by a collimator transmission function cutting off a strip in the sky following a close-to-polar great circle $\sim 15\degr$ wide. 
 
The IBEX-inertial frame is defined as an inertial frame that is moving with respect to the Sun with a vector velocity equal to an instantaneous velocity of IBEX, adopted as characteristic for a given IBEX orbit. IBEX-Lo integrates the ISN He flux over speed and a certain directional range in this reference frame. The change in reference frame from the solar-inertial (left column) to the IBEX-inertial right column results in a distortion of the ISN He distribution function maps, as illustrated in Figure~\ref{fig:difu2DMapsIBEX}. Because of the non-zero velocity of IBEX relative to the Sun, the features in the right-hand halves of the maps become smaller in size (beams become taller), while the features in the other halves of the maps become distorted and larger in size compared with the solar-inertial maps (left column). These effects are best visible in orbits 112 and 115 (the two lowest rows of Figure~\ref{fig:difu2DMapsIBEX}). The direct and indirect beams remain present in the maps, but the relations between their dimensions change. 

Due to observation geometry conditions, IBEX is not able to sample the entire distribution function. Rather, it samples close-to-polar strips, differently located in the sky between IBEX orbits. This is illustrated with blue shading in the right-hand column in Figure~\ref{fig:difu2DMapsIBEX}. They are shown as illustrative example. In orbit 105, IBEX entirely misses the peaks of the distribution function and the collected signal is entirely due to the secondary population of ISN He. In orbit 109, the visibility strips passes through a slope of the direct primary beam, but simultaneously, the detector collects plenty of secondary population atoms. In orbit 112, the boresight passes the maximum of the direct primary beam and the secondary population is visible only in the wings. In orbit 115, the visibility strip has moved so that it is at the other slope of the primary beam and the secondary population is little visible. In none of the orbits the indirect beam is anywhere near the field of view.

IBEX integrates the observations so that the ISN He flux is collected along nearly-polar strips. Therefore, an informative view is the ecliptic longitude-integrated distribution function, shown in the two approximations in Figure~\ref{fig:difuLongiIntIBEX}, which shows it in the solar-inertial and IBEX-inertial frames. The change to the IBEX-inertial frame results in the peaks being narrower, but the general conclusion holds. The largest differences between the full and two-Maxwellian model only appear in the wings of the distribution. It is interesting, however, that the ecliptic latitude profile portions where the deviations appear are different in the north and south hemispheres. This is especially well visible for orbits 112 and 115, i.e., those where the primary population is very strong. Generally, the two-Maxwellian approximation, while not perfect, is still a very reasonable one. These comparison conclusions qualitatively hold for the IBEX flux, shown in the right-hand panel of Figure~\ref{fig:difuLongiIntIBEX}. The model of IBEX flux is directly comparable \citep[after re-scaling by the geometric factor,][]{swaczyna_etal:18a} with the IBEX counting rates. Typically, the observation background level is at $\sim (8-10)\times 10^{-4}$ of the ISN peak magnitude \citep{galli_etal:15a}, so the figure presents the dynamical range of the usable signal. The differences between the results of the synthesis method and the two-Maxwellian model appear mostly in the wings of the signal. 

Finally, we compared the total He densities calculated using the distribution function synthesis method with those obtained from the two-Maxwellian approximation along the radial upwind and downwind directions up to the heliopause and along a 1~au circle around the Sun. In the two-Maxwellian approximation, the abundance of the secondary He population around 1 au is larger than that adopted at the source region, except a narrow conical region centered at the downwind axis. We found that the densities calculated using the synthesis method and the two-Maxwellian approximation agree with each other within $\sim 2$\%, i.e., within the numerical accuracy of the nWTPM model. Hence we conclude that the two-Maxwellian approximation is useful to obtain the ISN He densities inside the heliopause with an accuracy close to the accuracy of state of the art  models of ISN gas inside the heliosphere.

\section{Discussion and conclusions}
\label{sec:discussion}
In our previous papers we analyzed available direct-sampling observations of ISN He by IBEX \citep{bzowski_etal:12a, bzowski_etal:15a, kubiak_etal:14a, kubiak_etal:16a} and Ulysses \citep{bzowski_etal:14a} to determine macroscopic parameters of ISN He in the boundary region of the heliosphere and Very Local Interstellar Matter. From the differences between the inflow directions of the Warm Breeze and the ISN gas we concluded that the heliosphere must be asymmetric due to the action of the interstellar magnetic field. The plane in space determined by these directions turned out to coincide with the plane defined by the inflow velocity vector of ISN He and the direction of interstellar magnetic field (i.e., the B-V plane) determined by \citet{zirnstein_etal:16b}. In this paper, we reverse the approach: we assume a model of the heliosphere interaction with the VLISM with a distortion due to the interstellar magnetic field. We analyze features of the model distribution function of ISN He to understand their relation with features of the ISN He flux observable by direct sampling at 1~au. 

Within the adopted model, the secondary population of ISN He forms in the OHS and in a region within $\pm 60\degr$ around the upwind direction. We find that the distribution function of He in this region is symmetrical relative to the B-V plane but asymmetrical relative to both upwind direction and the direction of unperturbed interstellar magnetic field. The two-Maxwellian approximation overestimates the secondary population at far distances from the Sun, where the production of the secondary atoms is at a very small level, i.e., outside $\sim 250$~au. Between $\sim 250$ and $\sim 50$ au, the two-Maxwellian model drastically underestimates the abundance and angular distribution of the secondary population. The core of the distribution function, representing the primary population, is reproduced correctly by the two-Maxwellian model. Still, using this approximation for the distribution function of the secondary ISN He within the OHS is strongly discouraged. 

Inside the heliopause, where the production of secondary He atoms ceases, due to selection effect the two-Maxwellian approximation becomes increasingly good. The model distribution functions differ only in relatively far wings. We have verified that also the first moments of the distribution functions, i.e., the densities, agree within the numerical accuracy of the nWTPM model, which is the basis for our simulations. Hence for these purposes and in these regions, the two-Maxwellian approximation is handy for the calculation of moments of the ISN He distribution function. 

\citet{bzowski_etal:19a} showed that the model with production and losses of ISN He atoms in the OHS explicitly taken into account gives a better agreement with IBEX observations than the two-Maxwellian model. We showed that in some portions of the IBEX flux vs spin angle relation, significant differences between the two models exist. They can be verified using observations with a sufficiently high statistics. We speculate that they might be more easily resolved with a different geometry of observations, like this planned for the IMAP-Lo experiment \citep{mccomas_etal:18b}. This would largely facilitate studying the plasma flow in the OHS using direct-sampling observations of ISN He. 

The two-Maxwellian approximation is sufficient to calculate the density of ISN He in the inner heliosphere to model pickup ion distributions, heliospheric backscatter glow, and the production of energetic neutral atoms of helium. The inflow direction of the secondary population obtained in this approximation from analysis of direct-sampling measurements of ISN He is suitable to determine the plane defined by the unperturbed VLISM magnetic field and the direction of Sun's motion through the VLISM, but the fit temperature and speed are not directly comparable with the plasma temperature and speed in the OHS.

\acknowledgments{The authors are obliged to Jacob Heerikhuisen and Eric Zirnstein for sharing the model of the plasma in the OHS. This research was supported by Polish National Science Center grant 2015/18/M/ST9/00036.}

\bibliographystyle{aasjournal}
\bibliography{ms}
\end{document}